\documentclass[10pt, a4paper]{article}

\usepackage[british]{babel}

\usepackage[a4paper,top=2cm,bottom=2cm,left=2.5cm,right=2.5cm,marginparwidth=1.75cm]{geometry}

\usepackage[backend=bibtex, style=numeric-comp, sorting=none]{biblatex}
\usepackage{multirow}
\usepackage{mathrsfs}
\usepackage{amsmath}
\usepackage{graphicx}
\usepackage{adjustbox}
\usepackage[table]{xcolor}
\usepackage[colorlinks=true, allcolors=black]{hyperref}
\usepackage{hyperref}
\usepackage{orcidlink}
\usepackage[title]{appendix}
\usepackage{mathrsfs}
\usepackage{amsfonts}
\usepackage{mathtools}
\usepackage{booktabs} 
\usepackage{caption}  
\usepackage{comment}
\usepackage{threeparttable} 
\usepackage{algorithm}
\usepackage{algorithmicx}
\usepackage{algpseudocode}
\usepackage{listings}
\usepackage{enumitem}
\usepackage{chngcntr}
\usepackage{booktabs}
\usepackage{lipsum}
\usepackage{subcaption}
\usepackage{authblk}
\usepackage[T1]{fontenc}    
\usepackage{csquotes}       
\usepackage{diagbox}
\usepackage{physics}
\usepackage{amssymb}
\usepackage{tikz}
\usepackage{array}
\usepackage{orcidlink}
\usepackage{siunitx}
\usepackage{upgreek}

\usepackage{titlesec}
\titleformat{\section} 
  {\normalfont\Large\bfseries}{\thesection.}{1em}{}
  
\usepackage{lineno} 

\rightlinenumbers 

\usepackage{float}   
\usepackage{caption} 
\usepackage{textcomp}

\usepackage[acronym]{glossaries}
\newacronym{cfd}{CFD}{computational fluid dynamics}
\newacronym{cp}{CP}{conformal prediction}
\newacronym{fem}{FEM}{finite element method}
\newacronym[longplural={Gaussian processes}]{gp}{GP}{Gaussian process}
\newacronym{iga}{IGA}{isogeometric analysis}
\newacronym{igbt}{IGBT}{insulated gate bipolar transistor}
\newacronym{lar}{LAR}{least angle regression}
\newacronym{loo}{LOO}{leave-one-out}
\newacronym{ml}{ML}{machine learning}
\newacronym{nn}{NN}{neural network}
\newacronym{nurbs}{NURBS}{non-uniform B-splines}
\newacronym{ols}{OLS}{ordinary least squares}
\newacronym{pce}{PCE}{polynomial chaos expansion}
\newacronym{pdf}{PDF}{probability density function}
\newacronym[longplural={quantities of interest}]{qoi}{QoI}{quantity of interest}
\newacronym{rmse}{RMSE}{root mean square error}
\newacronym{uq}{UQ}{uncertainty quantification}

\usepackage{setspace}
\title{Conformalized Polynomial Chaos Expansion for Uncertainty-aware Surrogate Modeling}

\author[1,*]{Dimitrios Loukrezis \orcidlink{0000-0003-1264-1182}}
\author[2]{Dimitris G. Giovanis \orcidlink{0000-0003-2272-2584}}
\affil[1]{\small{Centrum Wiskunde \& Informatica, Scientific Computing, Amsterdam, The Netherlands}}
\affil[2]{\small{Johns Hopkins University, Department of Civil \& Systems Engineering, Baltimore, USA}}
\affil[*]{\small{Corresponding author: \texttt{d.loukrezis@cwi.nl}}}

\date{}  

\begin{document}
\maketitle

\begin{abstract}
\noindent This work introduces a method to equip data-driven polynomial chaos expansion surrogate models with intervals that quantify the predictive uncertainty of the surrogate. 
To that end, jackknife-based conformal prediction is integrated into regression-based polynomial chaos expansions. 
The jackknife algorithm uses leave-one-out residuals to generate prediction intervals around the predictions of the polynomial chaos surrogate. 
The jackknife+ extension additionally requires leave-one-out model predictions.
Both methods allow using the entire dataset for model training and do not require a hold-out dataset for prediction interval calibration.
The key to efficient implementation is to leverage the linearity of the polynomial chaos regression model, so that leave-one-out residuals and, if necessary, leave-one-out model predictions can be computed with analytical, closed-form expressions. 
This eliminates the need for repeated model re-training. 
The conformalized polynomial chaos expansion method is first validated on benchmark models and then applied to an engineering use-case. 
The method produces prediction intervals that provide the target coverages, even for low-accuracy models trained with small datasets. 
At the same time, training data availability plays a crucial role in improving the empirical coverage and narrowing the prediction interval, as well as in reducing their variability over different training datasets. \\

\noindent \textbf{Keywords}: conformal prediction, jackknife, polynomial chaos expansion, regression, surrogate modeling, uncertainty quantification.  
\end{abstract}

\section{Introduction}
\label{sec:intro}

Originally developed as an efficient \gls{uq} method \cite{spanos1989stochastic, xiu2002wiener, ghanem2003stochastic}, the \gls{pce} is in recent years routinely employed as a \gls{ml} regression model \cite{torre2019data}. 
Within this context, regression-based \glspl{pce} are often used as data-driven surrogate models in various science and engineering applications \cite{haghi2022surrogate, loukrezis2022power, campos2023polynomial, manfredi2024nonparametric, dai2025cost, lee2025surrogate}. 
Their use has also been extended to scientific \gls{ml} tasks like manifold learning \cite{soize2017polynomial, kontolati2022manifold, kontolati2022survey, giovanis2024polynomial}, operator learning \cite{sharma2025polynomial}, and physics-informed \gls{ml} \cite{novak2024physics, sharma2024physics}. 

In contrast to the use of \glspl{pce} for \gls{uq} purposes, \gls{pce}-based data-driven predictive models are purely deterministic. That is, they yield point predictions that lack any type of uncertainty or confidence metrics. 
\textcolor{black}{Bayesian regression models \cite{ranftl2021bayesian} natively provide such predictive uncertainty metrics. 
Examples include \glspl{gp} \cite{williams2006gaussian}, Bayesian \glspl{nn} \cite{jospin2022hands}, but also Bayesian \glspl{pce} \cite{shao2017bayesian, li2025new, rumsey2026bayesian}.
Considering frequentist approaches,} ensemble modeling and bootstrapping are common ways of providing predictive uncertainty intervals to data-driven models, including \glspl{pce} \cite{marelli2018active, loukrezis2022power, shang2024active}. 
\textcolor{black}{
Other relevant approaches include combined \gls{pce}-\gls{gp} models, where earlier works embed the \gls{pce} in the trend of a \gls{gp} model \cite{schobi2015polynomial, kersaudy2015new}, whereas more recently suggested hybrid \gls{pce}-\gls{gp} models \cite{manfredi2025hybrid, manfredi2026polynomial} allow for propagating the \gls{gp} uncertainty directly to the \gls{pce} coefficients. 
Other alternatives include stochastic \gls{pce} formulations \cite{zhu2023stochastic}.}
However, these approaches induce either specific modeling limitations (\textcolor{black}{e.g., Gaussianity assumptions in Bayesian approaches}) or significant computational overhead (\textcolor{black}{e.g., due to model retraining in bootstrapping and ensemble modeling}).

Considering predictive \gls{uq} for \gls{ml} models, \gls{cp} \cite{vovk2005algorithmic, fontana2023conformal} is an established method for producing statistically valid prediction intervals given a point predictor, only under the assumption of data exchangeability.
The resulting predictive models are commonly called ``conformalized'' \cite{burnaev2016conformalized, romano2019conformalized, huang2023uncertainty, podina2024conformalized, jaber2025conformal}.
\gls{cp} is a distribution-free \gls{uq} method, meaning that it does not require any assumptions about the underlying data distribution or the structure of the data-driven model. 
Additionally, certain \gls{cp} methods
provide statistically rigorous, finite-sample guarantees regarding prediction interval coverage.
On the downside, \gls{cp} does not yield full probability distributions like Bayesian methods do. 
Moreover, its coverage is marginal instead of conditional, hence possibly under- or overconfident in different regions of the input space.
Adaptive \gls{cp} methods that address the latter issue are actively pursued \cite{zaffran2022adaptive, deutschmann2024adaptive}.
Data demand can also be an issue, in particular considering so-called inductive or split \gls{cp} methods \cite{papadopoulos2002inductive, lei2018distribution}, which require that part of the available data is excluded from training and reserved for prediction interval estimation.
Remedies have been sought in so-called cross \gls{cp} approaches leveraging cross-validation techniques \cite{vovk2015cross, vovk2018cross}, including jackknife-based variants that utilize \gls{loo} residuals \cite{lei2018distribution, barber2021predictive}.  
Similar to bootstrapping and ensemble modeling, the downside of these methods is the need for repeated model retraining, which can result in undesired computational overhead.

\textcolor{black}{
This work proposes the combination of regression-based \glspl{pce} with \gls{cp} based on the jackknife method \cite{lei2018distribution} and its jackknife+ extension \cite{barber2021predictive}, such that the resulting conformalized \gls{pce} models are equipped with intervals quantifying their predictive uncertainty.
Its main contribution is demonstrating that the linearity of the \gls{pce} regression model allows the computation of the \gls{loo} residuals and \gls{loo} model evaluations used in jackknife-based \gls{cp} with analytic, closed-form expressions.
This bypasses the need to repeatedly re-train the \gls{pce}, as is typically necessary for jackknife-based \gls{cp}, thus effectively eliminating the computational overhead of predictive \gls{uq}.
Additionally, the suggested approach belongs to the class of cross \gls{cp} methods, and as such allows the full available dataset to be used for training, which is a crucial advantage in small-data learning regimes. 
We note that, combining \gls{cp} with data-driven, regression-based surrogate models has been explored before, e.g., for \glspl{gp} \cite{jaber2025conformal} or various \glspl{nn} architectures \cite{gopakumar2024uncertainty}.
To the best knowledge of the authors, this is the first work to develop a conformalized \gls{pce} method where \gls{pce} regression models are equipped with predictive \gls{uq} intervals based on \gls{cp}.}\footnote{In the meantime, another work on conformalized \glspl{pce} has appeared in the literature \cite{hatstatt2026conformal}. This work partially overlaps with ours, however, it was developed independently.}  
 
The method is implemented for total-degree \glspl{pce} utilizing \gls{ols} for the computation of the \gls{pce} coefficients.
\textcolor{black}{
Other types of bases, e.g., corresponding to sparse \glspl{pce}, can also be considered, however, for the aforementioned analytical computation of the jackknife-based \gls{cp} intervals to be possible, the final \gls{pce} basis must be held fixed and the coefficients must be computed via \gls{ols} (see section~\ref{sec:conf-pce}).}
The method is first validated on benchmark models and then applied to an engineering use-case concerning the design of a Stern-Gerlach electromagnet \cite{masschaele2011design}.
In all numerical experiments, we assess whether the conformalized \gls{pce} surrogate provides predictive uncertainty intervals that result in empirical coverages similar to the required target levels. 
Additionally, we investigate the impact of different conformalized \gls{pce} configurations, with an emphasis on training data availability.
 
The remainder of this paper is organized as follows.
Section \ref{sec:methodology} presents the methodology behind the suggested method, first recalling the necessary prerequisites of \gls{pce} and \gls{cp} methods, and then presenting the conformalized \gls{pce}.
The numerical experiments are presented in section~\ref{sec:num-exp}.
Section \ref{sec:conclusion} summarizes the findings of this work and discusses follow-up research opportunities.

\section{Methodology}
\label{sec:methodology}
We begin by presenting the \gls{pce} regression model in section~\ref{sec:pce}, followed by a 
presentation of \gls{cp} in section \ref{sec:cp}. 
The latter focuses on the jackknife and jackknife+ \gls{cp} methods. 
Section~\ref{sec:conf-pce} shows how \gls{loo} residuals and predictions of \gls{pce} regression models can be computed with analytical, closed-form expressions, as to enable the cost-effective estimation of prediction intervals based on jackknife and jackknife+ \gls{cp}.

\subsection{Regression-based polynomial chaos expansion}
\label{sec:pce}

In the following we consider a computational model, denoted as $\mu(\mathbf{x})$, which receives an input vector $\mathbf{x} \in \mathbb{R}^N$ and computes a scalar output $y \in \mathbb{R}$, such that $y = \mu\left(\mathbf{x}\right)$.
While scalar outputs are considered throughout this work, vector-valued outputs can also be considered by an element-wise application of all operations described next.
The form of the model can range from a simple analytical function to a complex numerical solver. 
The only assumption is that the model is deterministic, i.e., the same output $y$ is observed given the same input $\mathbf{x}$.

Next, we introduce a multivariate random variable (or random vector) $\mathbf{X} = \left(X_1, \dots, X_N\right)$, where the individual random variables $X_n$, $n=1,\dots,N$, are assumed to be mutually independent.
The random vector is defined on the probability space $\left(\Omega, \Sigma, P \right)$, where $\Omega$ is the sample space containing the outcomes $\omega \in \Omega$, $\Sigma$ denotes the $\sigma$-algebra of events (measurable grouped outcomes), and $P: \Sigma \rightarrow [0,1]$ is a probability measure. 
The latter is defined based on the assumed \gls{pdf} $f_{\mathbf{X}}: \Xi \rightarrow \mathbb{R}_{\geq 0}$, where $\Xi \subset \mathbb{R}^N$ denotes the state-space of the random vector. 
Then, the random vector is a map, $\mathbf{X}: \Omega \rightarrow \Xi$, with random realizations $\mathbf{x} = \mathbf{X}\left(\omega\right) \in \Xi$.

Next, we assume that the random realizations $\mathbf{x} = \mathbf{X}\left(\omega\right)$ coincide with the inputs of the computational model $\mu(\mathbf{x})$ and that the Doob-Dynkin lemma is satisfied \cite{rao2006probability}. 
Then, by propagation of uncertainty, the model's response is a random variable dependent on the input random vector, i.e., $Y = \mu\left(\mathbf{X}\right)$. 
Provided that its variance is finite, the random output can be represented in the form of a \gls{pce}, such that 
\begin{equation}
 \label{eq:pce-infinite}
y = \mu \left(\mathbf{x} \right) =  \sum_{\boldsymbol{\alpha} \in \mathbb{N}_0^N} c_{\boldsymbol{\alpha}} \Psi_{\boldsymbol{\alpha}} \left( \mathbf{x} \right), 
\end{equation}
where  $\Psi_{\boldsymbol{\alpha}}$ are multivariate polynomials, $c_{\boldsymbol{\alpha}} \in \mathbb{R}$ are the associated coefficients, and ${\boldsymbol{\alpha}} = \left(\alpha_1, \dots, \alpha_N\right)$ are unique multi-indices defined by the polynomial's partial degree in each input dimension.
The polynomials are chosen such that they form an orthogonal basis with respect to the input \gls{pdf}, i.e., they satisfy: 
\begin{equation}
\label{eq:orthogonality}
\mathbb{E}\left[\Psi_{\boldsymbol{\alpha}} \Psi_{\boldsymbol{\beta}}\right] = \int_{\Xi}  \Psi_{\boldsymbol{\alpha}}\left(\mathbf{x}\right) \Psi_{\boldsymbol{\beta}}\left(\mathbf{x}\right) f_{\mathbf{X}}\left(\mathbf{x}\right) \mathrm{d}\mathbf{x} =  \mathbb{E}\left[\Psi_{\boldsymbol{\alpha}}^2\right] \delta_{\boldsymbol{\alpha} \boldsymbol{\beta}}.
\end{equation}
For simplicity, we assume an orthonormal basis, such that $\mathbb{E}\left[\Psi_{\boldsymbol{\alpha}}^2\right] = 1$. 

In practice, the infinite series \eqref{eq:pce-infinite} must be truncated to a finite number of $K$ terms based on a set of selected multi-indices. This truncation yields a \gls{pce} approximation of the form
\begin{equation}
 \label{eq:pce-truncated}
y = \mu \left(\mathbf{x} \right) \approx \widehat{\mu} \left(\mathbf{x} \right) = \sum_{k=1}^K c_{k} \Psi_{k} \left( \mathbf{x} \right) = \sum_{\boldsymbol{\alpha} \in \Lambda} c_{\boldsymbol{\alpha}} \Psi_{\boldsymbol{\alpha}} \left( \mathbf{x} \right), 
\end{equation}
where $\Lambda \subset \mathbb{N}_0^N$ is a multi-index set with cardinality $\#\Lambda = K$. 
Each index $k=1,\dots,K$ corresponds to a unique multi-index $\boldsymbol{\alpha} \in \Lambda$. 
One common truncation method is to use a so-called total-degree basis, where $\Lambda = \left\{ \boldsymbol{\alpha}: \left| \boldsymbol{\alpha} \right|_1 \leq P \right\}$ and $P$ denotes the maximum polynomial degree.
The size of the total-degree basis is equal to $K = \frac{\left(N+P\right)!}{N! P!}$ and becomes intractable for high polynomial degrees or input dimensions.
To alleviate this problem, several methods have been suggested for constructing sparse bases \cite{luethen2021sparse}, e.g., by means of dimension-adaptive algorithms \cite{luethen2022automatic}.

Given a multi-index set $\Lambda$ and the corresponding polynomial basis, the remaining task is to estimate the \gls{pce} coefficients $\mathbf{c} = \left(c_{\boldsymbol{\alpha}}\right)_{\boldsymbol{\alpha} \in \Lambda}^\top \in \mathbb{R}^{K \times 1}$. 
We opt here for a regression-based approach \cite{migliorati2013approximation, migliorati2014analysis, hadigol2018least}. Alternatives include include pseudo-spectral projection \cite{constantine2012sparse, conrad2013adaptive} or interpolation \cite{buzzard2012global, porta2014inverse, galetzka2023hp}.
For the regression-based coefficient estimation, we assume a set of input parameter realizations $\left\{\mathbf{x}^{(m)}\right\}_{m=1}^{M}$, commonly called the experimental design, along with the corresponding model responses $\left\{y^{(m)}\right\}_{m=1}^{M}$. 
Put in an \gls{ml} context, the combined dataset $\left\{\mathbf{x}^{(m)}, y^{(m)}\right\}_{m=1}^{M}$ constitutes the training dataset. 
We then construct the so-called design matrix $\mathbf{D} \in \mathbb{R}^{M \times K}$, such that $d_{mk} = \Psi_k\left( \mathbf{x}^{(m)} \right)$.
Collecting the model responses in a vector $\mathbf{y} = \left(y^{(1)}, \dots, y^{(M)}\right)^\top \in \mathbb{R}^{M \times 1}$, the coefficients can be computed by solving the least squares minimization problem
\begin{equation} 
\label{eq:regression-matrix}
\mathbf{c} = \underset {\hat{\mathbf{c}} \in \mathbb{R}^K}{\arg\min} \left\{\frac{1}{M} \lVert \mathbf{y} - \mathbf{D} \hat{\mathbf{c}} \rVert_2^2 \right\},
\end{equation}
which admits the closed-form solution
\begin{equation} 
\label{eq:regression-solution}
\mathbf{c} = \left( \mathbf{D}^\top \mathbf{D}\right)^{-1} \mathbf{D}^\top \mathbf{y}.
\end{equation}
For the regression problem \eqref{eq:regression-matrix} to be well posed, the experimental design must be at minimum equal to the polynomial basis. 
Larger experimental designs are typically needed for well determined problems. 
In fact, the size of the experimental design is crucial for the stability and accuracy of the resulting \gls{pce} \cite{migliorati2013approximation, migliorati2014analysis}.
This limitation can be overcome with sparse \gls{pce} methods, where the problem \eqref{eq:regression-matrix} is complemented with a penalty term that induces sparsity in its solution. 
Common techniques include the least absolute shrinkage and selection operator (LASSO) and compressive sensing \cite{luethen2021sparse}. 

\subsection{Conformal prediction}
\label{sec:cp}

Following the notation introduced in section \ref{sec:pce}, we consider a regression model $\widehat{\mu}\left(\mathbf{x}\right)$ trained with the dataset $\mathcal{D}_{\text{train}} = \left\{\mathbf{x}^{(m)}, y^{(m)}\right\}_{m=1}^M$, where again the experimental design data points $\mathbf{x}^{(m)}$ are assumed to be realizations of random vector $\mathbf{X}$ characterized by a joint probability distribution $f_{\mathbf{X}}(\mathbf{x})$. 
Then, given a test point $\left\{ \mathbf{x}^*, y^*\right\} \not\in \mathcal{D}_{\text{train}}$, the trained model produces a prediction $\widehat{y}^* = \widehat{\mu}(\mathbf{x}^*)$ for the true value $y^*$.
The goal of \gls{cp} is
to complement the model's prediction with an interval which is likely to contain the true output $y^*$.
More formally, given a significance level $s \in \left(0,1\right)$  corresponding to the target coverage level $(1-s)$, \gls{cp} constructs a prediction interval $\widehat{C}_{1-s}\left(\mathbf{x}^*\right)$, such that 
\begin{equation}
\label{eq:coverage-guarantee}
\mathbb{P}\left\{y^* \in \widehat{C}_{1-s}\left(\mathbf{x}^*\right)\right\} \geq 1 - s.
\end{equation}
That is, the probability of the true output being included in the interval is at least equal to the target coverage level.
\textcolor{black}{For example, setting $s=0.05$ corresponds to a target coverage equal to $95\%$, meaning that the prediction interval has been constructed to capture the true test output values at least $95\%$ of the time.}
Note that \gls{cp} assumes that the training and test data are exchangeable, meaning that the joint data distribution is invariant to permutation.

A central concept of \gls{cp} is the so-called non-conformity score, which measures how well a given data point conforms to the fitted model.
To that end, a non-conformity function $\alpha\left(\mathbf{x}, y; \widehat{\mu}\right)$ is evaluated on a set of calibration points $\left\{\mathbf{x}^{(j)}, y^{(j)}\right\}_{j=1}^J$, thus producing the non-conformity scores $\alpha_j = \alpha\left(\mathbf{x}^{(j)}, y^{(j)}; \widehat{\mu}\right)$, $j=1,\dots,J$.
We denote the $(1-s)$-quantile of the empirical distribution of the non-conformity scores $\left\{\alpha_j\right\}_{j=1}^J$ as $\widehat{q}_{1-s}\left\{\alpha_j\right\}$. 
The prediction interval for a test input $\mathbf{x}^*$ is then defined in general by
\begin{equation}
\label{eq:predictive-interval-general}
\widehat C_{1-s}\left(\mathbf{x}^*\right) \;=\; \left\{\, y : \; \alpha\left(\mathbf{x}^*,y;\widehat{\mu}\right) \le \widehat q_{1-s}\left\{\alpha_j\right\}\,\right\},
\end{equation}
such that $y^* \in \widehat C_{1-s}\left(\mathbf{x}^*\right)$ implies that
$\alpha\!\left(\mathbf{x}^*,y^*;\widehat{\mu}\right)
\le \widehat q_{1-s}\!\left\{\alpha_j\right\}$ and vice versa.
A commonly used non-conformity score is the absolute residual, such that $\alpha_j = R_j = \left| r_j \right| = \left| y^{(j)} - \widehat{\mu}\left(\mathbf{x}^{(j)}\right)\right|$.
This choice results in the construction of the symmetric prediction interval 
\begin{equation}
\label{eq:predictive-interval-symmetric}
\widehat{C}_{1-s}\left(\mathbf{x}^*\right) = \left[ \widehat{\mu}\left(\mathbf{x}^*\right) - \widehat{q}_{1-s}\left\{ R_j \right\}, \widehat{\mu}\left(\mathbf{x}^*\right) + \widehat{q}_{1-s}\left\{ R_j \right\} \right].
\end{equation} 
Note that this prediction interval does not generally adapt to possible data heteroskedasticity. 

A naive approach to estimate the prediction interval \eqref{eq:predictive-interval-symmetric} would be to use the residuals on the training data $\mathcal{D}_{\text{train}} = \left\{\mathbf{x}^{(m)}, y^{(m)}\right\}_{m=1}^M$.
However, training data residuals are expected to be smaller compared to residuals obtained from previously unseen test data points. 
Hence, the resulting prediction interval  will most probably not provide the required coverage. 
Split or inductive \gls{cp} approaches \cite{papadopoulos2002inductive, lei2018distribution} use a separate validation dataset $\mathcal{D}_{\text{val}} \neq \mathcal{D}_{\text{train}}$ to compute the non-conformity scores and construct the prediction interval \eqref{eq:predictive-interval-symmetric}. 
Split \gls{cp} guarantees the required finite-sample coverage \eqref{eq:coverage-guarantee}, however, this comes at the cost of a larger data demand.
To avoid the need for additional data, cross \gls{cp} \cite{vovk2015cross, vovk2018cross} utilizes $K$-fold cross-validation, such that the training dataset $\mathcal{D}_{\text{train}}$ is split into $K$ disjoint subsets $\mathcal{D}_1, \dots, \mathcal{D}_K$ of (approximately) equal size. 
Then, the regression models $\widehat{\mu}_{\sim k}\left(\mathbf{x}\right)$ are trained using the datasets $\mathcal{D}_{\sim k} = \mathcal{D}_{\text{train}} \setminus \mathcal{D}_k$, $k=1,\dots,K$, and the absolute cross-validation residuals
\begin{equation}
\label{eq:cv-residuals}
R_m^{\text{cv}} =  \left| r_m^{\text{cv}} \right| = \left| y^{(m)} - \widehat{\mu}_{\sim k(m)}\left(\mathbf{x}^{(m)}\right) \right|,
\end{equation}
are used as non-conformity scores, where the index $k(m)$ denotes that $\left(\mathbf{x}^{(m)}, y^{(m)}\right) \in \mathcal{D}_k$. 
Cross \gls{cp} typically yields an empirical coverage close to the target level $(1-s)$, but does not provide formal coverage guarantees.
A major downside of cross \gls{cp} is that $K$ different regression models must be trained, which can lead to an undesired computational cost.

A special case of cross-validation is the jackknife method \cite{chernick2012jackknife}. 
Using the jackknife in the context of \gls{cp} \cite{lei2018distribution}, the non-conformity scores coincide with the absolute \gls{loo} residuals 
\begin{equation}
\label{eq:loo-residual}
R_m^{\text{LOO}} = \left| r_m^{\text{LOO}} \right| = \left| y^{(m)} - \widehat{\mu}_{\sim m}\left(\mathbf{x}^{(m)}\right) \right|,
\end{equation}
where the regression model $\widehat{\mu}_{\sim m}\left(\mathbf{x}\right)$ is trained using the dataset $\mathcal{D}_{\sim m} = \mathcal{D}_{\text{train}} \setminus \left\{\left(\mathbf{x}^{(m)}, y^{(m)}\right)\right\}$.
Then, the prediction interval \eqref{eq:predictive-interval-symmetric} takes two equivalent forms, i.e.,
\begin{subequations}
\label{eq:predictive-interval-jackknife}
\begin{align}
\widehat{C}_{1-s}^{\text{jack}}\left(\mathbf{x}^*\right) &= \left[ \widehat{\mu}\left(\mathbf{x}^*\right) - \widehat{q}_{1-s}\left\{R_m^{\text{LOO}}\right\}, \widehat{\mu}\left(\mathbf{x}^*\right) + \widehat{q}_{1-s}\left\{R_m^{\text{LOO}}\right\}\right] \label{eq:predictive-interval-jackknife-1}\\ 
&= \left[ \widehat{q}_{s}\left\{\widehat{\mu}\left(\mathbf{x}^*\right) - R_m^{\text{LOO}}\right\}, \widehat{q}_{1-s}\left\{\widehat{\mu}\left(\mathbf{x}^*\right) + R_m^{\text{LOO}}\right\}\right] \label{eq:predictive-interval-jackknife-2}.
\end{align}
\end{subequations} 
Being a cross \gls{cp} variant, jackknife \gls{cp} also yields an empirical coverage close to the target level, without formal coverage guarantees.
Barber et al. \cite{barber2021predictive} suggested an extension of jackknife \gls{cp}, called jackknife+, where the prediction interval is constructed as
\begin{equation}
\label{eq:predictive-interval-jackknife+}
\widehat{C}_{1-s}^{\text{jack+}}\left(\mathbf{x}^*\right) = \left[ \widehat{q}_{s}\left\{\widehat{\mu}_{\sim m}\left(\mathbf{x}^*\right) - R_m^{\text{LOO}}\right\}, \widehat{q}_{1-s}\left\{\widehat{\mu}_{\sim m}\left(\mathbf{x}^*\right) + R_m^{\text{LOO}}\right\}\right].
\end{equation}
In contrast to the jackknife interval \eqref{eq:predictive-interval-jackknife} which is centered at the prediction $\widehat{\mu}\left(\mathbf{x}^*\right)$ of the fully trained model, the jackknife+ interval \eqref{eq:predictive-interval-jackknife+} is based on the ensemble of the \gls{loo} predictions $\widehat{\mu}_{\sim m}\left(\mathbf{x}^*\right)$, $m=1,\dots,M$.
Hence, the jackknife+ interval \eqref{eq:predictive-interval-jackknife+} is asymmetric.
This slight modification results in a formal coverage guarantee of $\left(1-2s\right)$, while the empirical coverage remains close to the target level $(1-s)$, similar to cross \gls{cp} and jackknife \gls{cp}.

In practice, both the jackknife and jackknife+ \gls{cp} methods achieve an empirical coverage close to the target level $(1-s)$ and produce almost identical prediction intervals \cite{barber2021predictive}.
Moreover, both approaches entail training $M$ regression models, thus being extreme cases of cross \gls{cp} in terms of computational cost. 
However, for linear regression models like the \gls{pce}, the \gls{loo} residuals $r_m^{\text{LOO}}$ and the \gls{loo} predictions $\widehat{\mu}_{\sim m}\left(\mathbf{x}^*\right)$ can be obtained without model retraining, as shown next.

\subsection{Conformalized polynomial chaos regression}
\label{sec:conf-pce}
We now assume that $\widehat{\mu}\left(\mathbf{x}\right)$ is a regression-based \gls{pce} in the form of \eqref{eq:pce-truncated} and trained with a dataset $\mathcal{D}_{\text{train}} = \left\{\mathbf{x}^{(m)}, y^{(m)}\right\}_{m=1}^M$.
Then, the prediction intervals \eqref{eq:predictive-interval-jackknife} (for jackknife) and \eqref{eq:predictive-interval-jackknife+} (for jackknife+) can be estimated without the need to retrain the \gls{pce} $M$ times.

Starting with the jackknife method, only the \gls{loo} residuals defined in \eqref{eq:loo-residual} are needed to estimate the prediction interval \eqref{eq:predictive-interval-jackknife}.
Recalling from section \ref{sec:pce} the design matrix $\mathbf{D} \in \mathbb{R}^{M \times K}$, the so-called hat matrix $\mathbf{H} \in \mathbb{R}^{M \times M}$ is defined as 
\begin{equation} 
\label{eq:hat-matrix}
\mathbf{H} = \mathbf{D} \left(\mathbf{D}^\top \mathbf{D}\right)^{-1} \mathbf{D}^\top.
\end{equation}
The \gls{loo} residuals can then be computed analytically via the closed-form expression 
\begin{equation}
\label{eq:loo-residual-analytical}
r_m^{\text{LOO}} = \frac{y^{(m)} - \widehat{\mu}\left(\mathbf{x}^{(m)}\right)}{1 - h_{mm}},
\end{equation}
where $h_{mm} \in \mathbb{R}$ is the $m$-th diagonal term of $\mathbf{H}$.

For the prediction interval \eqref{eq:predictive-interval-jackknife+} based on the jackknife+, the \gls{loo} predictions $\left\{ \widehat{\mu}_{\sim m}\left(\mathbf{x}^*\right) \right\}_{m=1}^M$ for the test input $\mathbf{x}^*$ are additionally needed. 
Let us denote with $\mathbf{d}_m^\top \in \mathbb{R}^{1 \times K}$ the $m$-th row of the design matrix $\mathbf{D} \in \mathbb{R}^{M \times K}$, which contains the evaluation of the \gls{pce} basis on the training input $\mathbf{x}^{(m)}$, such that $\mathbf{d}_m^\top = \left(\Psi_1\left(\mathbf{x}^{(m)}\right), \dots, \Psi_K\left(\mathbf{x}^{(m)}\right)\right)$.
Accordingly, let us denote with $\mathbf{d}_*^\top \in \mathbb{R}^{1 \times K}$ the vector of basis evaluations at the test input $\mathbf{x}^*$, such that $\mathbf{d}_*^\top = \left(\Psi_1\left(\mathbf{x}^*\right), \dots, \Psi_K\left(\mathbf{x}^*\right)\right)$.
Then, the \gls{loo} predictions can be analytically computed as 
\begin{align}
\label{eq:loo-prediction-analytical}
\widehat{\mu}_{\sim m}\left(\mathbf{x}^*\right) &= \widehat{\mu}\left(\mathbf{x}^*\right) - \mathbf{d}_*^\top \left(\mathbf{D}^\top \mathbf{D}\right)^{-1} \mathbf{d}_m \frac{y^{(m)} - \widehat{\mu}\left(\mathbf{x}^{(m)}\right)}{1 - h_{mm}} \nonumber \\
&= \widehat{\mu}\left(\mathbf{x}^*\right) - \mathbf{d}_*^\top \left(\mathbf{D}^\top \mathbf{D}\right)^{-1} \mathbf{d}_m r_m^{\text{LOO}}. 
\end{align} 

\textcolor{black}{
The derivations of the analytical expressions \eqref{eq:loo-residual-analytical} and \eqref{eq:loo-prediction-analytical} are provided in appendix~\ref{sec:appendixA} and appendix~\ref{sec:appendixB}, respectively. 
We note that these analytical expressions and their derivation rely on standard linear regression theory \cite{seber2003linear} and are already known \cite{blatman2009phd}. 
They are given again using the notation of this work strictly for the reader's convenience. 
}

\textcolor{black}{
\paragraph{Remark:} As also noted in section~\ref{sec:pce}, sparse \glspl{pce} are often employed to overcome computational limitations in problem settings featuring high-dimensional inputs or limited data. 
These models are typically computed using $L_1$-minimization and sparsity-promoting algorithms \cite{luethen2021sparse, luethen2022automatic}. 
In this case, explicit model refitting is required during the basis selection procedure, as the closed-form expressions presented above are no longer valid.
However, once the final sparse basis has been identified, it is common practice to compute the final \gls{pce} coefficients by solving the \gls{ols} problem \eqref{eq:regression-matrix} \cite{blatman2011adaptive}.
Therefore, once the final basis is fixed, the analytical, closed-form expressions for the \gls{loo} residuals and model evaluations are again applicable.
This approach should however be used with caution, as the robustness of the identified basis (or lack thereof) is critical to the statistical efficiency of the resulting \gls{cp} intervals \cite{hatstatt2026conformal}.
}

\section{Numerical experiments}
\label{sec:num-exp}
The conformalized \gls{pce} method is first applied to benchmark models that are commonly used for the validation of surrogate modeling methods (section~\ref{sec:benchmarks}). 
Then, the method is applied to an electrical engineering use-case concerning the design of a Stern-Gerlach electromagnet (section~\ref{sec:stern-gerlach}).
Different conformalized \gls{pce} configurations are examined, depending on jackknife method, maximum polynomial degree, and, most importantly, training data availability.

The conformalized \gls{pce} models are assessed with respect to their ability to construct prediction intervals that provide the required target coverage.
\textcolor{black}{To that end, separate test datasets are used to calculate the empirical coverage as the fraction of the test samples for which the true model output successfully falls within the prediction interval.
The empirical coverage is then compared against the pre-defined target coverage to establish the statistical efficiency of the provided \gls{cp} intervals.}

Each experiment is repeated for $100$ random seeds to estimate coverage and prediction interval statistics.
\textcolor{black}{Note that the use of different random seeds is performed strictly to validate the statistical robustness of the empirical coverage, and is not a part of the methodology itself.}

Last, note that the abbreviation ``PI'' used in the parity plots refers to the prediction intervals provided by the conformalized \glspl{pce}.

\subsection{\textcolor{black}{Validation on benchmark models}}
\label{sec:benchmarks}
A univariate model is used in section~\ref{sec:meromorphic}, while a model with ten inputs is considered in section~\ref{sec:wing}.
A discussion based on the numerical results obtained from these benchmark models is available in section~\ref{sec:num-exp-discussion}.
The discussed trends and observations are further supported by the numerical results corresponding to two further multivariate benchmark models, which are presented in appendix~\ref{sec:appendixC}.
The multivariate models in section~\ref{sec:wing} and appendix~\ref{sec:appendixC} are taken from the ``Virtual Library of Simulation Experiments: Test Functions and Datasets'' by Surjanovic and Bingham \cite{surjanovic2013virtual}.

For all examined benchmark models, the experimental design used to train the \gls{pce} consists of samples drawn randomly from the distribution of the input parameters. 
The target coverage is fixed at $95\%$ and the test datasets consist of $10^4$ random samples.

\subsubsection{Meromorphic function}
\label{sec:meromorphic}
We consider the one-dimensional meromorphic function 
\begin{equation}
\label{eq:meromorphic}
\mu(x) = \frac{1}{a + b x},
\end{equation}
with $a=1$ and $b=0.5$. 
We assume that the input $x$ takes values as uniformly distributed in $\left[-1,1\right]$. 
The conformalized \gls{pce} is constructed with maximum polynomial degrees $P \in \left\{2,3\right\}$.
Since the input is one-dimensional, the polynomial basis consists of $P+1$ terms.
The experimental design has size $M = C \left(P+1\right)^2$, where $C \in \left\{2,3,5,10\right\}$ is an oversampling coefficient. 
Note that the quadratic scaling of the experimental design in relation to the polynomial basis is necessary to ensure a well conditioned least squares problem \eqref{eq:regression-matrix} in the one-dimensional input case \cite{migliorati2014analysis}.

\begin{figure}[t!]
\centering
\begin{subfigure}[b]{0.32\textwidth}
    \centering
    \includegraphics[width=\textwidth]{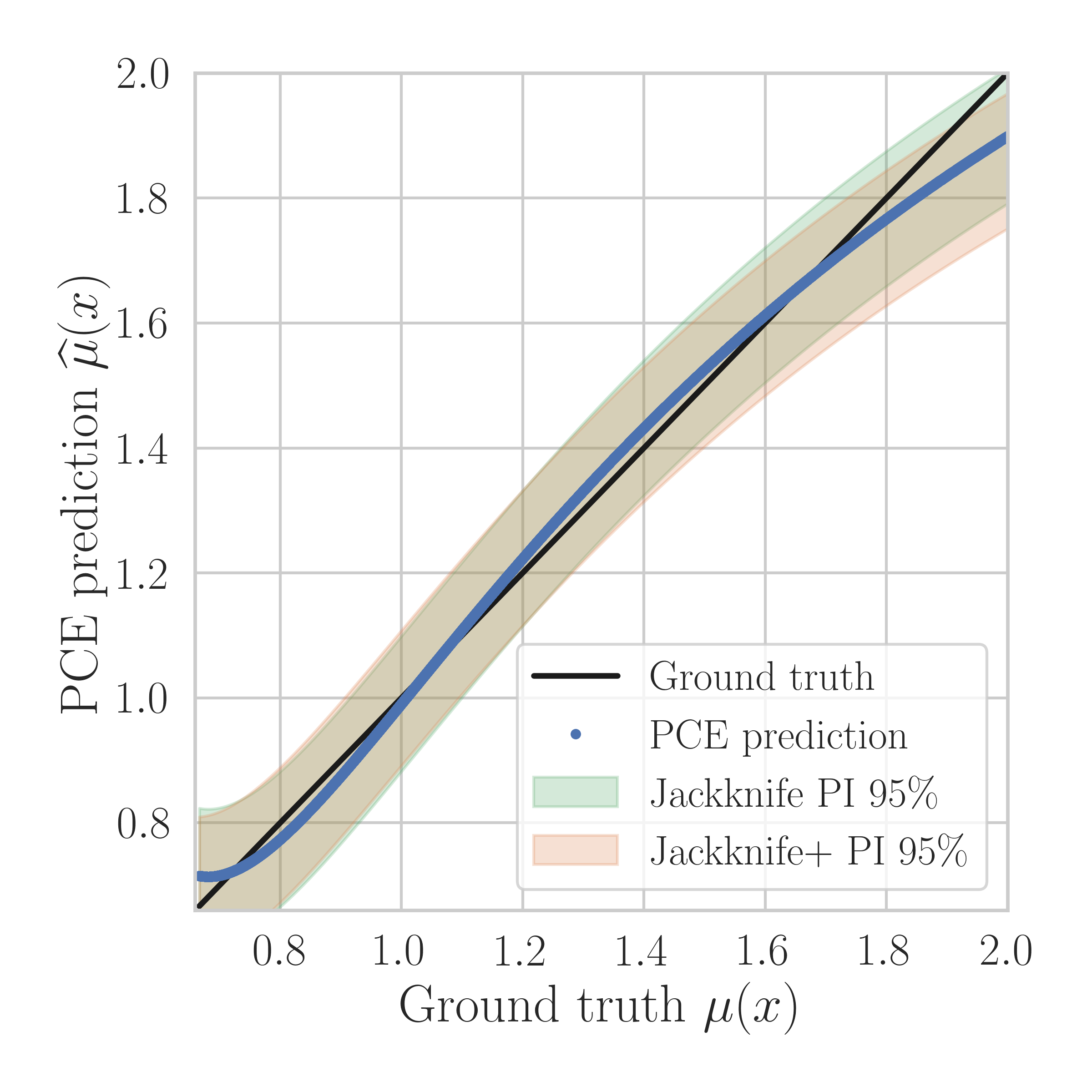}
    \caption{ $P=2$, $C=2$.}
\end{subfigure}
\hfill
\begin{subfigure}[b]{0.32\textwidth}
    \centering
    \includegraphics[width=\textwidth]{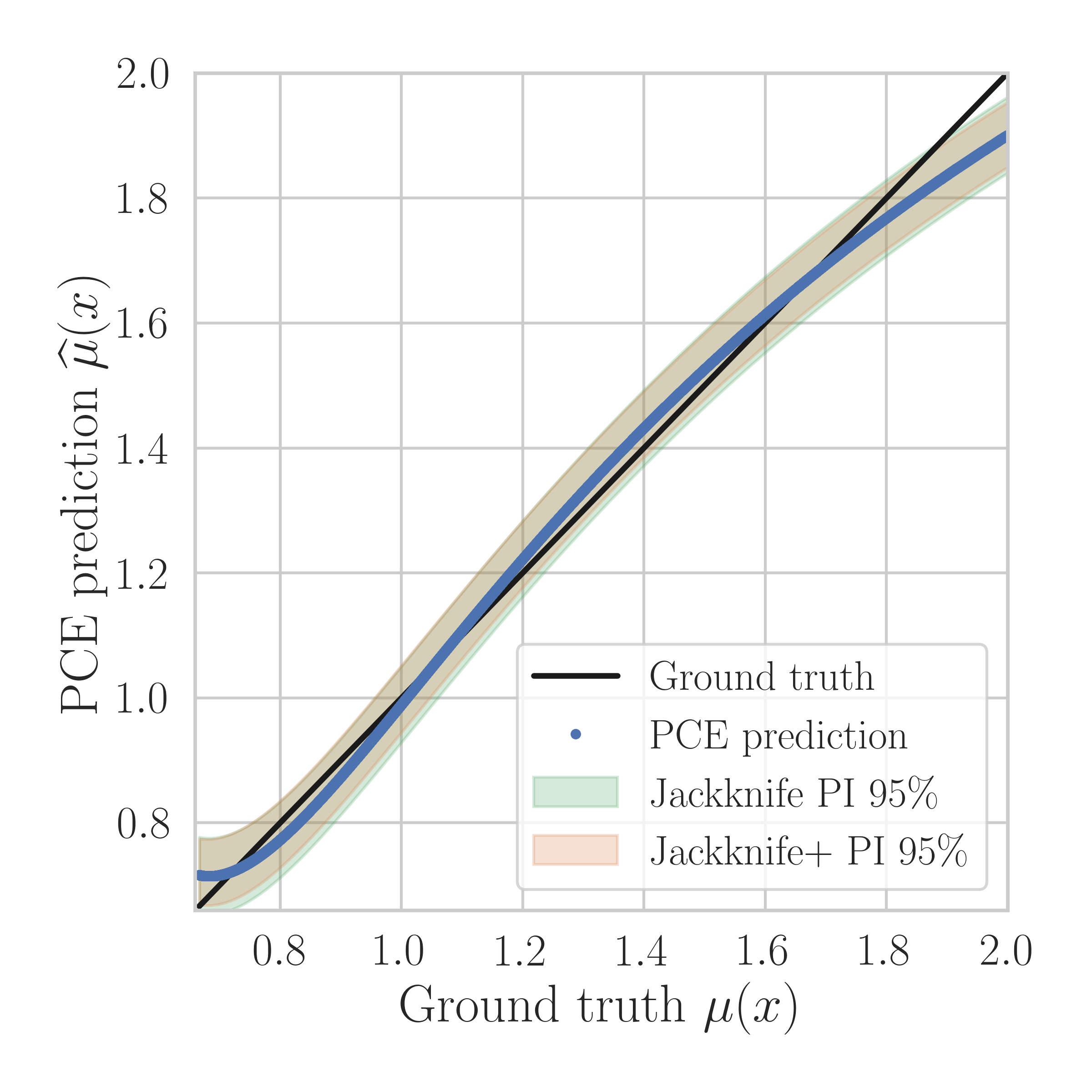}
    \caption{ $P=2$, $C=5$.}
\end{subfigure}
\hfill
\begin{subfigure}[b]{0.32\textwidth}
    \centering
    \includegraphics[width=\textwidth]{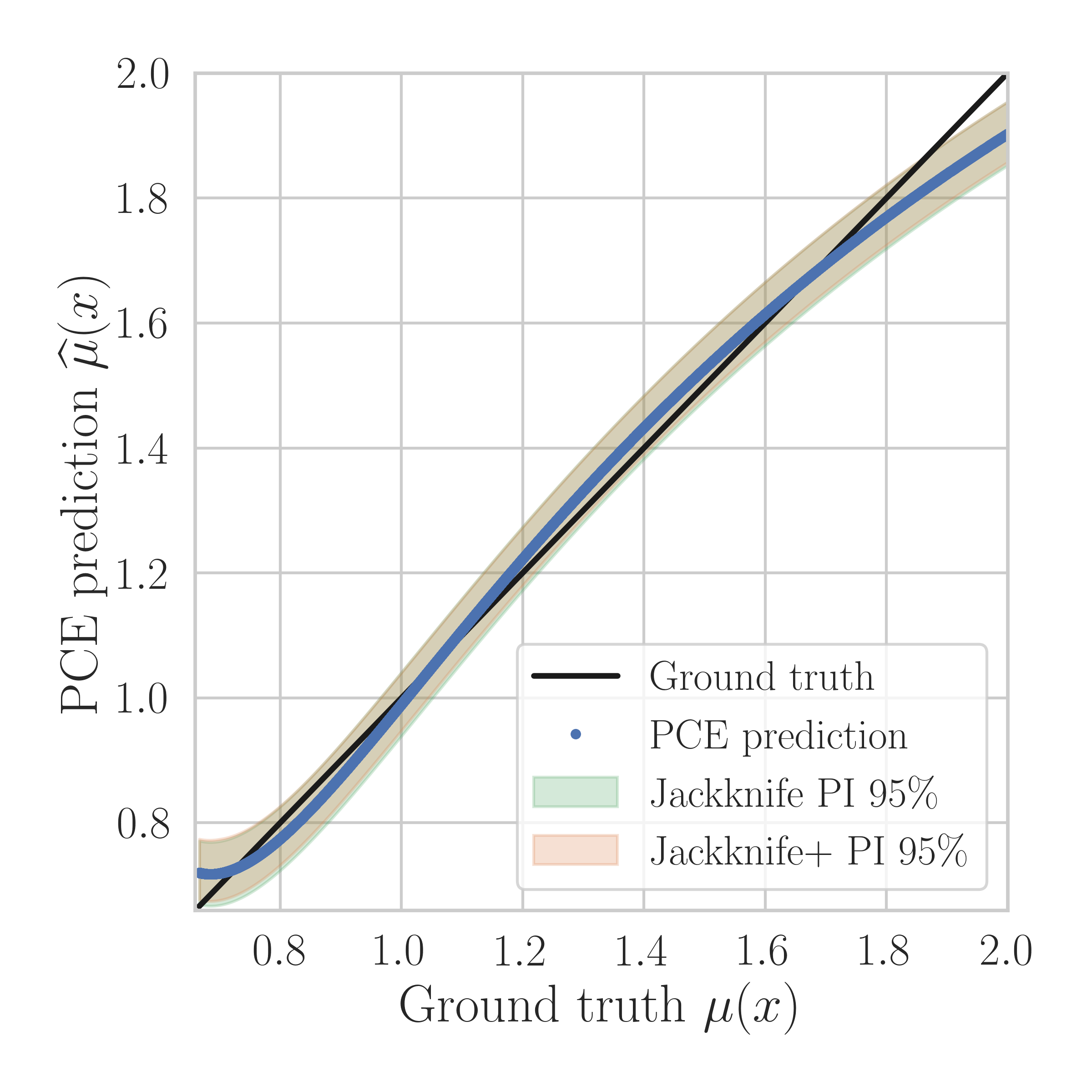}
    \caption{ $P=2$, $C=10$.}
\end{subfigure}
\\
\begin{subfigure}[b]{0.32\textwidth}
    \centering
    \includegraphics[width=\textwidth]{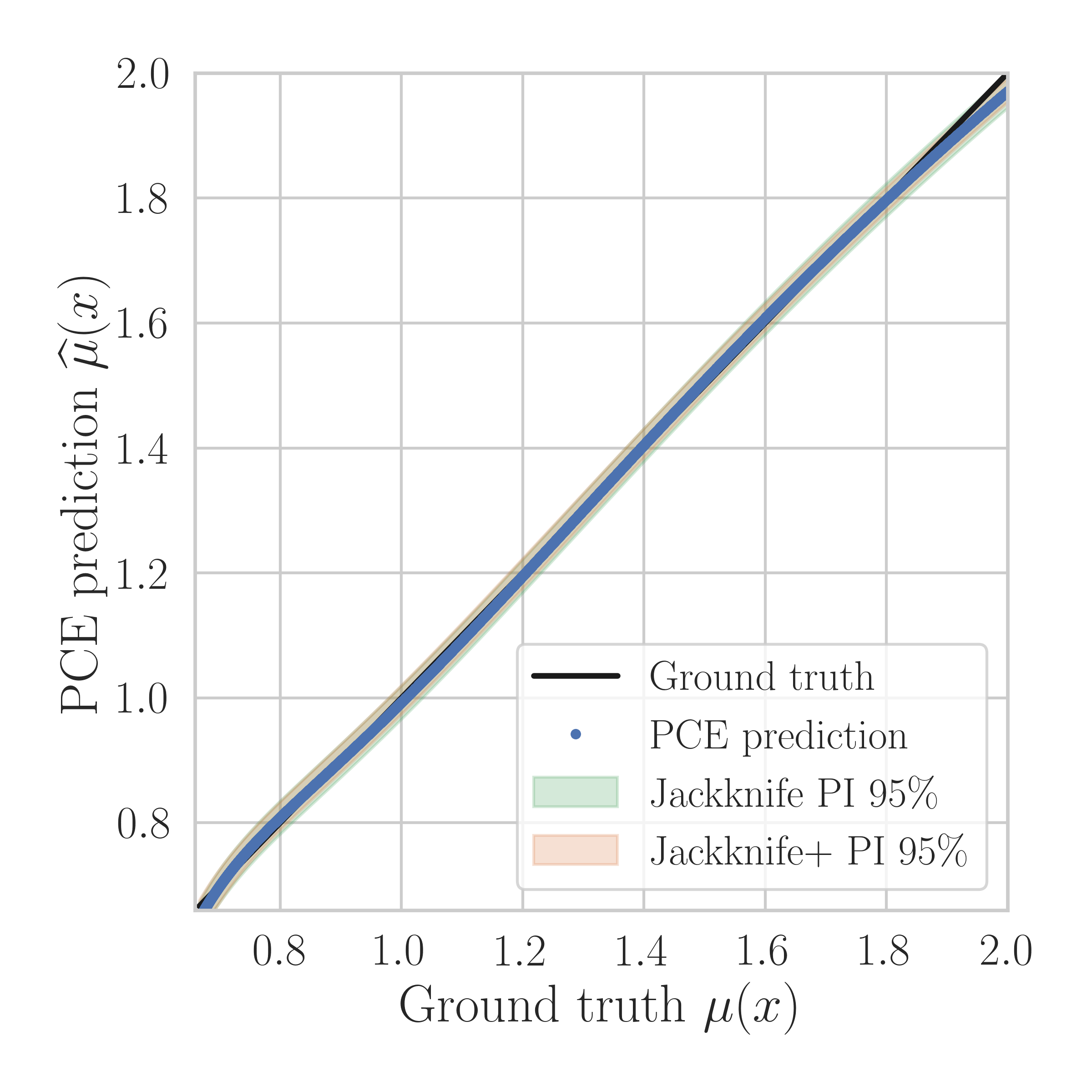}
    \caption{ $P=3$, $C=2$.}
\end{subfigure}
\begin{subfigure}[b]{0.32\textwidth}
    \centering
    \includegraphics[width=\textwidth]{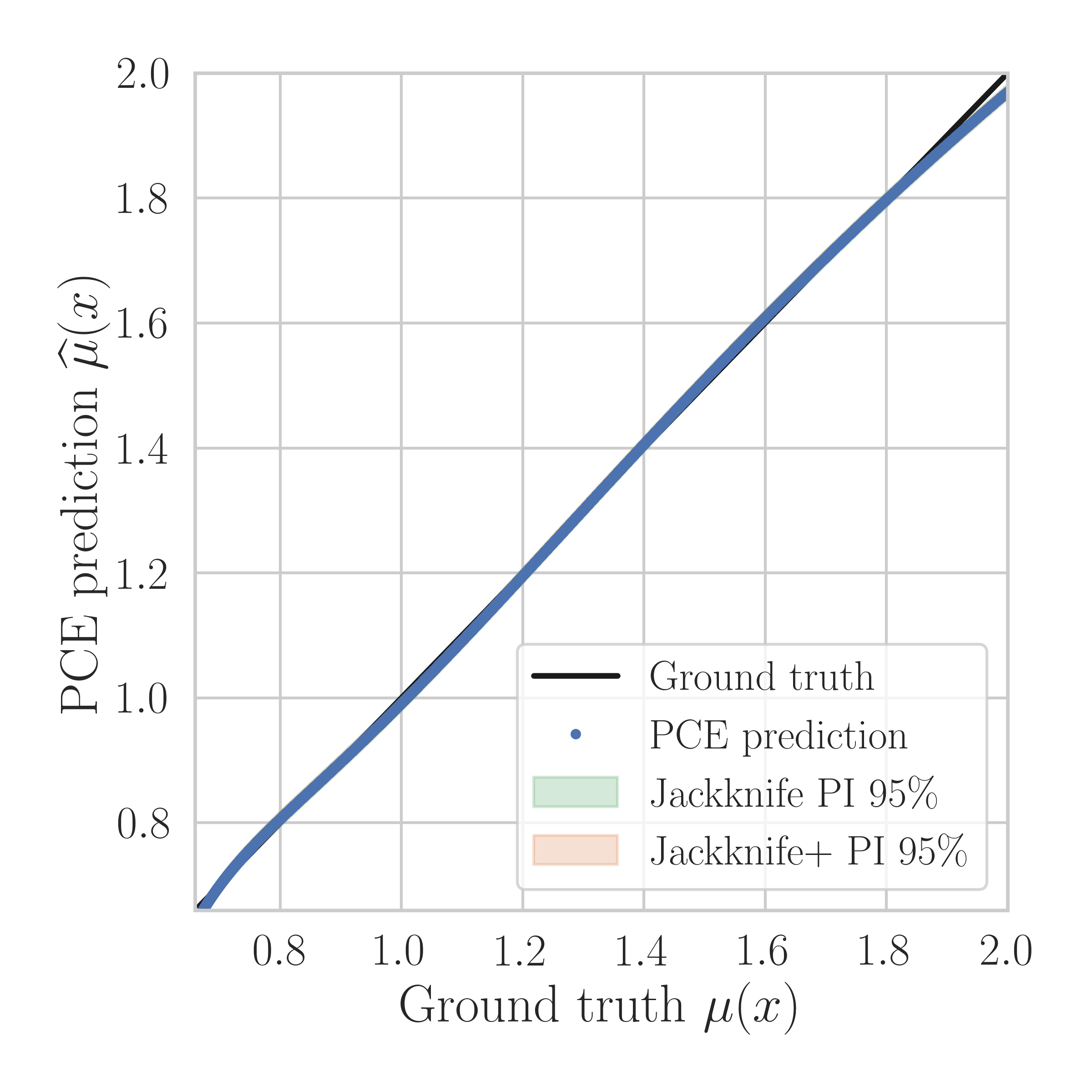}
    \caption{ $P=3$, $C=5$.}
\end{subfigure}
\hfill
\begin{subfigure}[b]{0.32\textwidth}
    \centering
    \includegraphics[width=\textwidth]{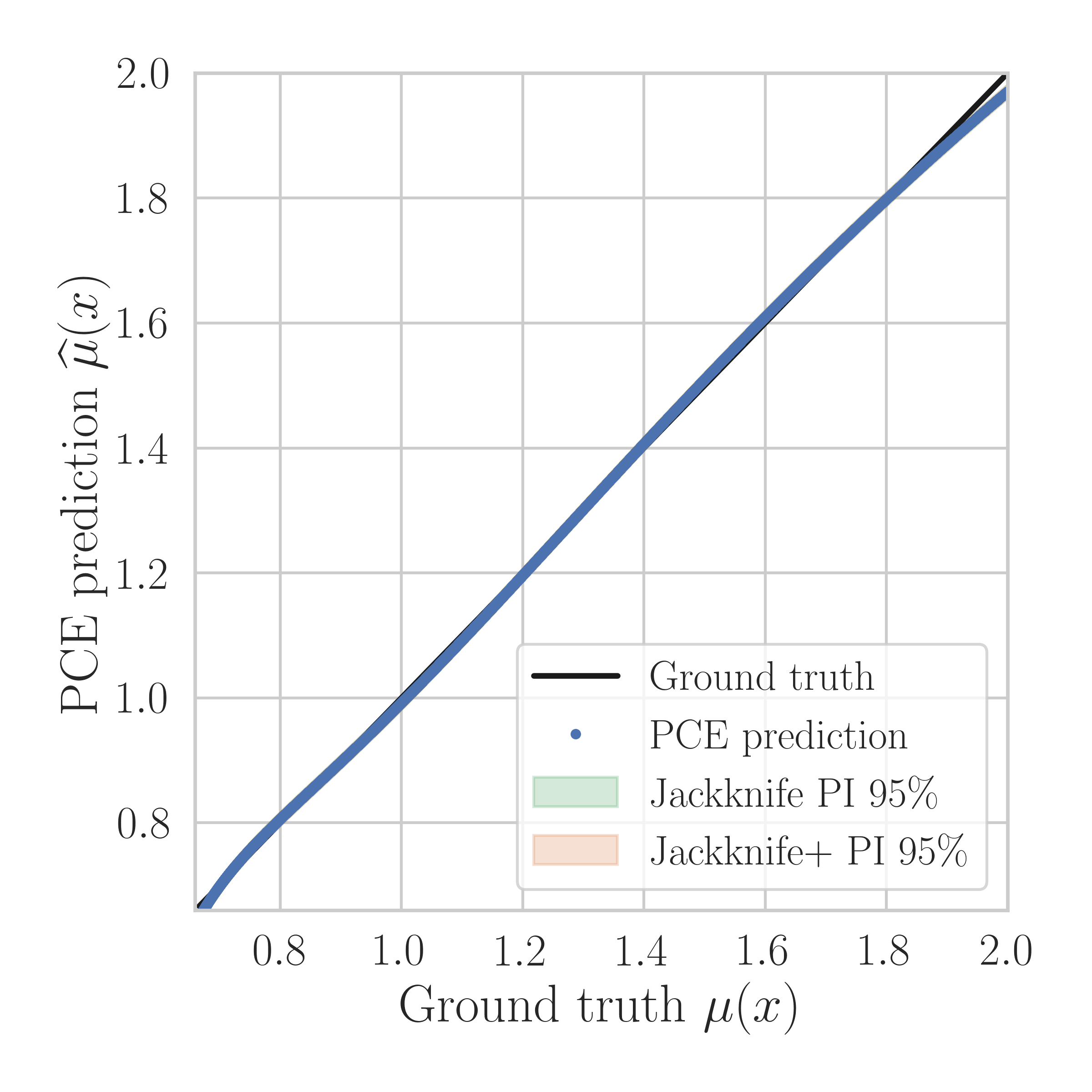}
    \caption{ $P=3$, $C=10$.}
\end{subfigure}
\caption{Parity plots comparing ground truth values of the meromorphic function against conformalized \gls{pce} predictions for different combinations of polynomial degree $P$ and oversampling coefficient $C$. The results correspond to a single random seed. \textcolor{black}{The target coverage is fixed at $95\%$, corresponding to a significance level of $s=0.05$.}}
\label{fig:meromorhic-parity-plots}
\end{figure}

\begin{figure}[t!]
\centering
\begin{subfigure}[b]{0.45\textwidth}
    \centering
    \includegraphics[width=\textwidth]{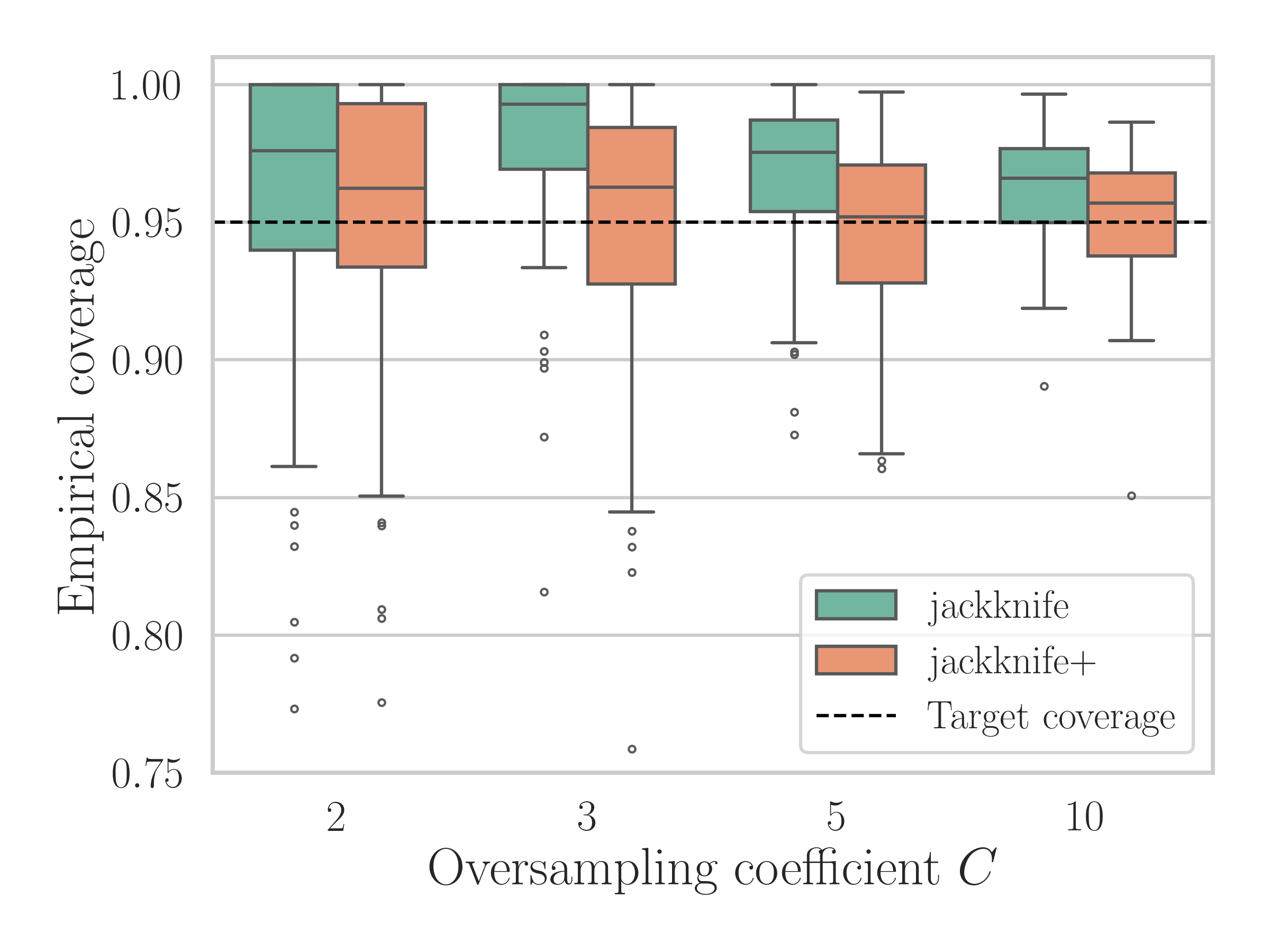}
    \caption{$P=2$.}
\end{subfigure}
\hfill
\begin{subfigure}[b]{0.45\textwidth}
    \centering
    \includegraphics[width=\textwidth]{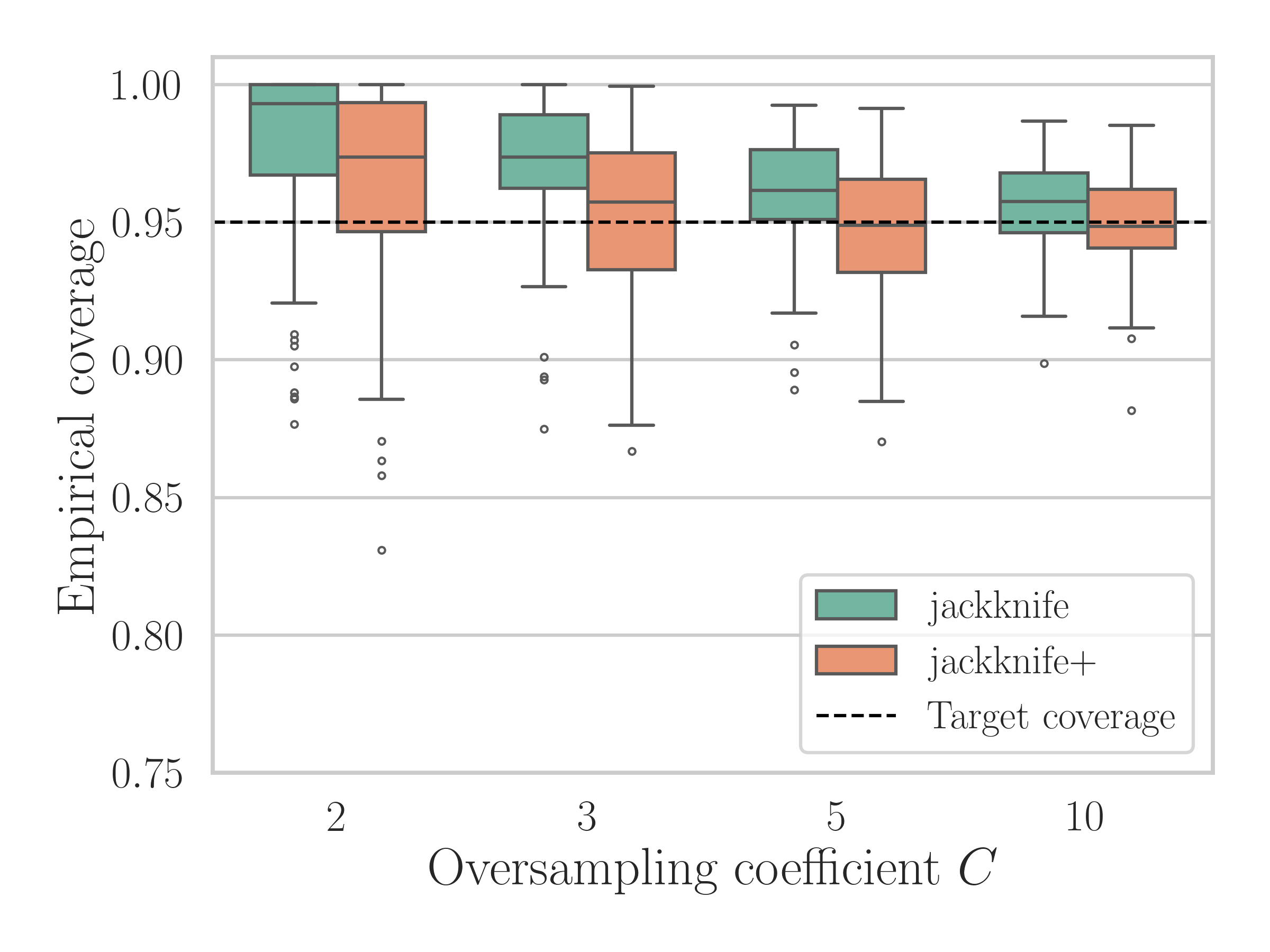}
    \caption{$P=3$.}
\end{subfigure}
\caption{Box plots of the empirical coverage provided by conformalized \gls{pce} surrogates of the meromorphic function, for different combinations of polynomial degree $P$ and oversampling coefficient $C$. \textcolor{black}{The target coverage is fixed at $95\%$, corresponding to a significance level of $s=0.05$.}}
\label{fig:meromorhic-coverage-boxplots}
\end{figure}

\begin{figure}[t!]
\centering
\begin{subfigure}[b]{0.45\textwidth}
    \centering
    \includegraphics[width=\textwidth]{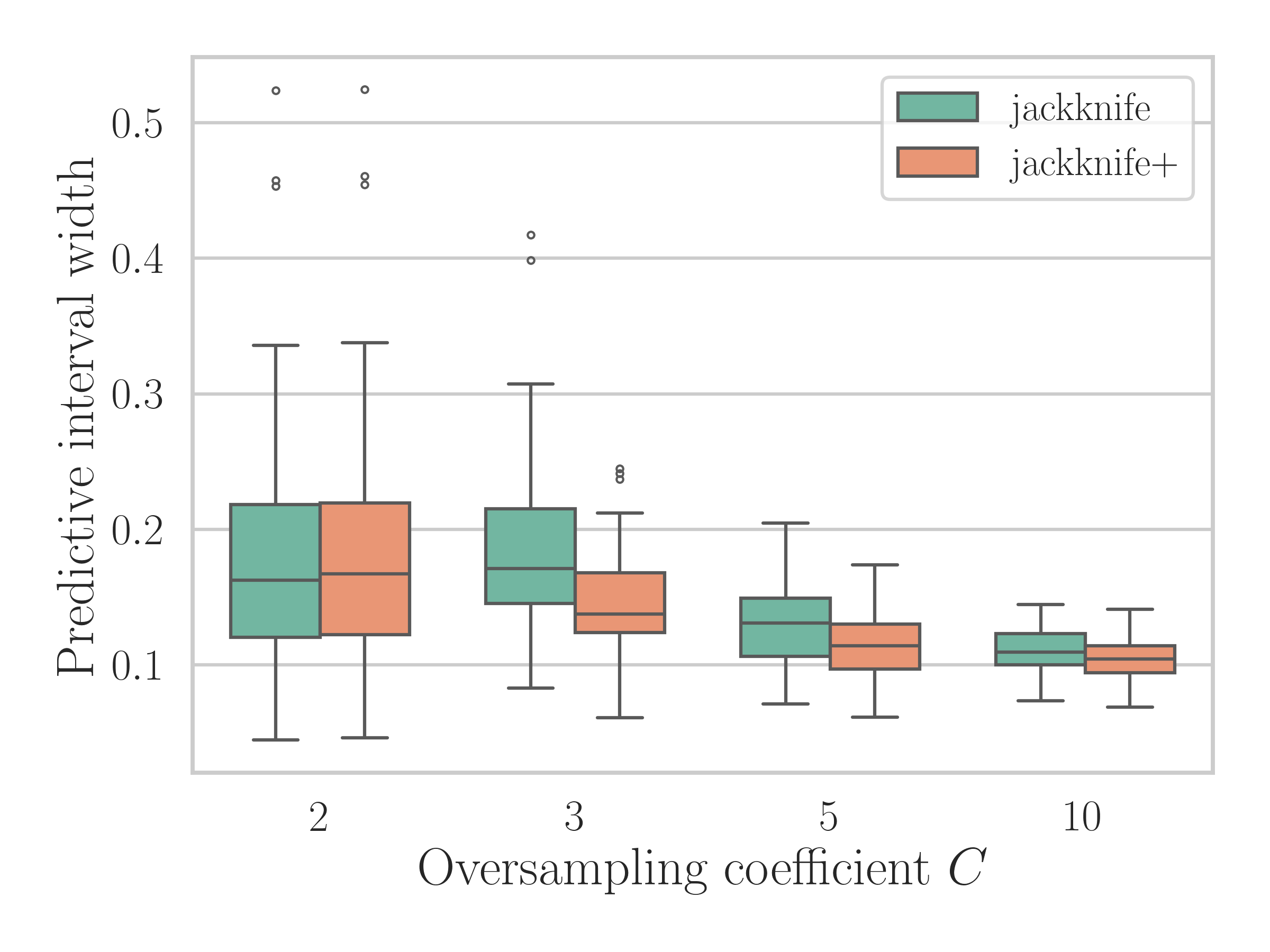}
    \caption{$P=2$.}
\end{subfigure}
\hfill
\begin{subfigure}[b]{0.45\textwidth}
    \centering
    \includegraphics[width=\textwidth]{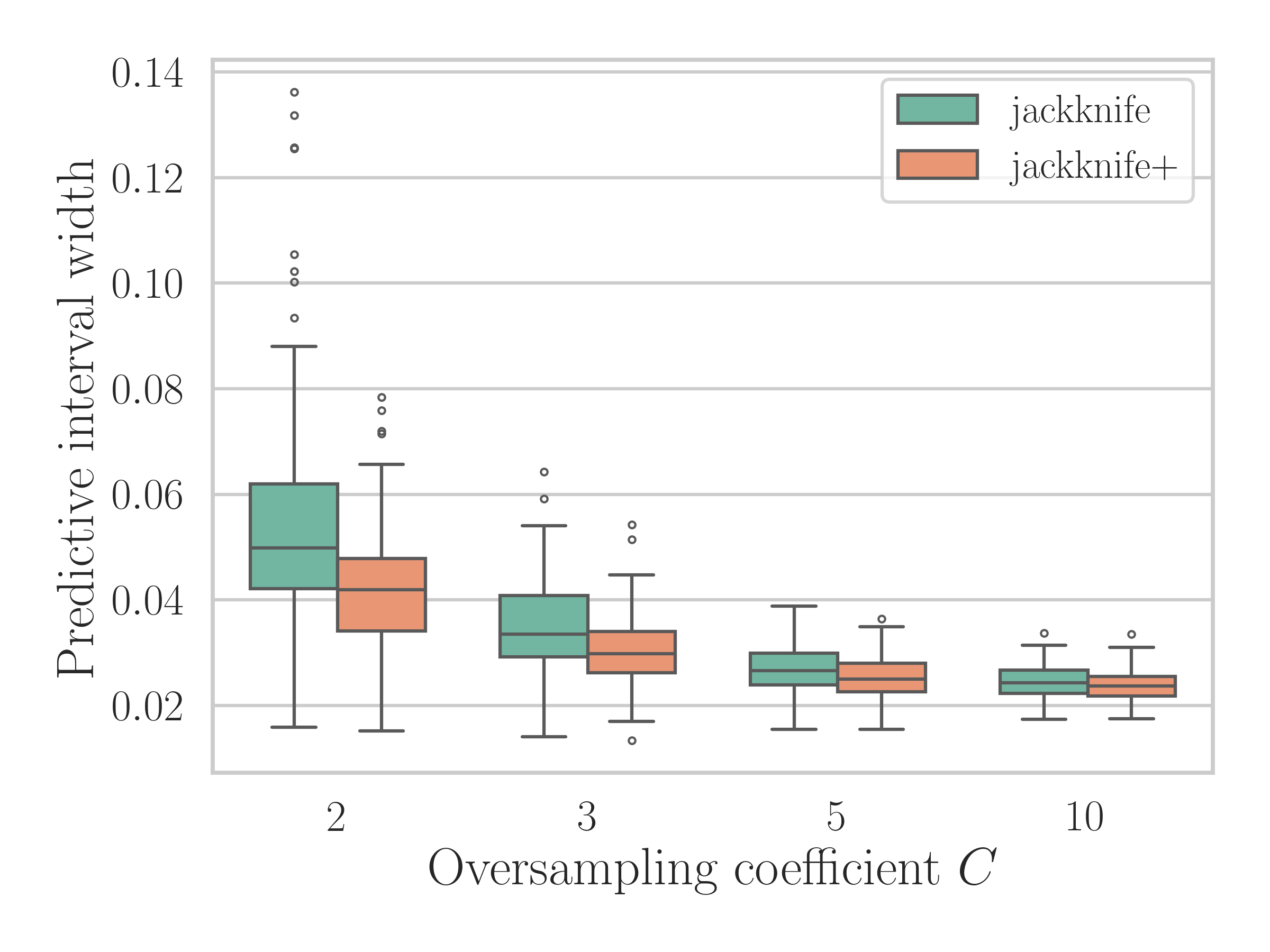}
    \caption{$P=3$.}
\end{subfigure}
\caption{Box plots of the prediction interval widths provided by conformalized \gls{pce} surrogates of the meromorphic function, for different combinations of polynomial degree $P$ and oversampling coefficient $C$. \textcolor{black}{The target coverage is fixed at $95\%$, corresponding to a significance level of $s=0.05$.}}
\label{fig:meromorhic-interval-boxplots}
\end{figure}

Figure~\ref{fig:meromorhic-parity-plots} compares the different configurations of the conformalized \gls{pce} against the true meromorphic function for a single random seed, where the results for $C=3$ have been omitted for better readability. 
Figure~\ref{fig:meromorhic-coverage-boxplots} shows the empirical coverage results, where the box plots summarize the distribution of the empirical coverage over $100$ random seeds.
Similarly, Figure~\ref{fig:meromorhic-interval-boxplots} shows box plots regarding the corresponding prediction intervals. 

\subsubsection{Wing weight function}
\label{sec:wing}
We consider a function that models the weight of light aircraft wing, given as
\begin{equation}
\mu(\mathbf{x}) =   0.036 S_{\text{w}}^{0.758} W_{\text{fw}}^{0.0035} \left(\frac{A}{\cos^2(\Lambda)}\right)^{0.6}  q^{0.006} \lambda^{0.04} \left(\frac{100 t_{\text{c}}}{\cos(\Lambda)}\right)^{-0.3} (N_{\text{z}} \ W_{\text{dg}})^{0.49} + S_{\text{w}} W_{\text{p}},
\end{equation}
where the input vector $\mathbf{x} \in \mathbb{R}^{10}$ consists of the parameters listed in Table~\ref{tab:wing-parameters}, along with their value ranges. 
All parameters are assumed to be uniformly distributed within their ranges.
\begin{table}[h!]
\centering
\caption{Input parameters of the wing weight function.}
\label{tab:wing-parameters}
\begin{threeparttable}
\begin{tabular}{c c c l}
\toprule
Parameter & Description & Units & Range \\
\midrule 
$S_{\text{w}}$ & wing area & ft$^2$  & $\left[150, 200\right]$ \\  
$W_{\text{fw}}$ & wing fuel weight & lb & $\left[220, 300\right]$ \\
$A$ & aspect ratio & $-$ & $\left[6, 10\right]$ \\
$\Lambda$     & quarter-chord sweep & $(^o)$ & $\left[-10, 10\right]$ \\
$q$ & dynamic pressure at cruise & lb/ft$^2$ & $\left[16, 45\right]$ \\
$\lambda$  & taper ratio & $-$ & $\left[0.5, 1\right]$ \\
$t_{\text{c}}$   & airfoil thickness to chord ratio & $-$ & $\left[0.08, 0.18\right]$\\
$N_{\text{z}}$   & ultimate load factor & $-$ & $\left[2.5, 6\right]$\\
$W_{\text{dg}}$   & flight design gross weight & lb & $\left[1700, 2500\right]$\\
$W_{\text{p}}$   & paint weight & lb/ft$^2$ & $\left[0.025, 0.08\right]$\\
\bottomrule 
\end{tabular}
\end{threeparttable}
\end{table}

\begin{figure}[t!]
\centering
\begin{subfigure}[b]{0.32\textwidth}
    \centering
    \includegraphics[width=\textwidth]{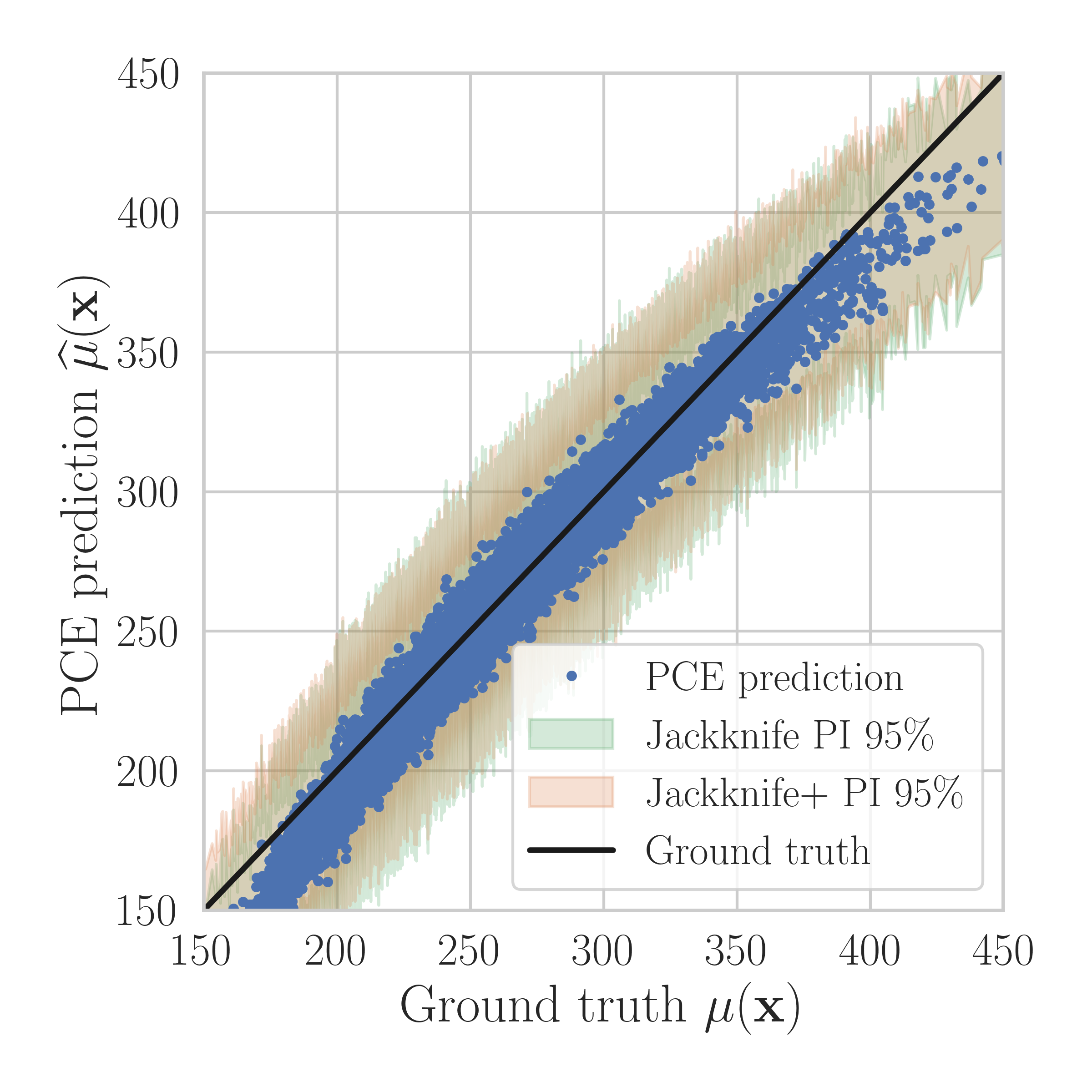}
    \caption{ $P=1$, $C=2$.}
\end{subfigure}
\hfill
\begin{subfigure}[b]{0.32\textwidth}
    \centering
    \includegraphics[width=\textwidth]{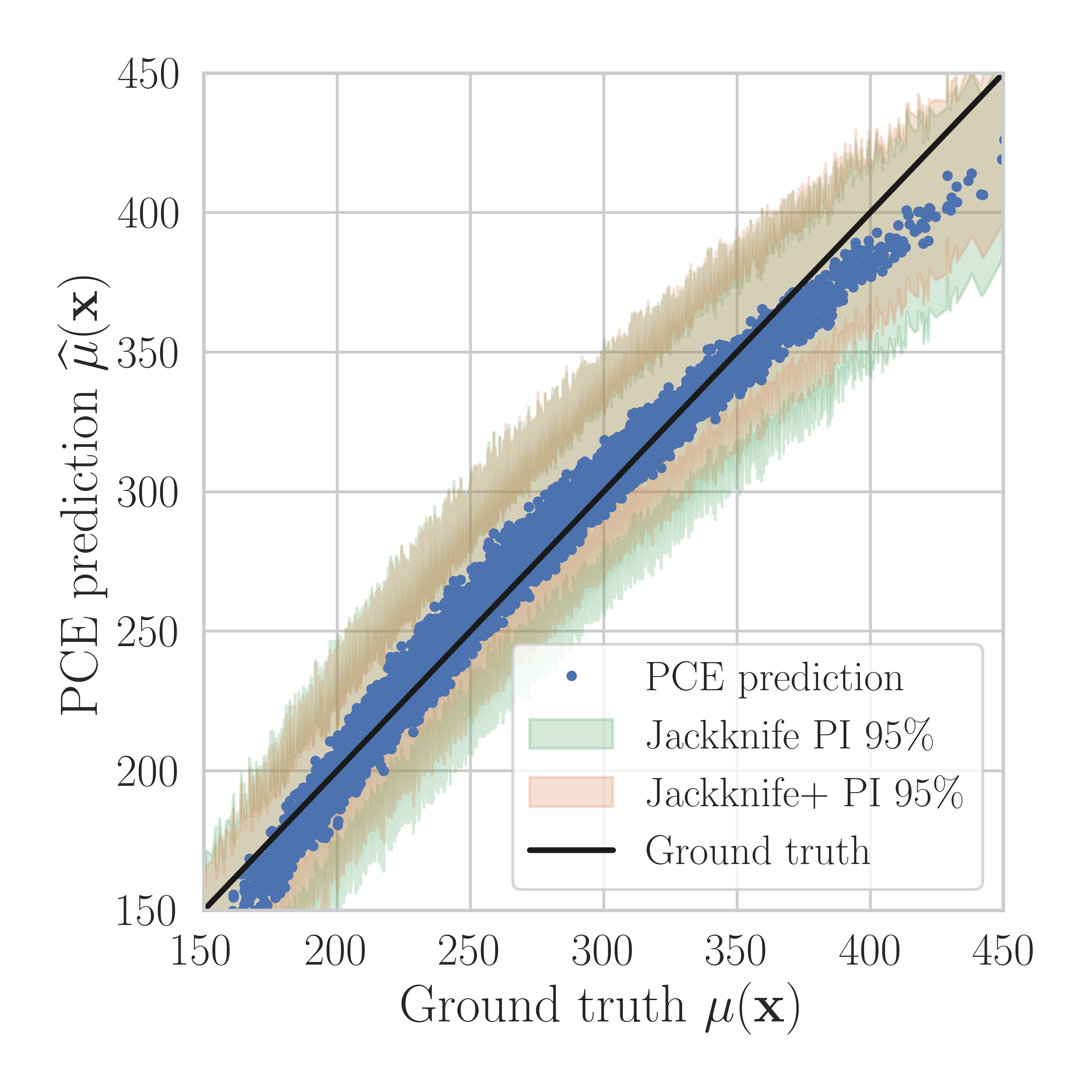}
    \caption{ $P=1$, $C=5$.}
\end{subfigure}
\hfill
\begin{subfigure}[b]{0.32\textwidth}
    \centering
    \includegraphics[width=\textwidth]{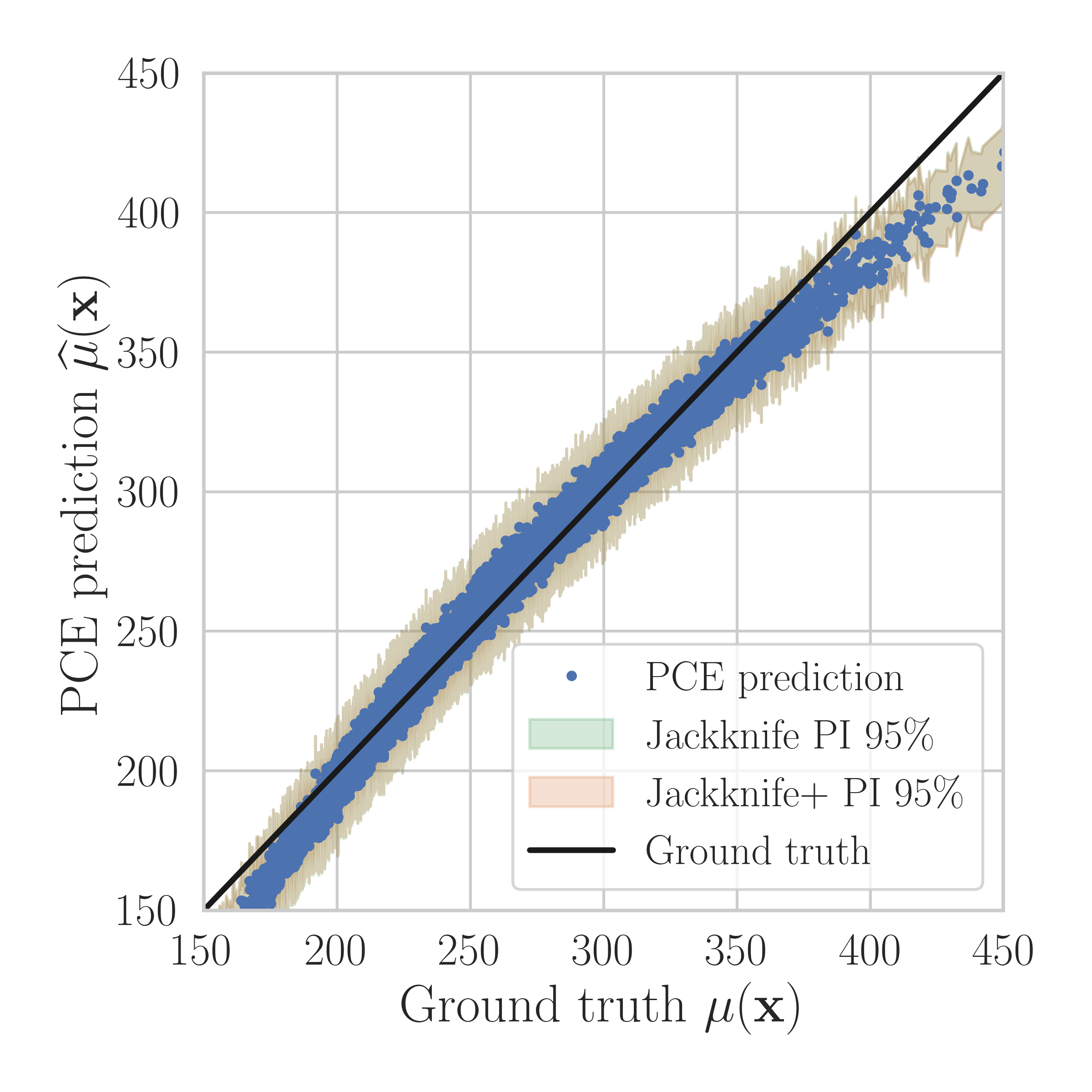}
    \caption{ $P=1$, $C=10$.}
\end{subfigure}
\\
\begin{subfigure}[b]{0.32\textwidth}
    \centering
    \includegraphics[width=\textwidth]{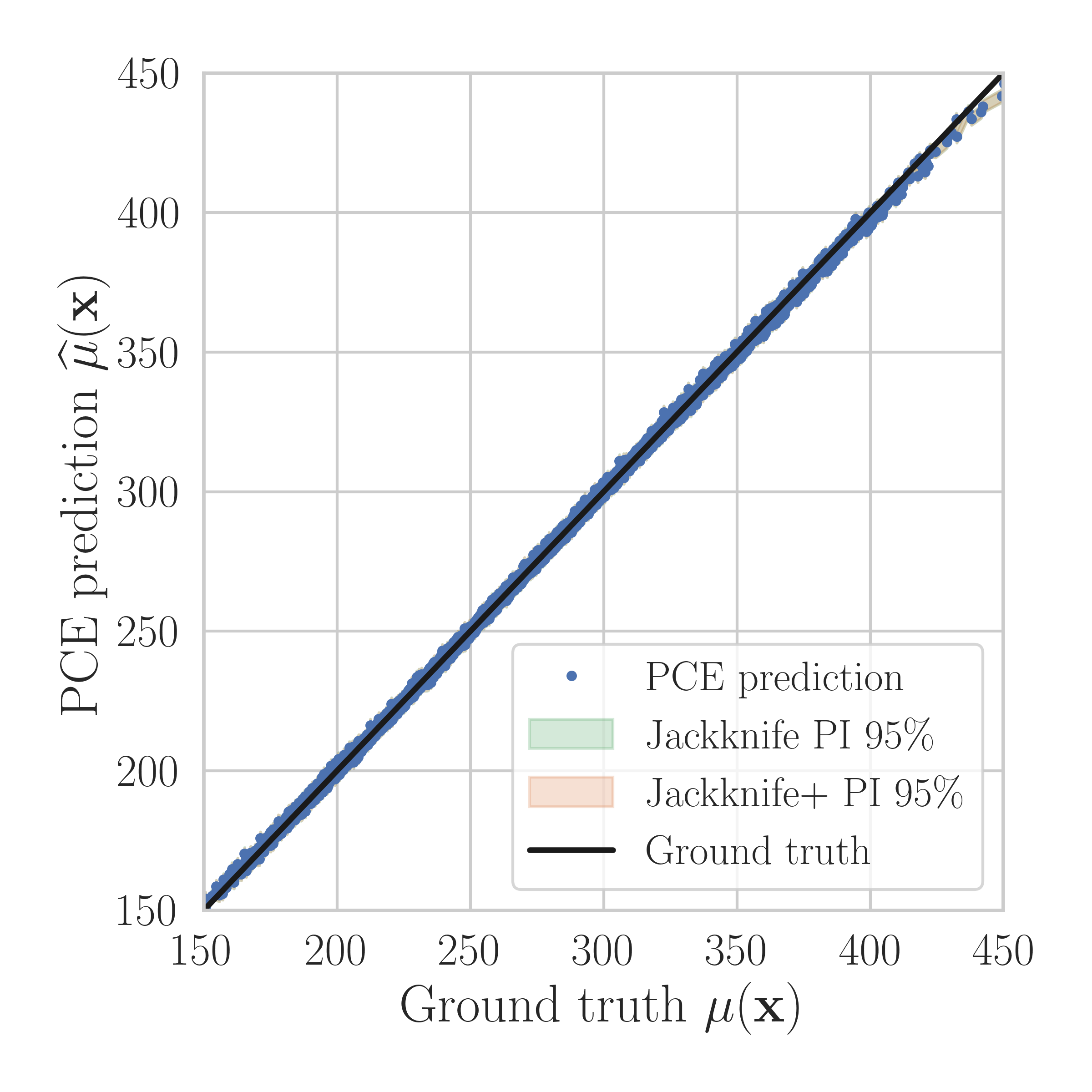}
    \caption{ $P=2$, $C=2$.}
\end{subfigure}
\hfill
\begin{subfigure}[b]{0.32\textwidth}
    \centering
    \includegraphics[width=\textwidth]{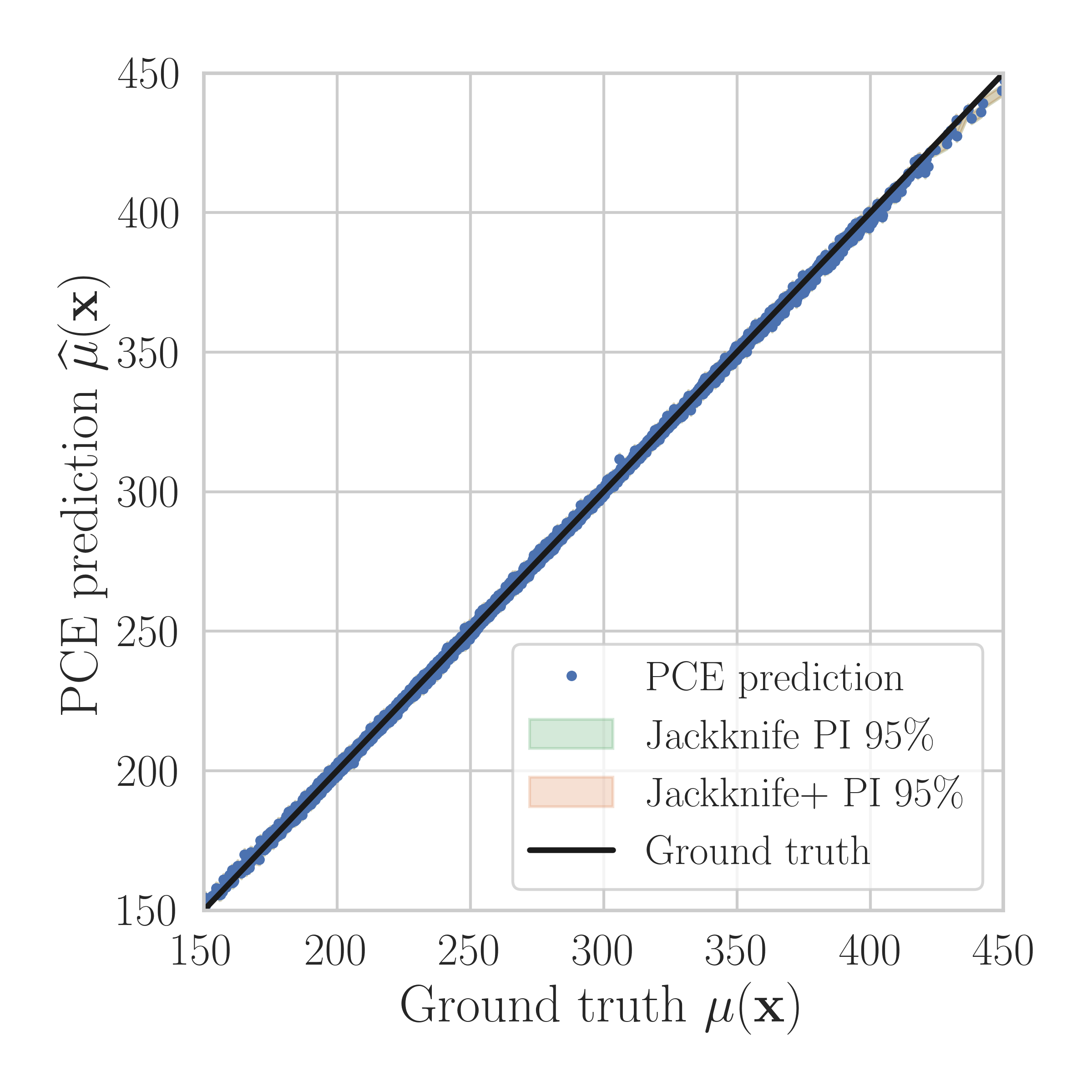}
    \caption{ $P=2$, $C=5$.}
\end{subfigure}
\hfill
\begin{subfigure}[b]{0.32\textwidth}
    \centering
    \includegraphics[width=\textwidth]{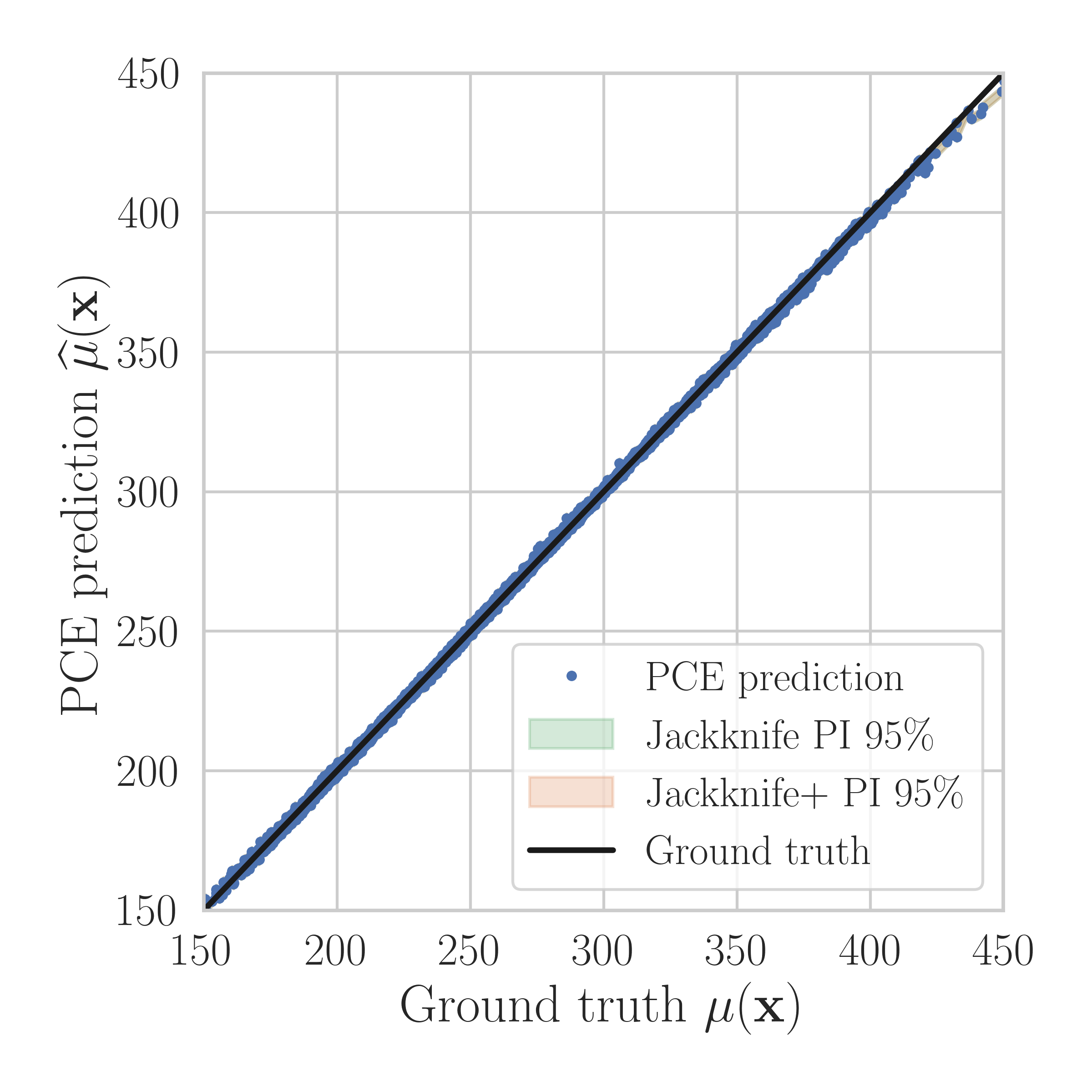}
    \caption{ $P=2$, $C=10$.}
\end{subfigure}
\caption{Parity plots comparing ground truth values of the wing weight function against conformalized \gls{pce} predictions for different combinations of polynomial degree $P$ and oversampling coefficient $C$. \textcolor{black}{The target coverage is fixed at $95\%$, corresponding to a significance level of $s=0.05$.}}
\label{fig:wing-weight-parity-plots}
\end{figure}

\begin{figure}[t!]
\centering
\begin{subfigure}[b]{0.45\textwidth}
    \centering
    \includegraphics[width=\textwidth]{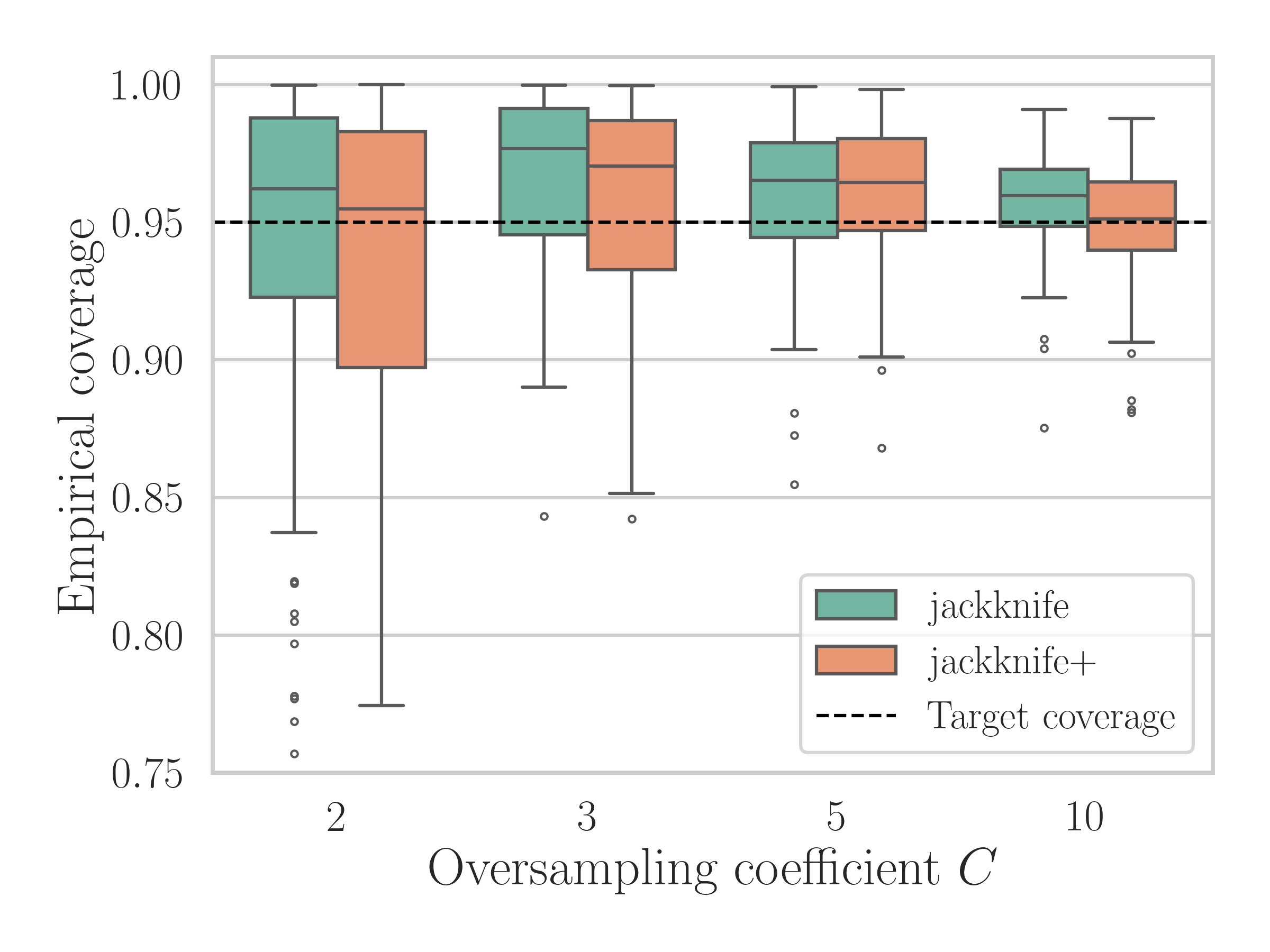}
    \caption{$P=1$.}
\end{subfigure}
\hfill 
\begin{subfigure}[b]{0.45\textwidth}
    \centering
    \includegraphics[width=\textwidth]{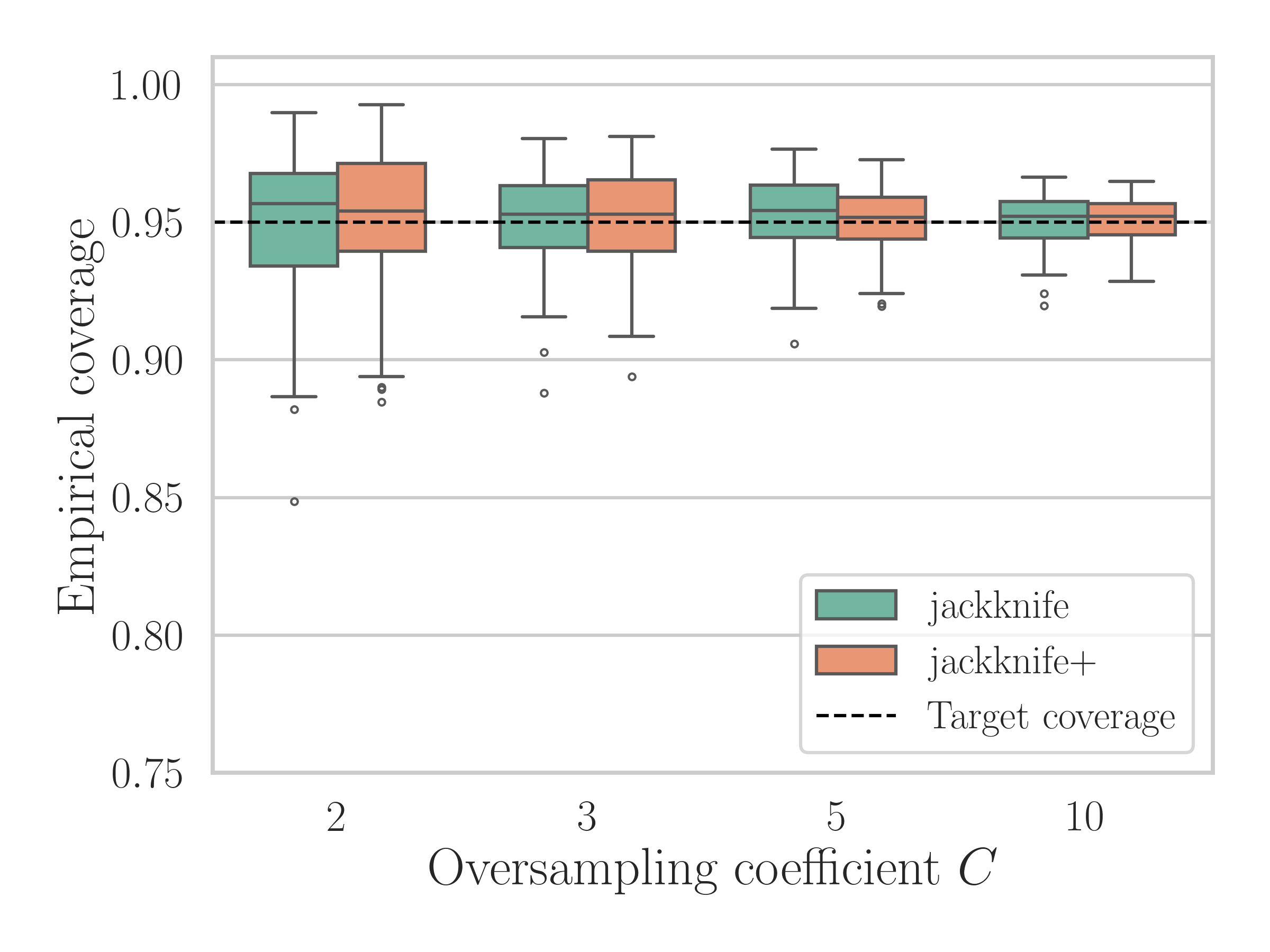}
    \caption{$P=2$.}
\end{subfigure}
\caption{Box plots of the empirical coverage provided by conformalized \gls{pce} surrogates of the wing weight function, for different combinations of polynomial degree $P$, and oversampling coefficient $C$. \textcolor{black}{The target coverage is fixed at $95\%$, corresponding to a significance level of $s=0.05$.}}
\label{fig:wing-weight-coverage-boxplots}
\end{figure}

\begin{figure}[t!]
\centering
\begin{subfigure}[b]{0.45\textwidth}
    \centering
    \includegraphics[width=\textwidth]{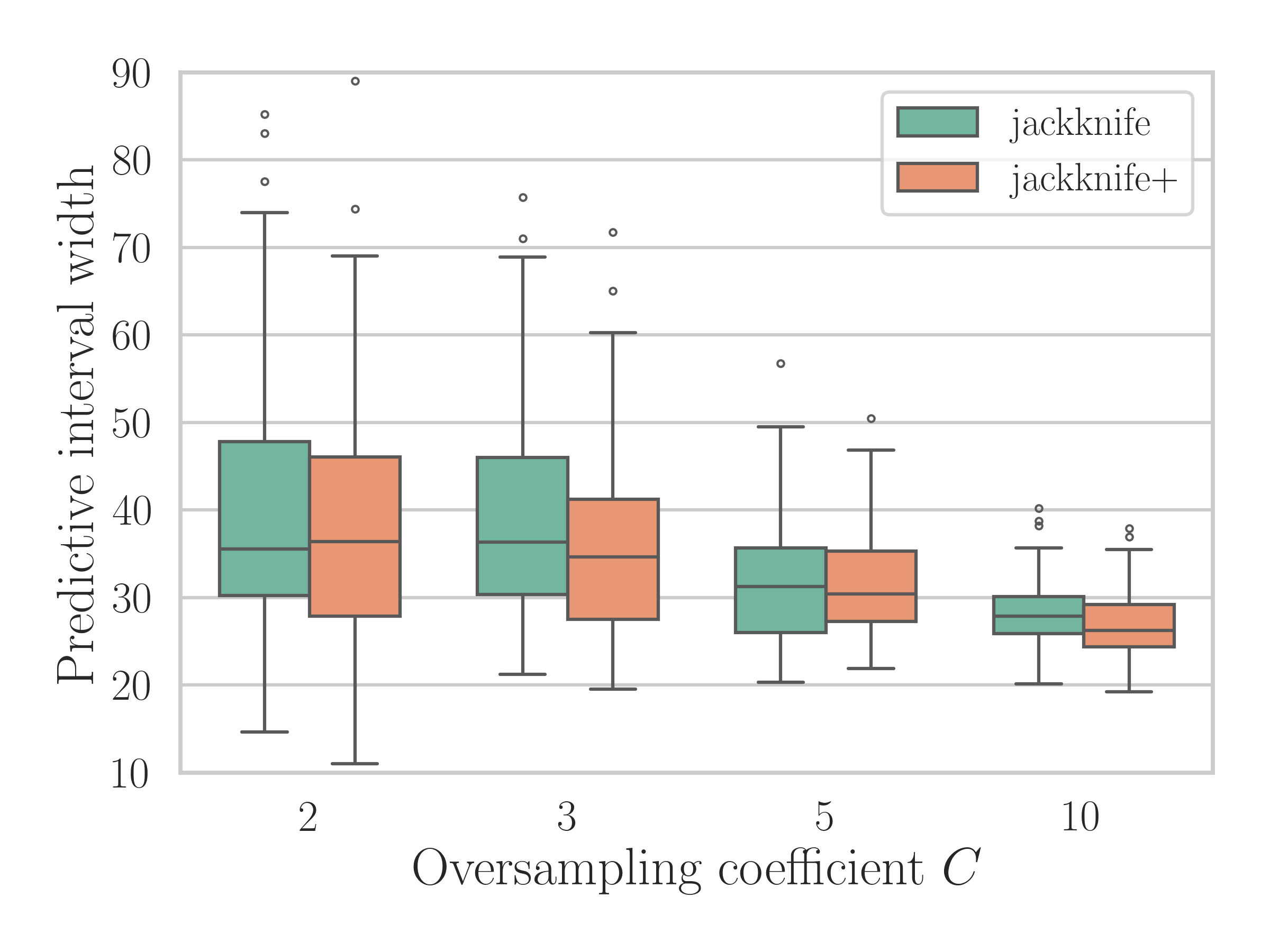}
    \caption{$P=1$.}
\end{subfigure}
\hfill
\begin{subfigure}[b]{0.45\textwidth}
    \centering
    \includegraphics[width=\textwidth]{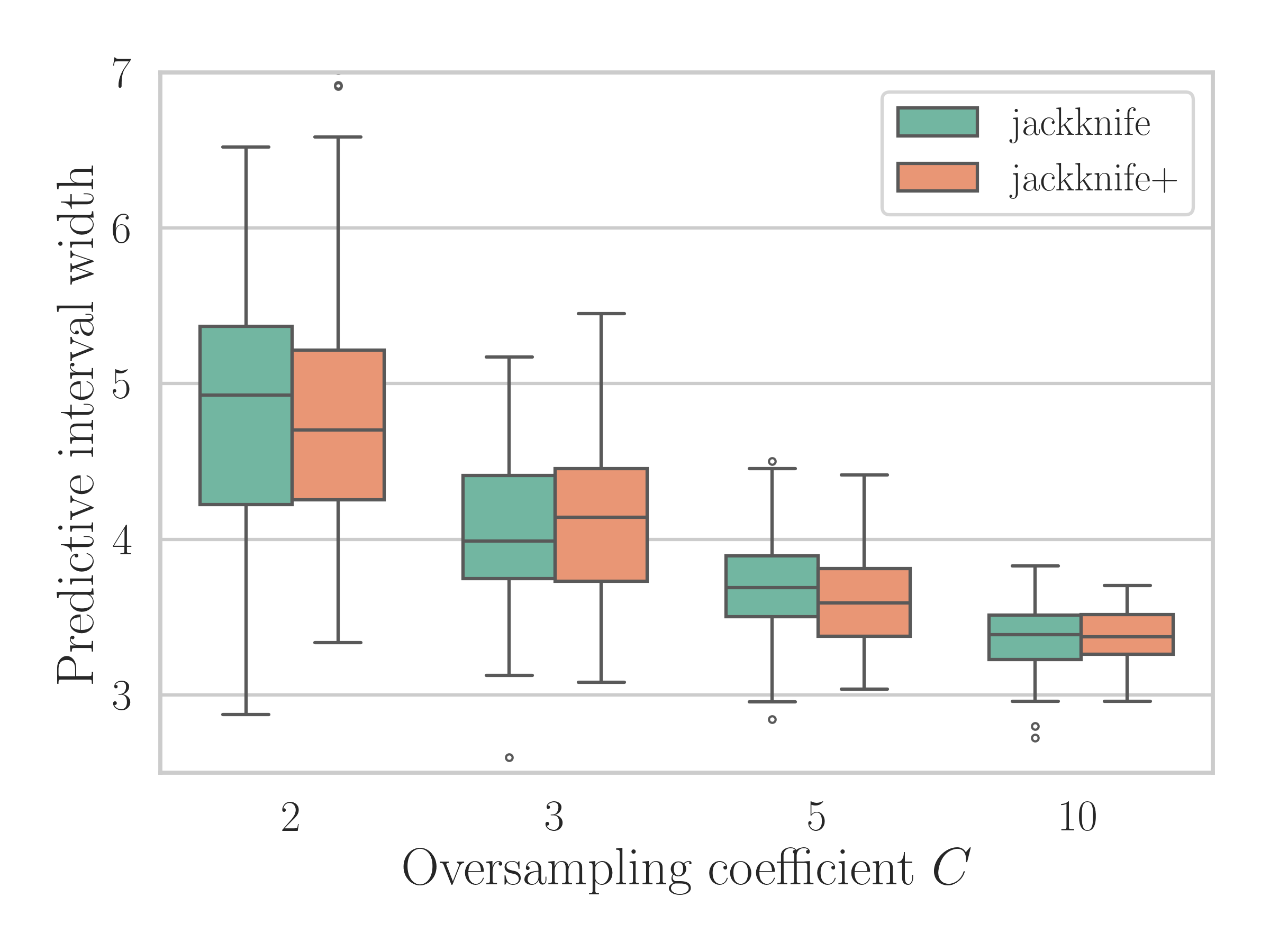}
    \caption{$P=2$.}
\end{subfigure}
\caption{Box plots of the prediction interval widths provided by conformalized \gls{pce} surrogates of the wing weight function, for different combinations of polynomial degree $P$, and oversampling coefficient $C$. \textcolor{black}{The target coverage is fixed at $95\%$, corresponding to a significance level of $s=0.05$.}}
\label{fig:wing-weight-interval-boxplots}
\end{figure}

Total-degree conformalized \glspl{pce} with maximum polynomial degrees $P \in \left\{1,2\right\}$ and oversampling coefficients $C \in \left\{2,3,5,10\right\}$ are employed. 
The experimental design scales linearly with the size of the \gls{pce} basis, i.e., $M = C K = C \#\Lambda$. 
Figure~\ref{fig:wing-weight-parity-plots} compares the different conformalized \gls{pce} configurations against the true wing weight function for a single random seed, where the results for $C=3$ have been omitted for better readability.
Figures~\ref{fig:wing-weight-coverage-boxplots} and \ref{fig:wing-weight-interval-boxplots} present coverage and prediction interval statistics over 100 random seeds.

\subsubsection{Discussion on benchmarks' results}
\label{sec:num-exp-discussion}
The numerical results presented in sections~\ref{sec:meromorphic} and \ref{sec:wing}, but also those available in appendix~\ref{sec:appendixC}, show a number of patterns and similarities.
First, the conformalized \gls{pce} is in general capable of providing empirical coverages close to the target level, more commonly above it.
For less accurate \glspl{pce}, e.g., ones corresponding to low polynomial degrees and/or small experimental designs, the coverage over 100 random seeds varies significantly.
As the accuracy of the \gls{pce} increases with larger experimental designs and higher polynomial degrees, the variability in coverage is reduced and the coverage moves closer to the target level. 
The size of the experimental design is particularly important, as a sufficiently large training dataset enables even low-accuracy conformalized \glspl{pce} to reach the target coverages.

The coverage provided by the conformalized \gls{pce} is naturally connected to the corresponding prediction intervals. 
For less accurate \glspl{pce}, the required coverage is accomplished by larger prediction intervals, which also show a large variability over 100 random seeds. 
As expected, the prediction intervals become narrower as the accuracy of the \gls{pce} increases.
Importantly, even for low-accuracy \glspl{pce}, increasing the size of the experimental design leads not only to a reduced variability in the prediction interval, but also to a reduced interval width.
\textcolor{black}{
It is worth noting that interval widths for different polynomial orders can differ significantly and do not necessarily converge to the same median value, even for large training datasets. 
This behavior is expected and reflects the irreducible structural approximation bias of \gls{pce} models, which is directly affected by the choice of polynomial order. 
That is, the residuals remain bounded below by the structural approximation error of the model, regardless of training data availability. 
Higher-degree models capture the underlying function with higher fidelity, resulting in smaller residuals and correspondingly narrower, more efficient prediction intervals.
}

Comparing the jackknife and jackknife+ methods, their prediction intervals are in general very similar, especially for larger experimental designs and more accurate \glspl{pce}. 
The jackknife method is typically more conservative and results in comparatively larger intervals and to over-coverage, especially for smaller experimental designs. 
On the other hand, jackknife+ produces empirical coverages closer to the target level, in most cases.
For both jackknife and jackknife+, it is clear that training data availability plays a significant role.

\subsection{Application to electromagnet design}
\label{sec:stern-gerlach}
\begin{figure}[t!]
\centering
\begin{subfigure}[b]{0.48\textwidth}
    \centering
    \includegraphics[width=\textwidth]{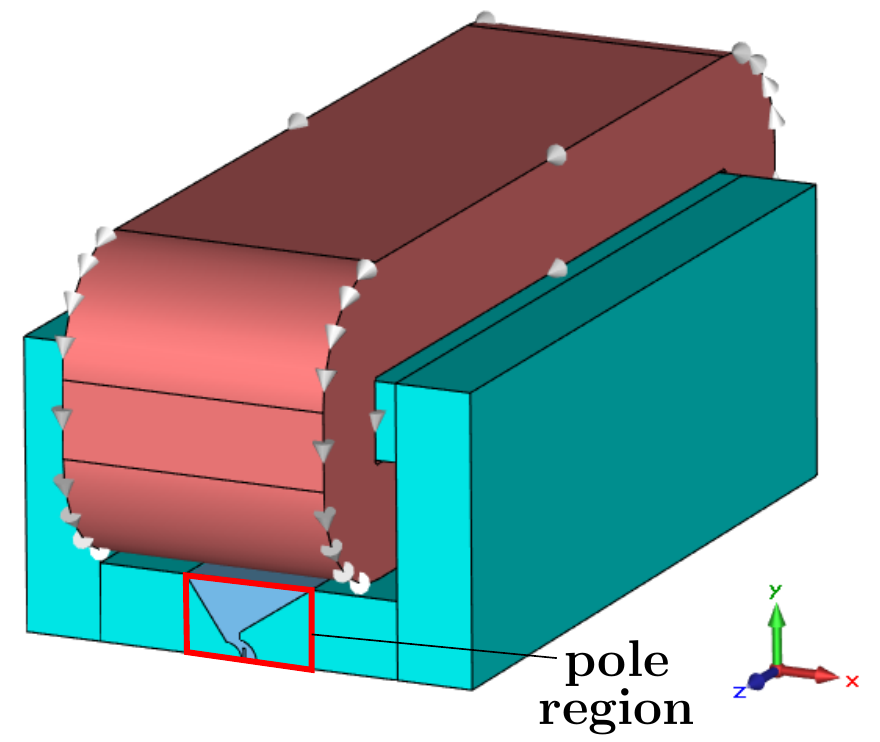}
    \caption{Stern-Gerlach magnet model (half geometry).}
    \label{fig:stern-gerlach-magnet-3D}
\end{subfigure}
\hfill
\begin{subfigure}[b]{0.48\textwidth}
    \centering
    \includegraphics[width=\textwidth]{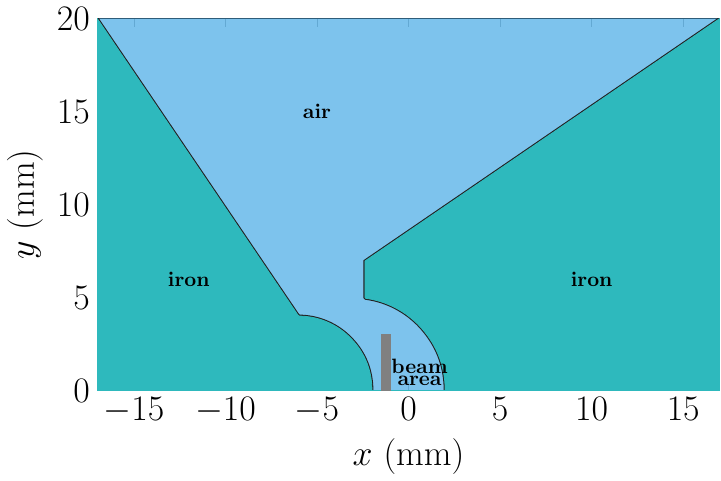}
    \caption{Pole region geometry.}
    \label{fig:stern-gerlach-magnet-zoom}
\end{subfigure}
\caption{Illustration of the Stern-Gerlach magnet model.}
\label{fig:stern-gerlach-magnet}
\end{figure}

As an engineering design use-case, we consider a so-called Stern-Gerlach electromagnet, typically employed for the magnetic separation of atom beams \cite{masschaele2011design}. 
A 3D model of the magnet is depicted in Figure~\ref{fig:stern-gerlach-magnet-3D}, where half of the geometry is shown.
This model is generated with the CST Studio Suite$^\text{\textregistered}$ commercial software\footnote{www.3ds.com/products/simulia/cst-studio-suite}.
A key design requirement of such magnets is a magnetic field with a homogeneous and strong gradient. 
Field homogeneity is particularly important in the magnet's beam area which is included in the pole region of the magnet, as shown in Figure~\ref{fig:stern-gerlach-magnet-zoom}.

The target model output is the average magnetic field gradient in the beam area, denoted as $\overline{\tau}_{\text{b}}$ and measured in \si{\tesla\per\meter}.
For computational efficiency, $\overline{\tau}_{\text{b}}$ is computed with a 2D \gls{iga} model of the pole region, while the coils and yoke of the original 3D model are replaced by equivalent circuit models \cite{pels2015optimization}.
Then, $\overline{\tau}_{\text{b}}$ is estimated as 
\begin{equation}
\overline{\tau}_{\text{b}} = \frac{1}{\Omega_\text{b}} \int_{\Omega_\text{b}} \tau(x,y) \mathrm{d}\Omega,
\end{equation}
where $\Omega$ denotes the full 2D computational domain, $\Omega_\text{b}$ corresponds to the beam area, and $\tau(x,y) = \frac{\partial \left| \vec{B} \right|}{\partial x}$ is the magnetic field gradient in the $x$-direction, where $\vec{B}$ denotes the magnetic flux density. 
The latter is obtained by the \gls{iga} model's field solution.

While much more efficient compared to the original 3D model, the circuit-coupled \gls{iga} model is still too computationally expensive for design optimization \cite{masschaele2011design, pels2015optimization} or \gls{uq} \cite{loukrezis2019assessing} studies, hence the need for a surrogate model. 
Here, the surrogate model is used to approximate $\overline{\tau}_{\text{b}}$ under random shape deformations in the pole region. 
The geometry of the pole region is described using \gls{nurbs} $\mathbf{R}\left(\xi\right)$, $\mathbf{\xi} \in \left[0,1\right]$, defined by control points  $\mathbf{P}_i$, $i=1,\dots,N_{\text{cp}}$, given as
\begin{equation}
\mathbf{R}\left(\xi\right) = \sum_{i=1}^{N_{\text{cp}}} \mathbf{P}_i \frac{w_i B_i^p\left(\xi\right)}{\sum_{j=1}^{N_{\text{cp}}} w_j B_j^p\left(\xi\right)},
\end{equation} 
where $B_i^p\left(\xi\right)$ are B-spline basis functions of order $p$ and $w_i$ are their corresponding weights.
Random shape deformations are introduced by modeling the coordinates of five control points and four of the corresponding weights as uniform random variables. 
All input parameters and their ranges are listed in Table~\ref{tab:stern-gerlach-parameters}.

\begin{table}[b!]
\centering
\caption{Input parameters of the Stern-Gerlach magnet model.}
\label{tab:stern-gerlach-parameters}
\begin{threeparttable}
\begin{tabular}{c c c l}
\toprule 
Parameter & Description & Unit & Range \\ 
\midrule 
$x_1$ & $x$-coordinate, control point 1 &  \si{\milli\meter} & $\left[-3.38, -1.38\right]$ \\ 
$y_1$ & $y$-coordinate, control point 1 & \si{\milli\meter} & $\left[5.96, 7.96\right]$ \\ 
$x_2$ & $x$-coordinate, control point 2 &  \si{\milli\meter} & $\left[-3.38, -1.38\right]$ \\ 
$y_2$ & $y$-coordinate, control point 2 & \si{\milli\meter} & $\left[3.96, 5.96\right]$ \\
$x_3$ & $x$-coordinate, control point 3 &  \si{\milli\meter} & $\left[16, 18\right]$ \\ 
$y_3$ & $y$-coordinate, control point 3 & \si{\milli\meter} & $\left[19, 21\right]$ \\ 
$x_4$ & $x$-coordinate, control point 4 &  \si{\milli\meter} & $\left[-18, -16\right]$ \\ 
$y_4$ & $y$-coordinate, control point 4 & \si{\milli\meter} & $\left[19, 21\right]$ \\
$x_5$ & $x$-coordinate, control point 5 &  \si{\milli\meter} & $\left[-7, -5\right]$ \\ 
$y_5$ & $y$-coordinate, control point 5 & \si{\milli\meter} & $\left[3, 5\right]$ \\ 
$w_1$ & weight, control point 1 &  -- & $\left[0, 1\right]$ \\ 
$w_2$ & weight, control point 2 &  -- & $\left[0, 1\right]$ \\ 
$w_3$ & weight, control point 3 &  -- & $\left[0, 1\right]$ \\ 
$w_4$ & weight, control point 4 &  -- & $\left[0, 1\right]$ \\
\bottomrule
\end{tabular}
\end{threeparttable}
\end{table}

Two significance levels $s=0.05$ and $s=0.01$ are considered regarding the predictive uncertainty of the surrogate model.  
The conformalized \gls{pce} is applied with maximum polynomial degrees $P \in \left\{1,2\right\}$. 
A dataset with $1000$ different shape deformations of the pole region is available, along with the corresponding $\overline{\tau}_{\text{b}}$ values. 
Training datasets of increasing size $M \in \left\{150, 200, \dots, 500\right\}$ are employed. 
The remaining $M'=500$ data points are used for validation.
The dataset is shuffled $100$ times to estimate coverage and prediction interval statistics.

\begin{figure}[t!]
\centering
\begin{subfigure}[b]{0.32\textwidth}
    \centering
    \includegraphics[width=\textwidth]{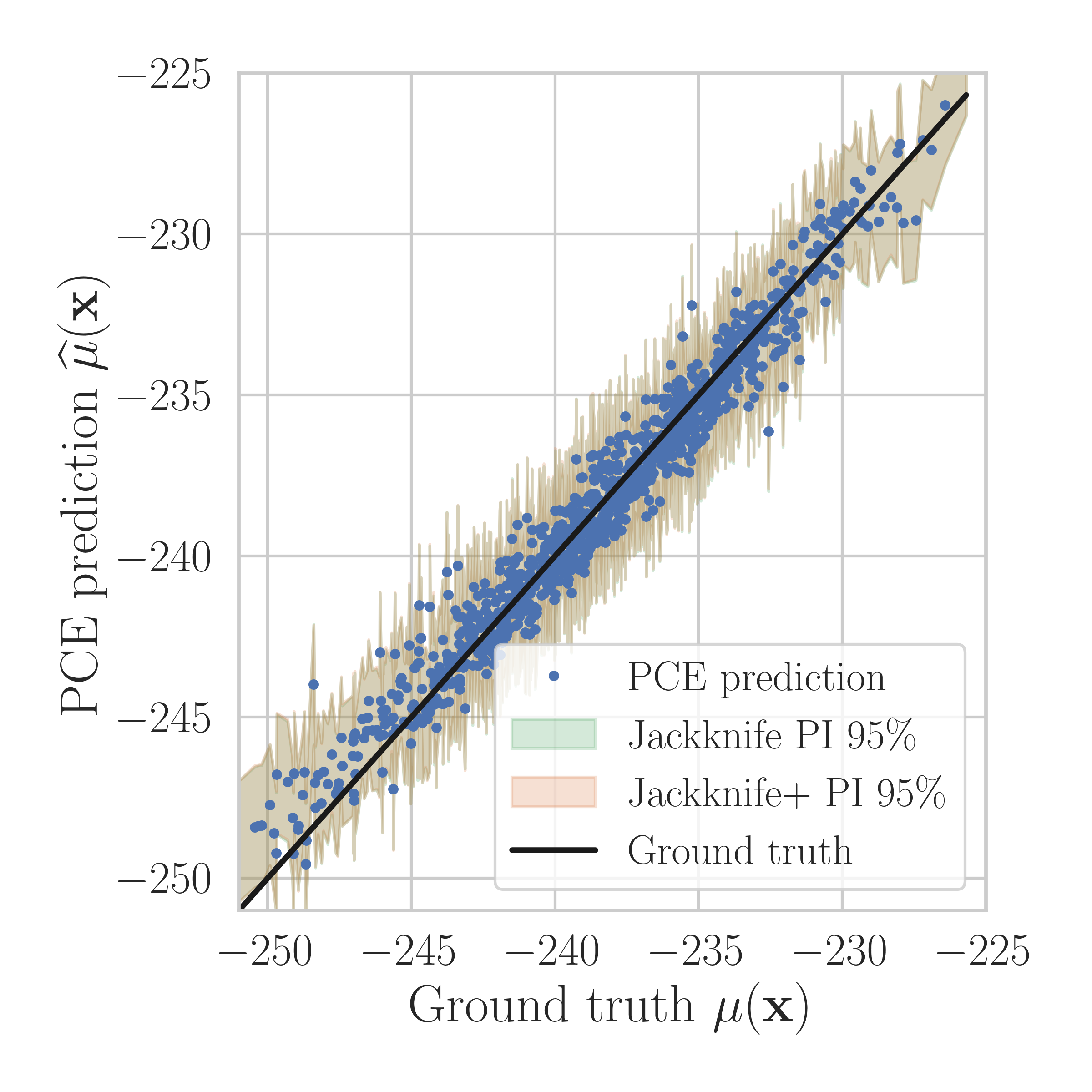}
    \caption{$P=1$, $M=150$.}
\end{subfigure}
\hfill
\begin{subfigure}[b]{0.32\textwidth}
    \centering
    \includegraphics[width=\textwidth]{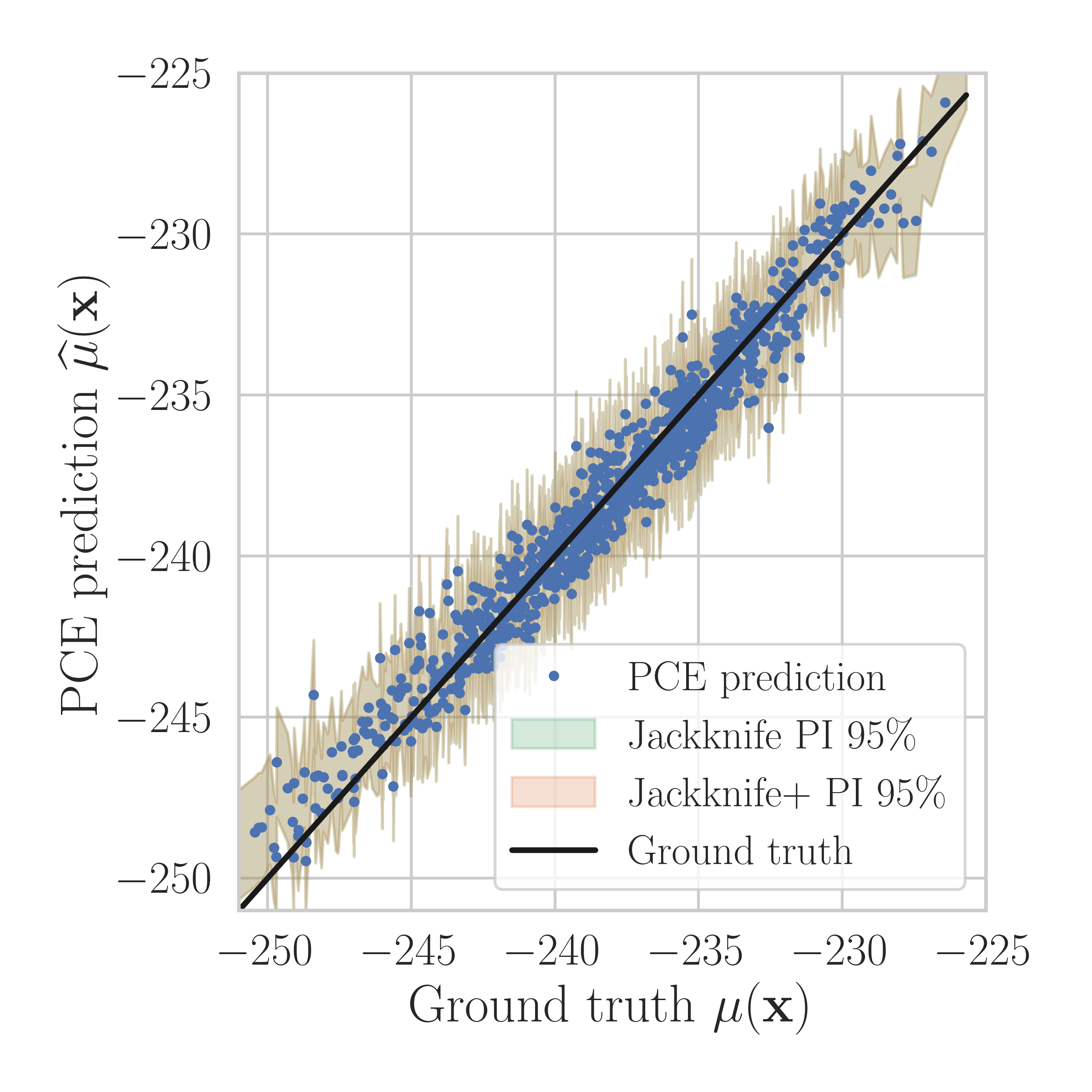}
    \caption{ $P=1$, $M=300$.}
\end{subfigure}
\hfill
\begin{subfigure}[b]{0.32\textwidth}
    \centering
    \includegraphics[width=\textwidth]{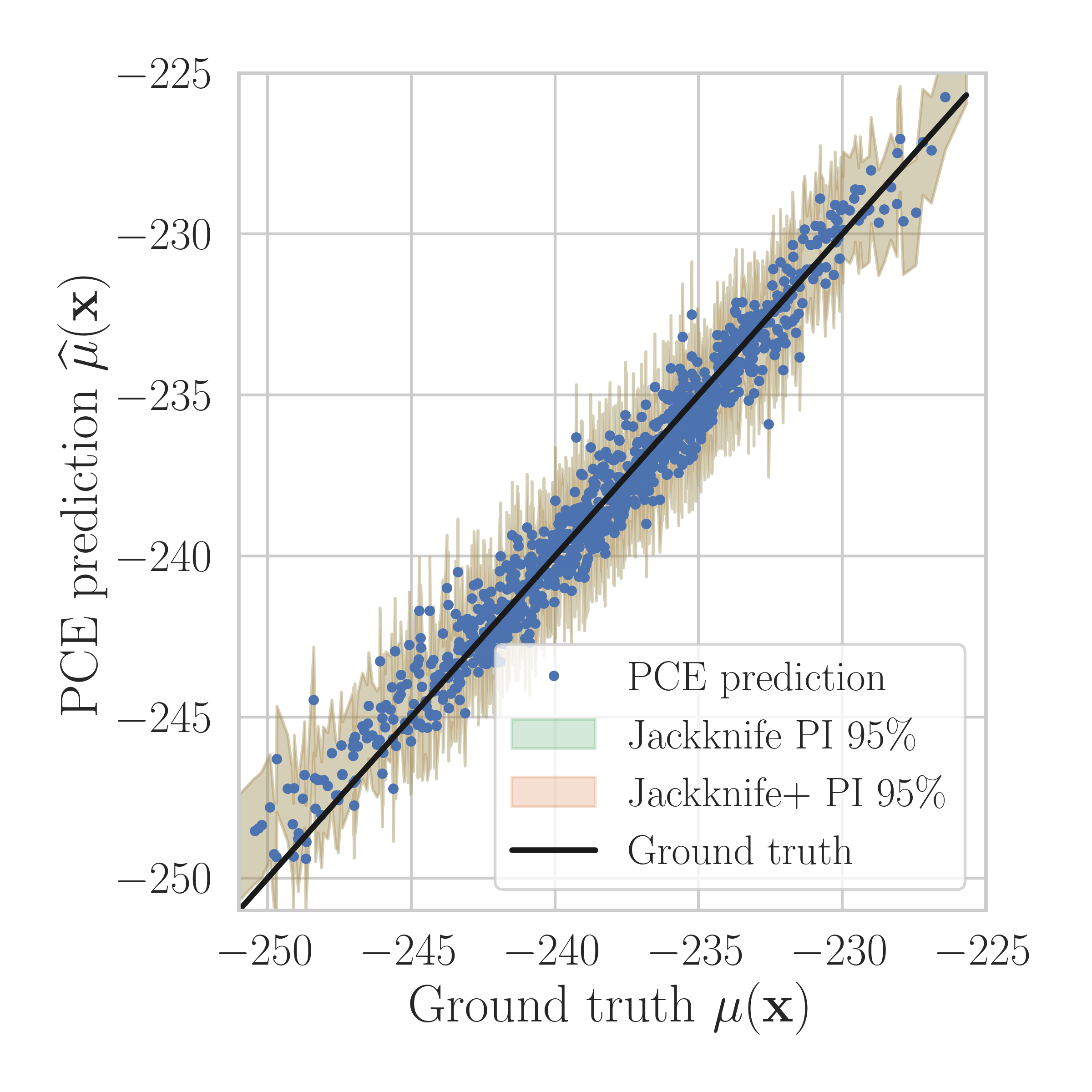}
    \caption{ $P=1$, $M=450$.}
\end{subfigure}
\\
\begin{subfigure}[b]{0.32\textwidth}
    \centering
    \includegraphics[width=\textwidth]{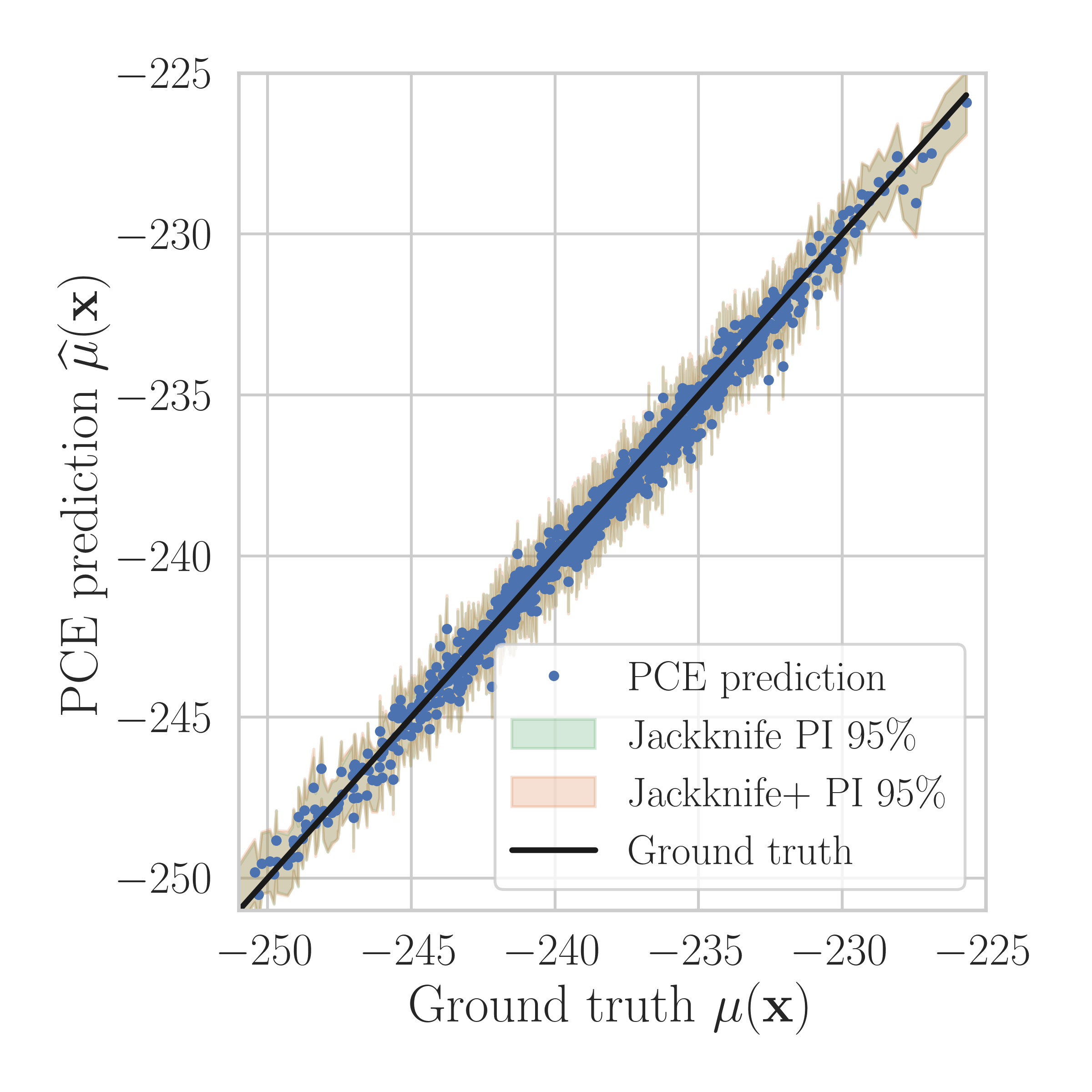}
    \caption{ $P=2$, $M=150$.}
\end{subfigure}
\hfill
\begin{subfigure}[b]{0.32\textwidth}
    \centering
    \includegraphics[width=\textwidth]{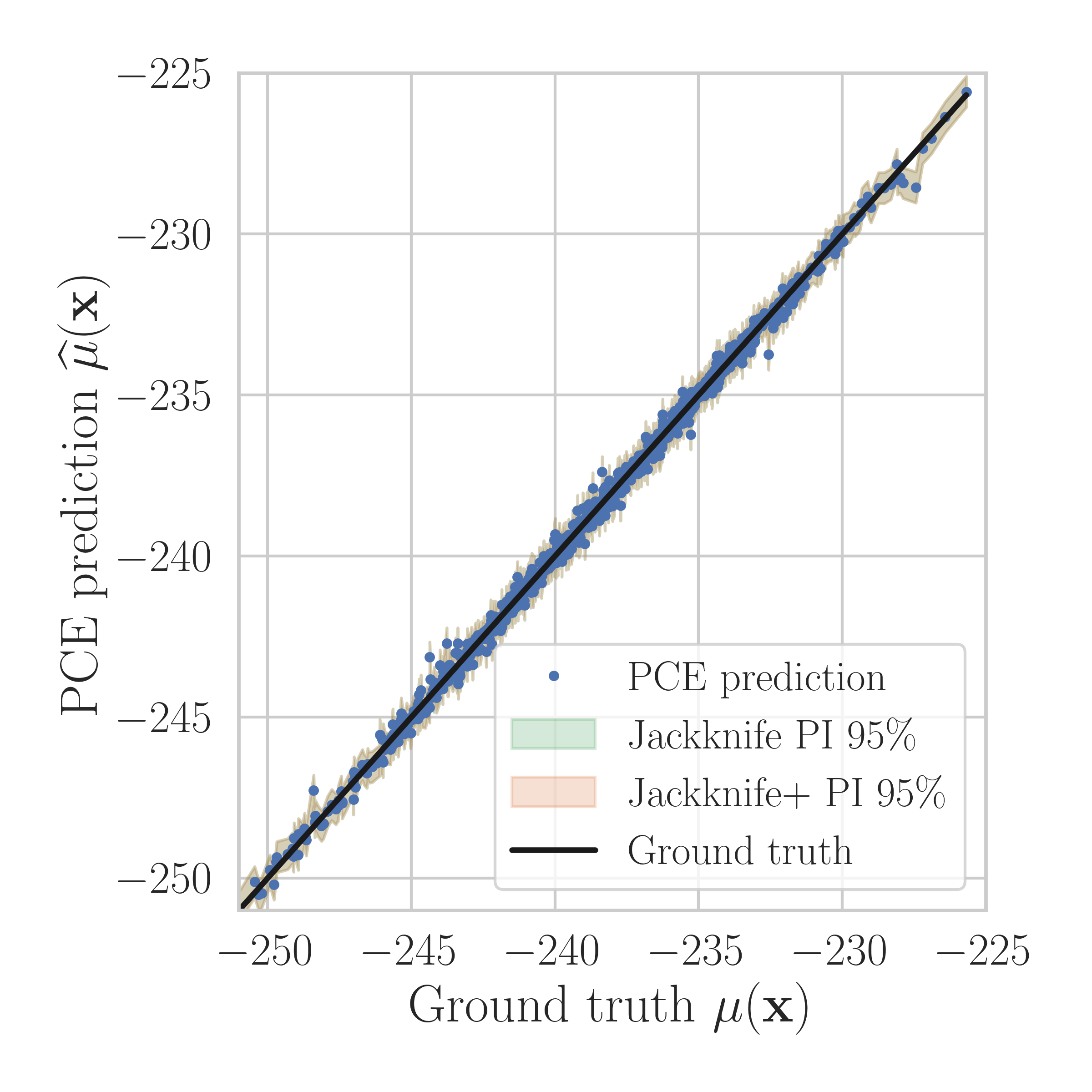}
    \caption{ $P=2$, $M=300$.}
\end{subfigure}
\hfill
\begin{subfigure}[b]{0.32\textwidth}
    \centering
    \includegraphics[width=\textwidth]{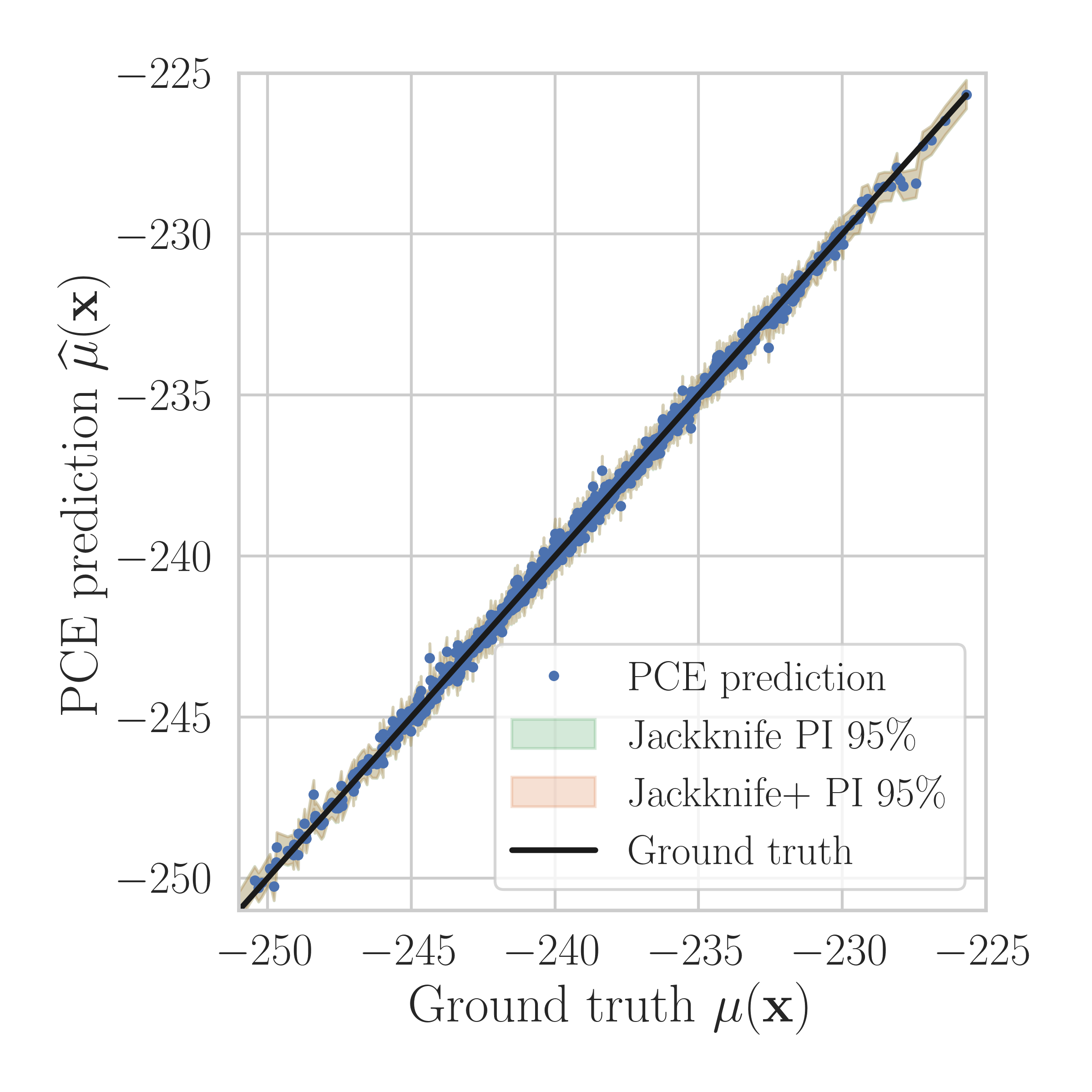}
    \caption{ $P=2$, $M=450$.}
\end{subfigure}
\caption{Parity plots comparing ground truth $\overline{\tau}_{\text{b}}$ values given by the Stern-Gerlach magnet model against conformalized \gls{pce} predictions for different combinations of polynomial degree $P$, training dataset size $M$ and significance level $s$. The results correspond to a single partition of the available design dataset. \textcolor{black}{The target coverage is set to $95\%$, corresponding to a significance level of $s=0.05$.}}
\label{fig:stern-gerlach-parity-plots-005}
\end{figure}

\begin{figure}[t!]
\centering
\begin{subfigure}[b]{0.32\textwidth}
    \centering
    \includegraphics[width=\textwidth]{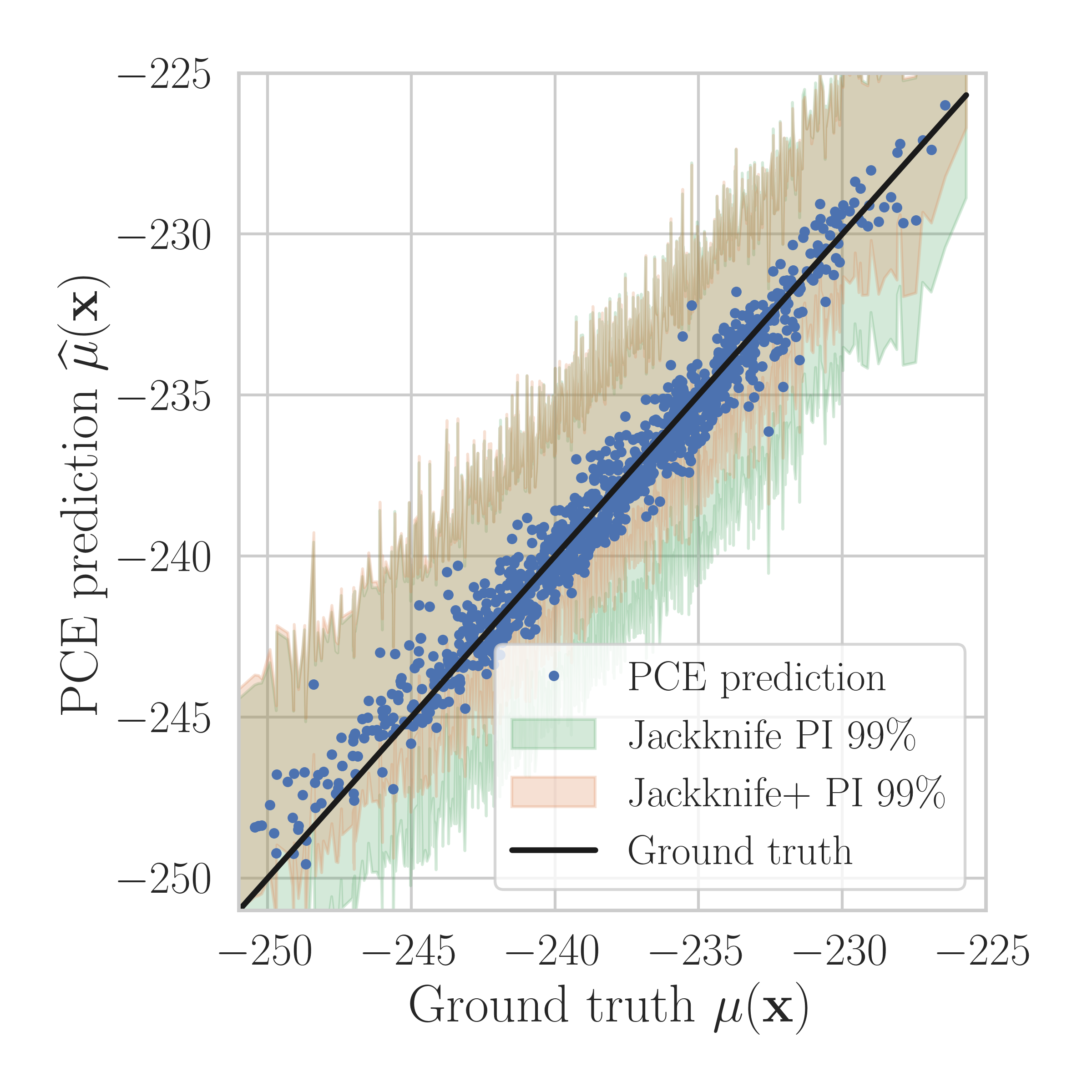}
    \caption{ $P=1$, $M=150$.}
\end{subfigure}
\hfill
\begin{subfigure}[b]{0.32\textwidth}
    \centering
    \includegraphics[width=\textwidth]{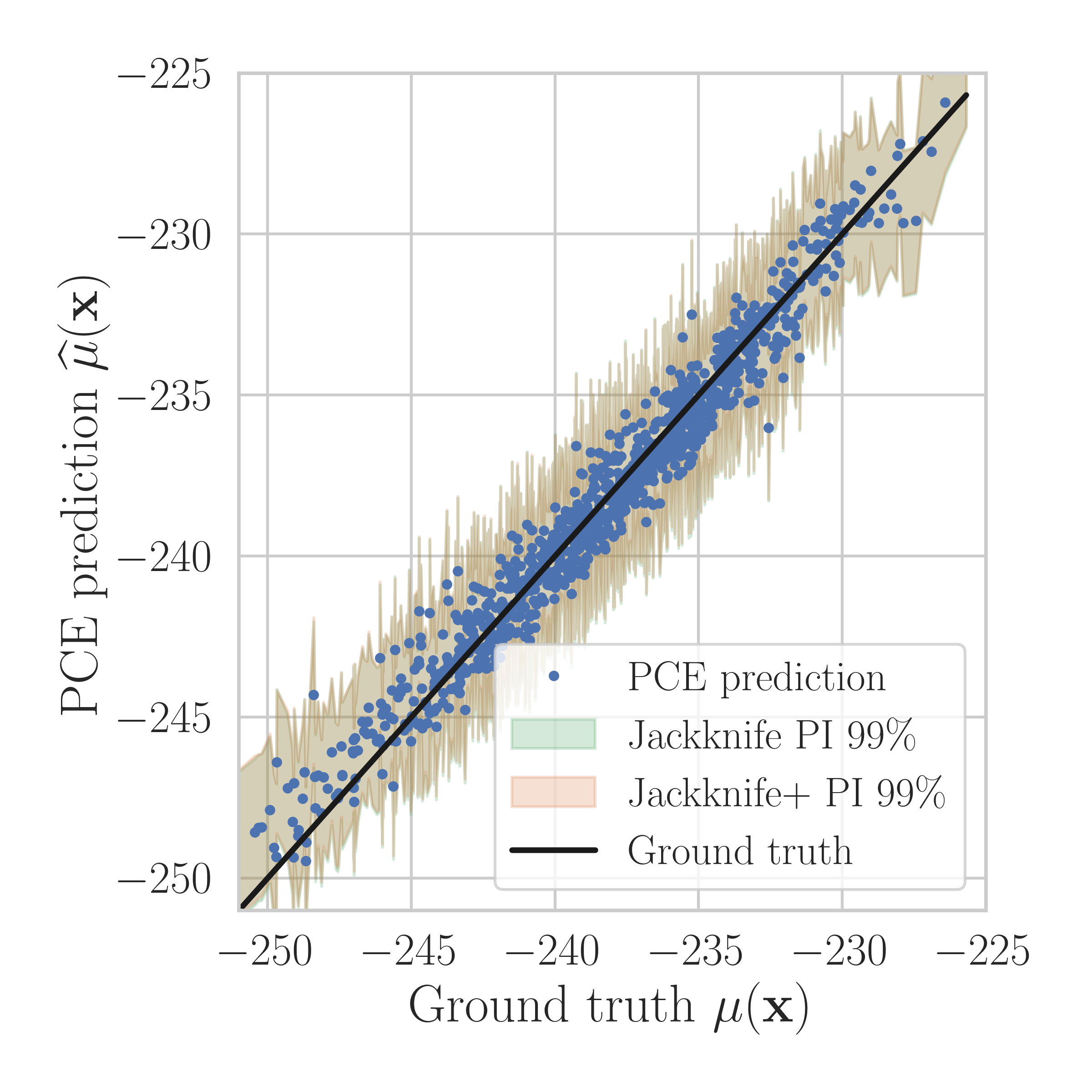}
    \caption{ $P=1$, $M=300$.}
\end{subfigure}
\hfill
\begin{subfigure}[b]{0.32\textwidth}
    \centering
    \includegraphics[width=\textwidth]{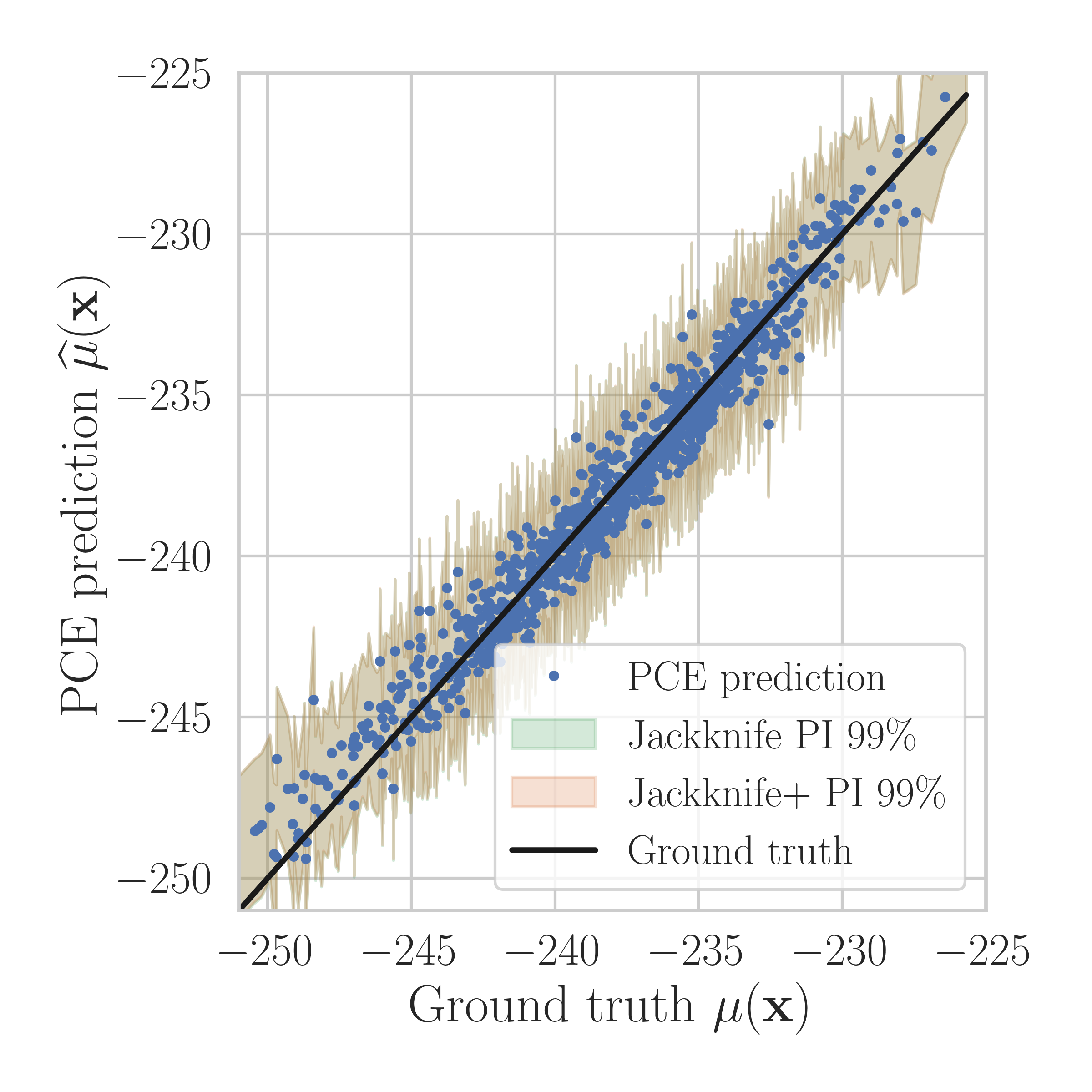}
    \caption{ $P=1$, $M=450$.}
\end{subfigure}
\\
\begin{subfigure}[b]{0.32\textwidth}
    \centering
    \includegraphics[width=\textwidth]{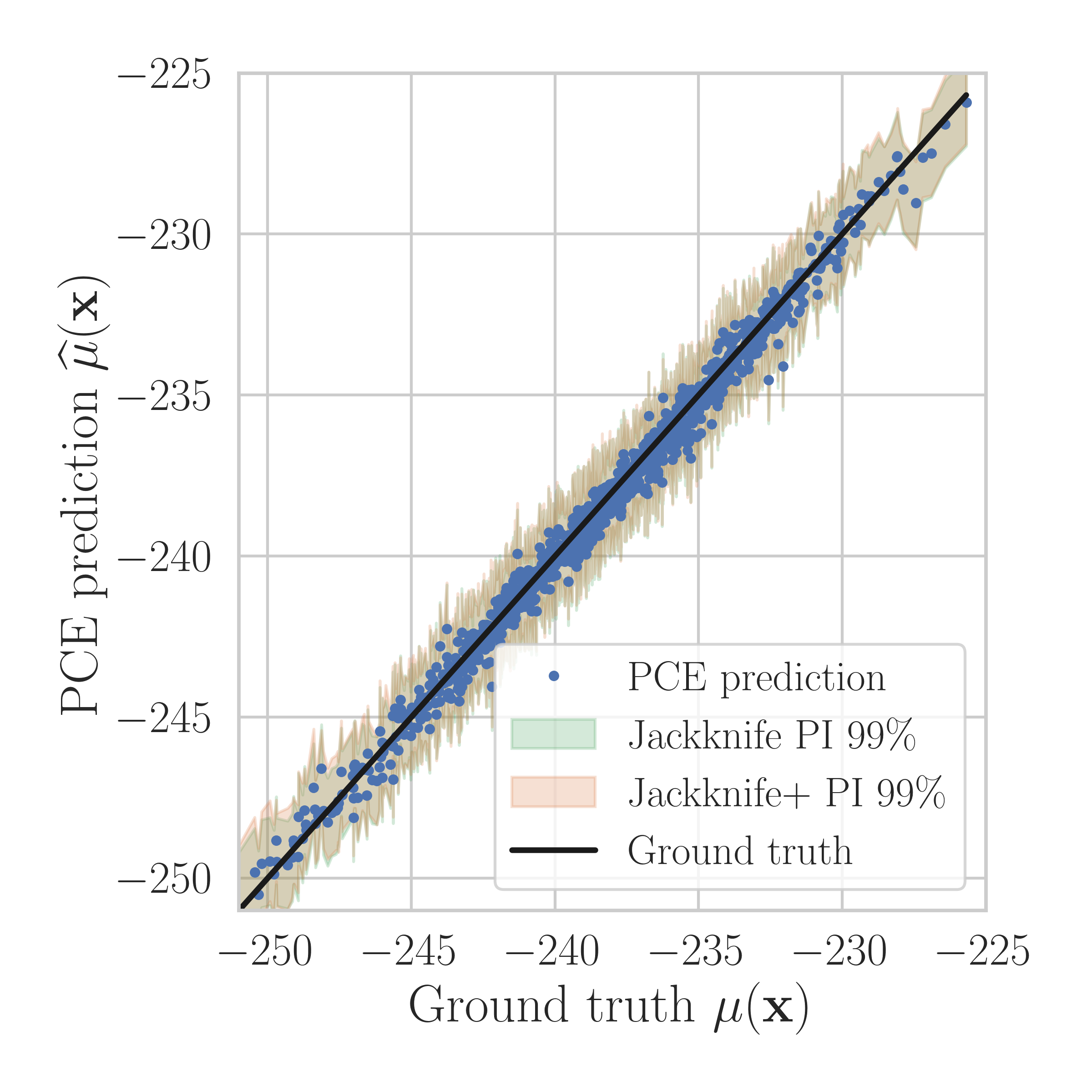}
    \caption{ $P=2$, $M=150$.}
\end{subfigure}
\hfill
\begin{subfigure}[b]{0.32\textwidth}
    \centering
    \includegraphics[width=\textwidth]{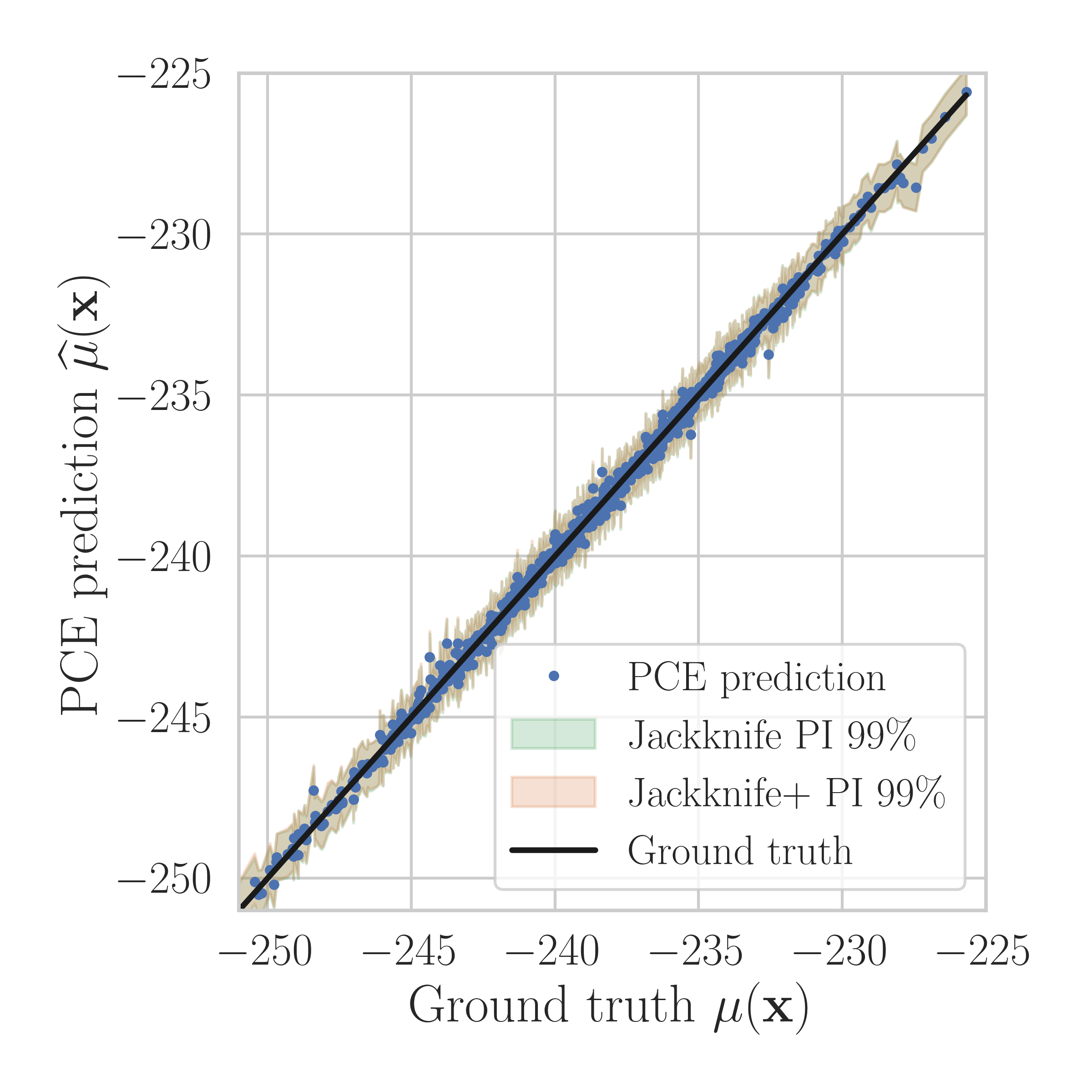}
    \caption{ $P=2$, $M=300$.}
\end{subfigure}
\hfill
\begin{subfigure}[b]{0.32\textwidth}
    \centering
    \includegraphics[width=\textwidth]{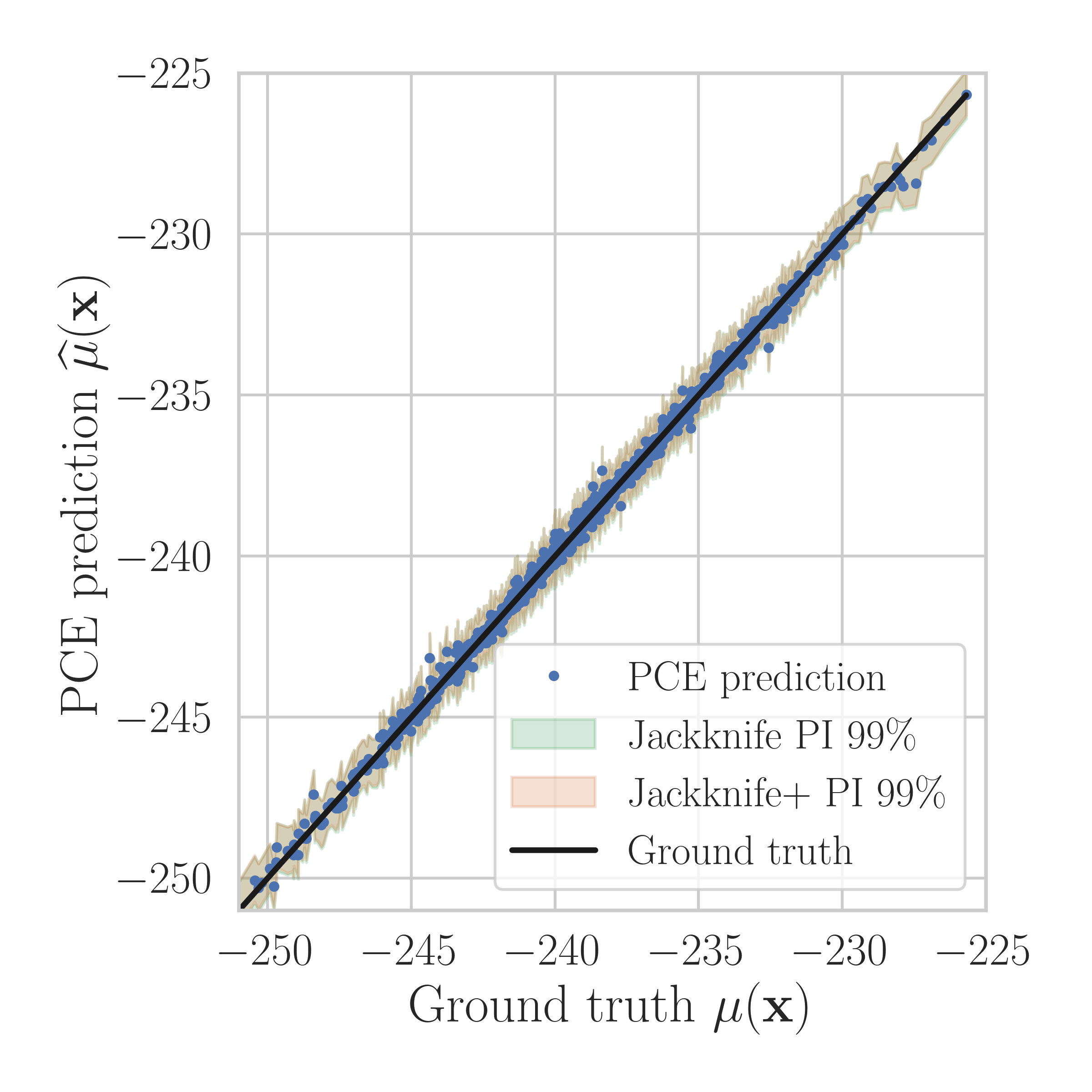}
    \caption{ $P=2$, $M=450$.}
\end{subfigure}
\caption{Parity plots comparing ground truth $\overline{\tau}_{\text{b}}$ values given by the Stern-Gerlach magnet model against conformalized \gls{pce} predictions for different combinations of polynomial degree $P$, training dataset size $M$ and significance level $s$. The results correspond to a single partition of the available design dataset. \textcolor{black}{The target coverage is set to $99\%$, corresponding to a significance level of $s=0.01$.}}
\label{fig:stern-gerlach-parity-plots-001}
\end{figure}

\begin{figure}[t!]
\centering
\begin{subfigure}[b]{0.45\textwidth}
    \centering
    \includegraphics[width=\textwidth]{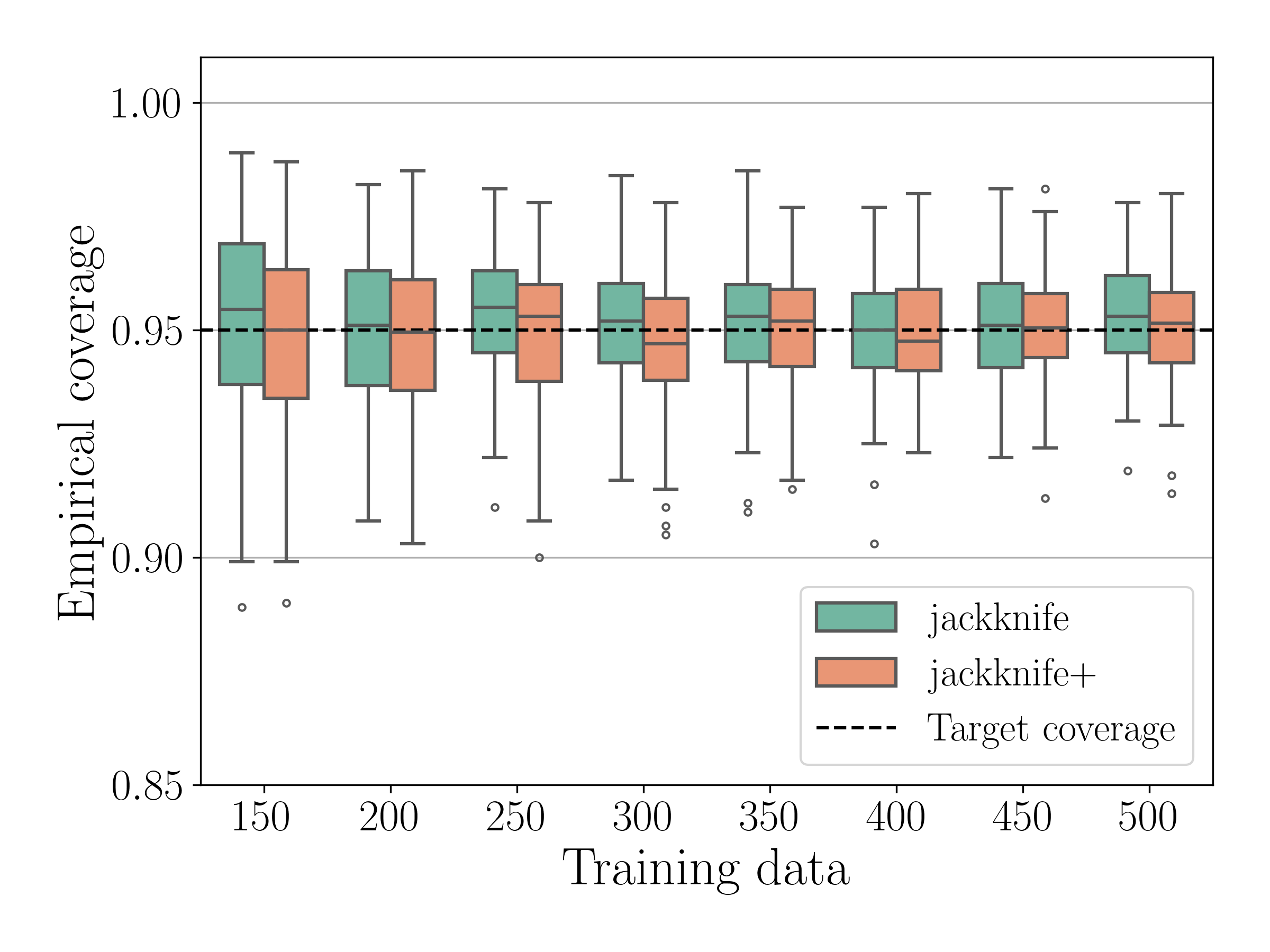}
    \caption{$P=1$, $s=0.05$.}
\end{subfigure}
\hfill
\begin{subfigure}[b]{0.45\textwidth}
    \centering
    \includegraphics[width=\textwidth]{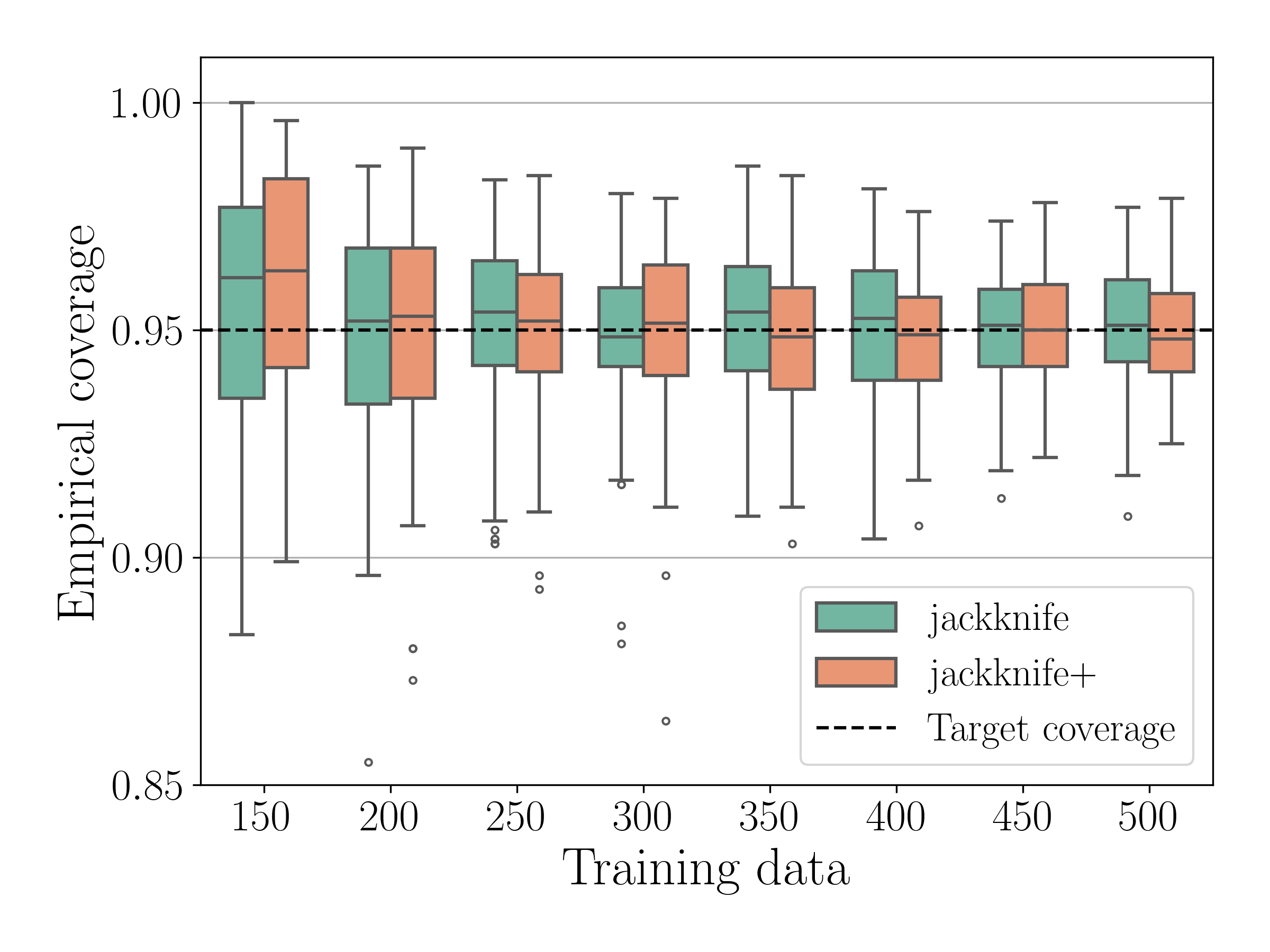}
    \caption{$P=2$, $s=0.05$.}
\end{subfigure}
\\
\begin{subfigure}[b]{0.45\textwidth}
    \centering
    \includegraphics[width=\textwidth]{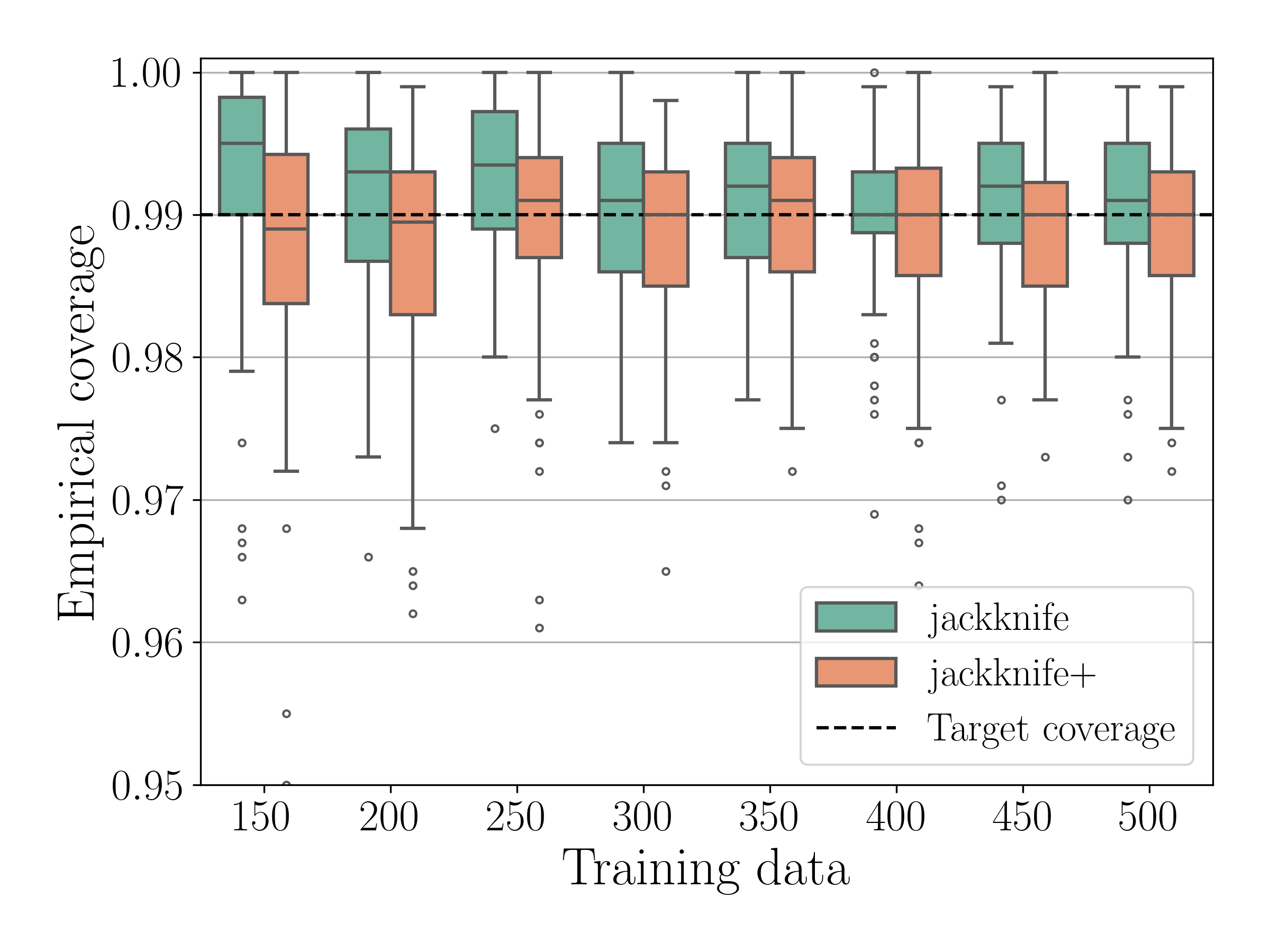}
    \caption{$P=1$, $s=0.01$.}
\end{subfigure}
\hfill
\begin{subfigure}[b]{0.45\textwidth}
    \centering
    \includegraphics[width=\textwidth]{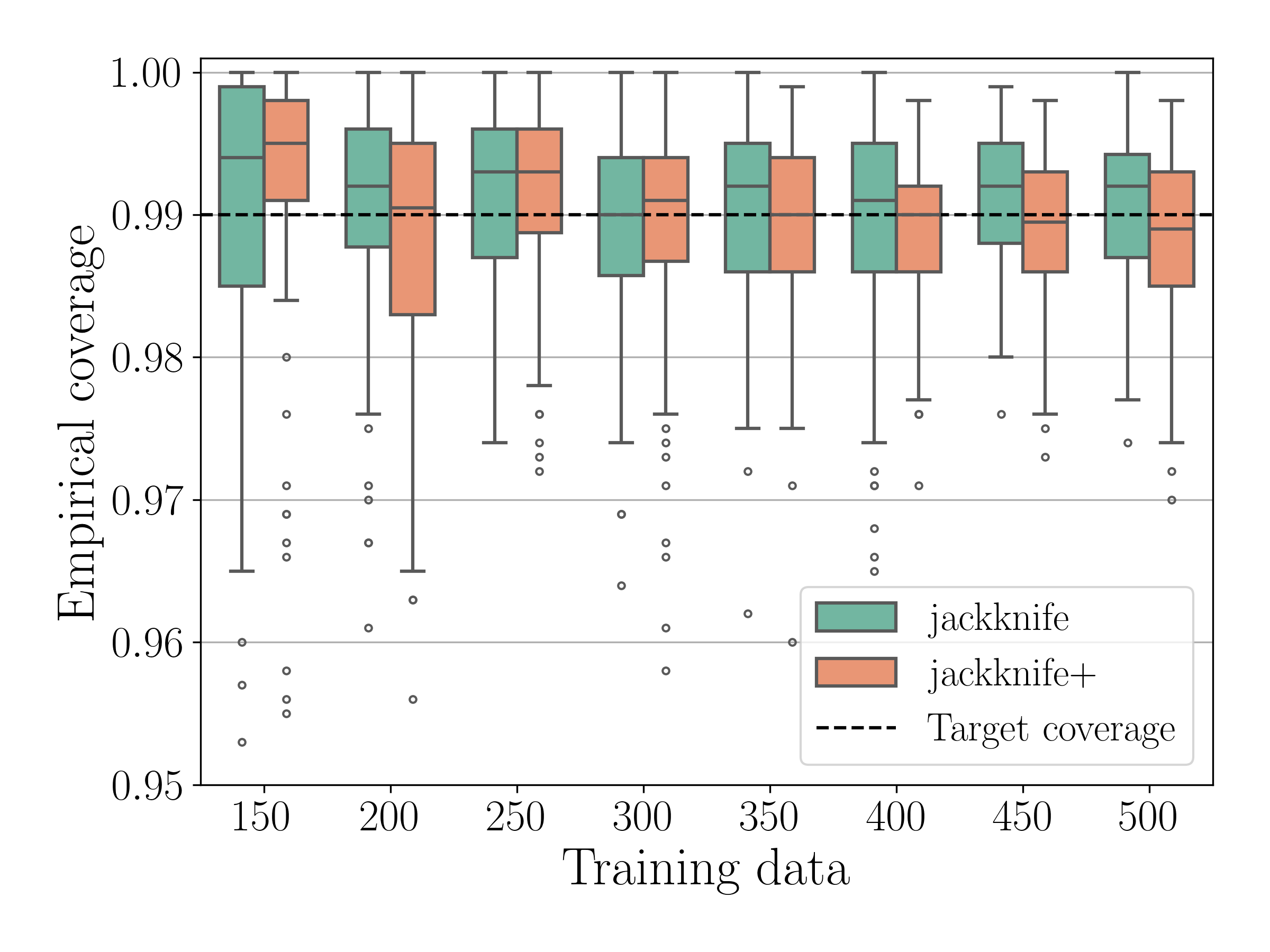}
    \caption{$P=2$, $s=0.01$.}
\end{subfigure}
\caption{Box plots of the empirical coverage provided by conformalized \gls{pce} surrogates of the Stern-Gerlach magnet model, for different combinations of polynomial degree $P$, training dataset size $M$, and significance level $s$.}
\label{fig:stern-gerlach-coverage-boxplots}
\end{figure}

\begin{figure}[t!]
\centering
\begin{subfigure}[b]{0.45\textwidth}
    \centering
    \includegraphics[width=\textwidth]{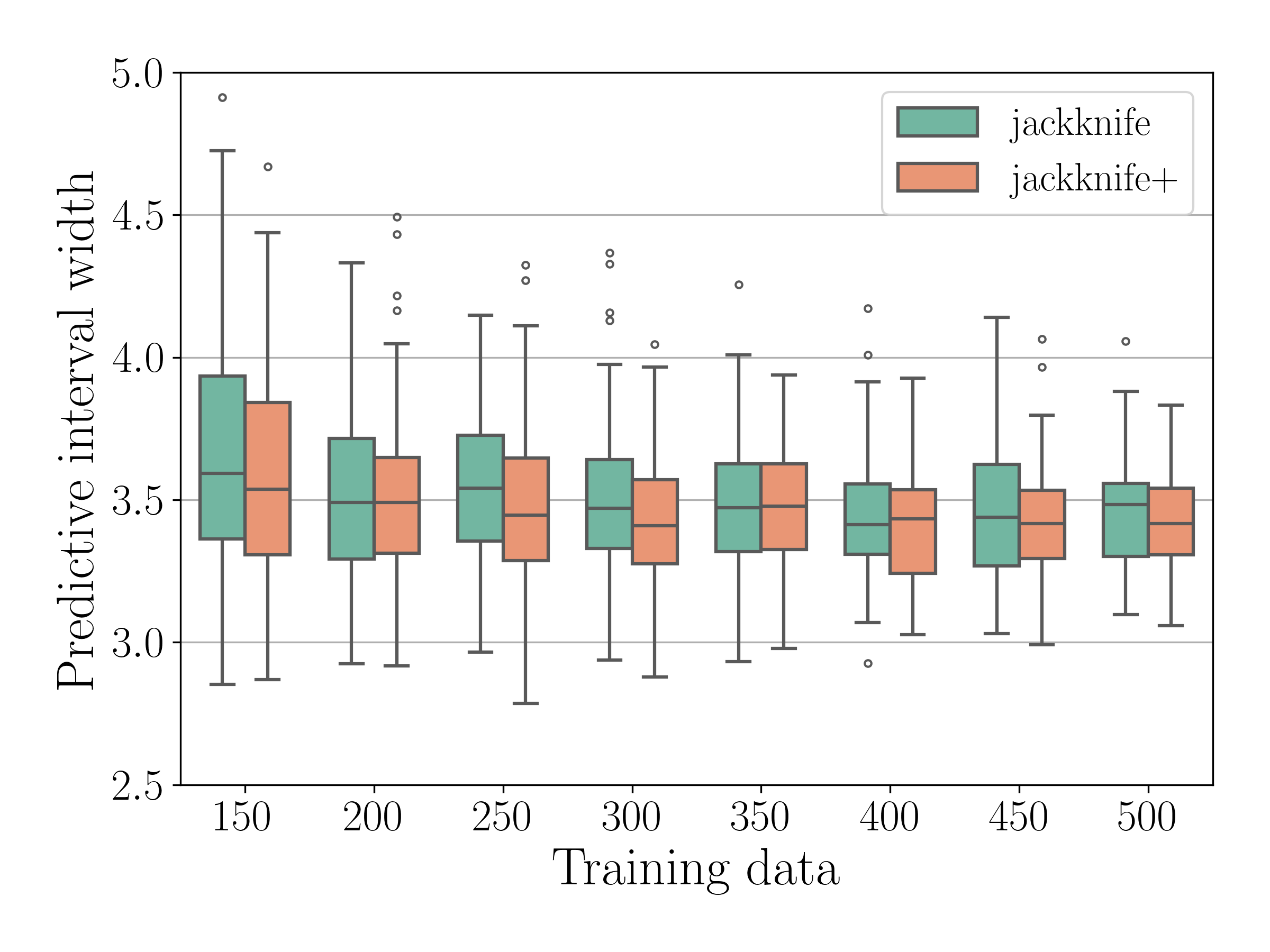}
    \caption{$P=1$, $s=0.05$.}
\end{subfigure}
\hfill
\begin{subfigure}[b]{0.45\textwidth}
    \centering
    \includegraphics[width=\textwidth]{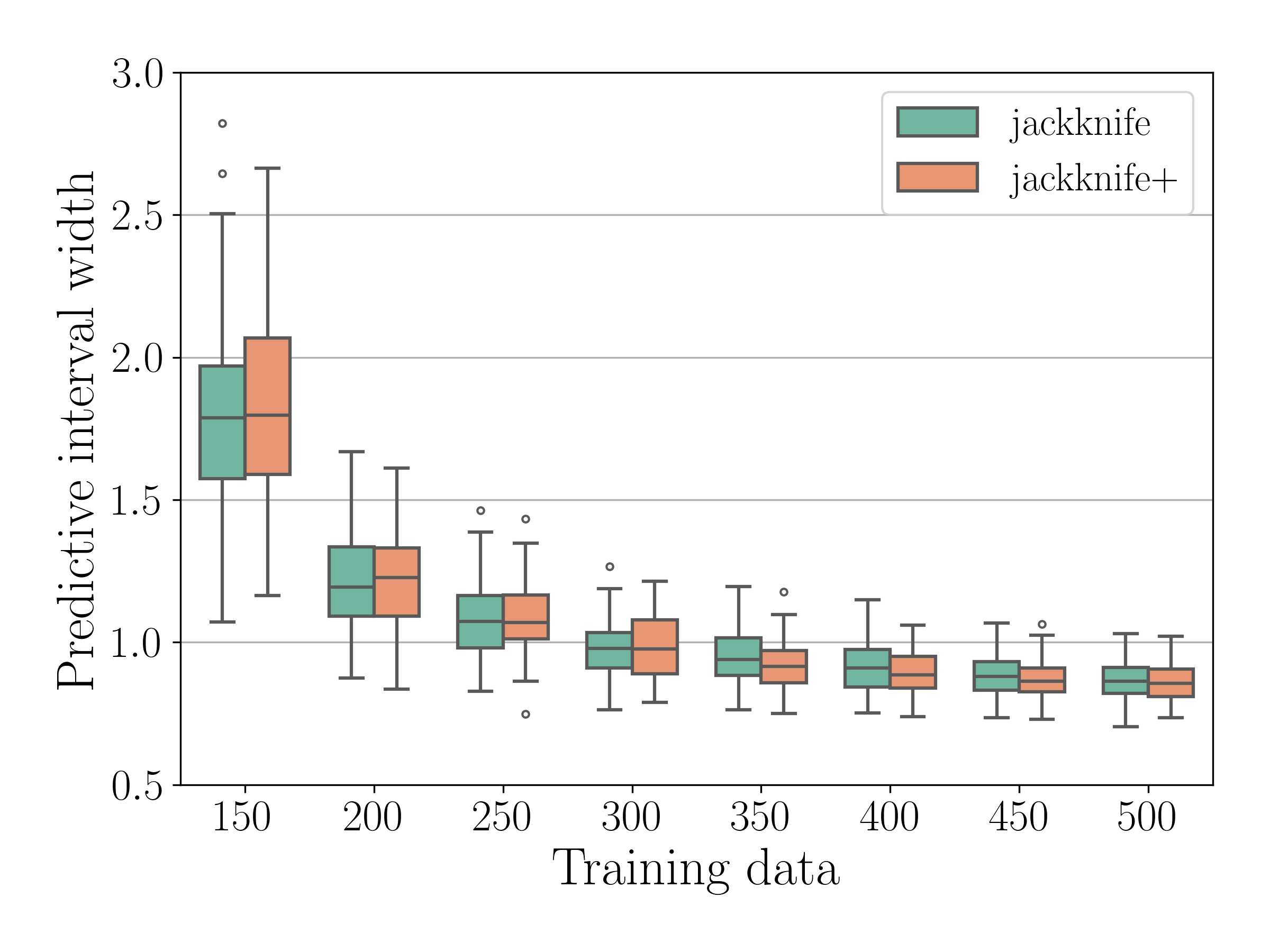}
    \caption{$P=2$, $s=0.05$.}
\end{subfigure}
\\
\begin{subfigure}[b]{0.45\textwidth}
    \centering
    \includegraphics[width=\textwidth]{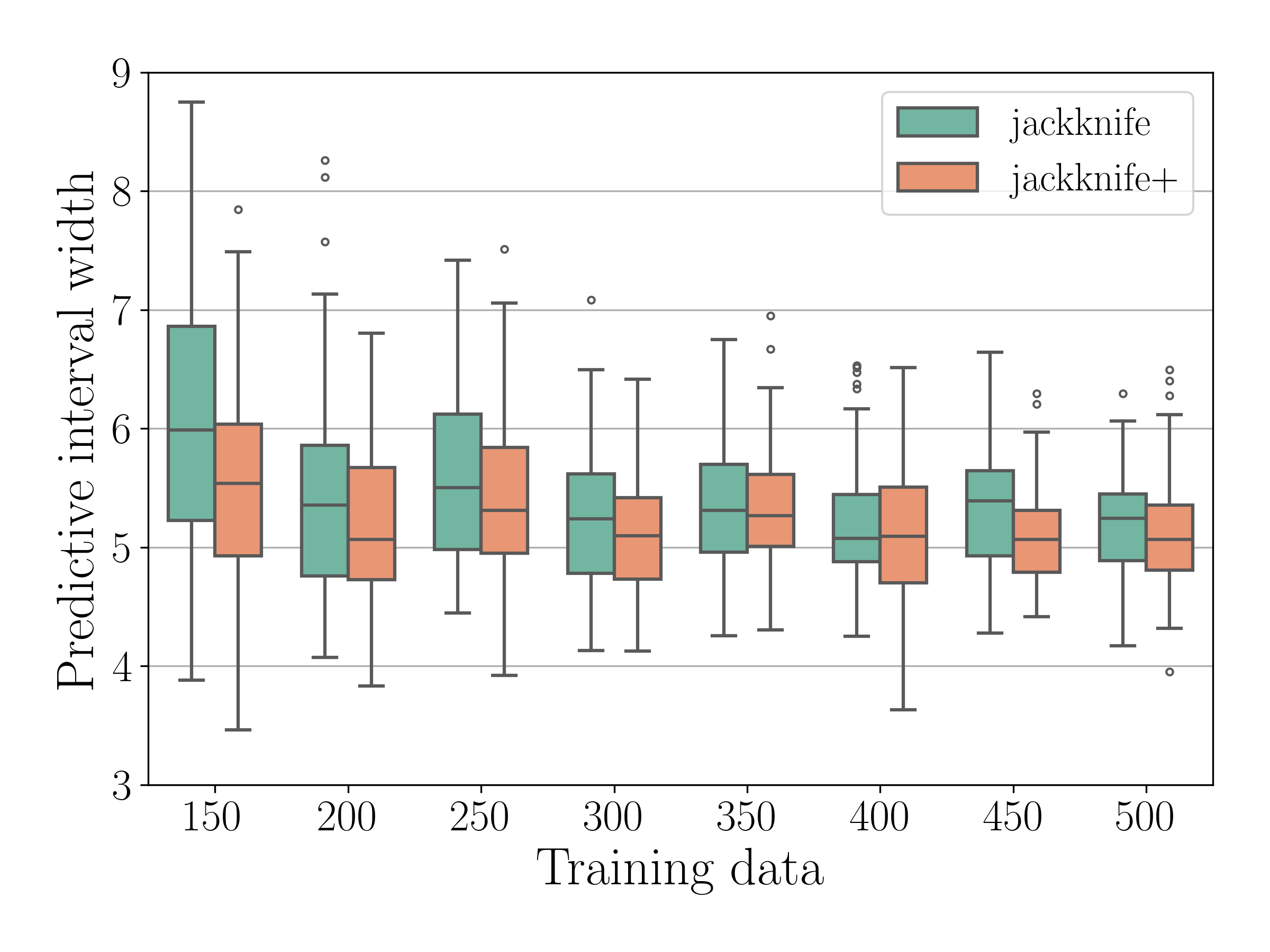}
    \caption{$P=1$, $s=0.01$.}
\end{subfigure}
\hfill
\begin{subfigure}[b]{0.45\textwidth}
    \centering
    \includegraphics[width=\textwidth]{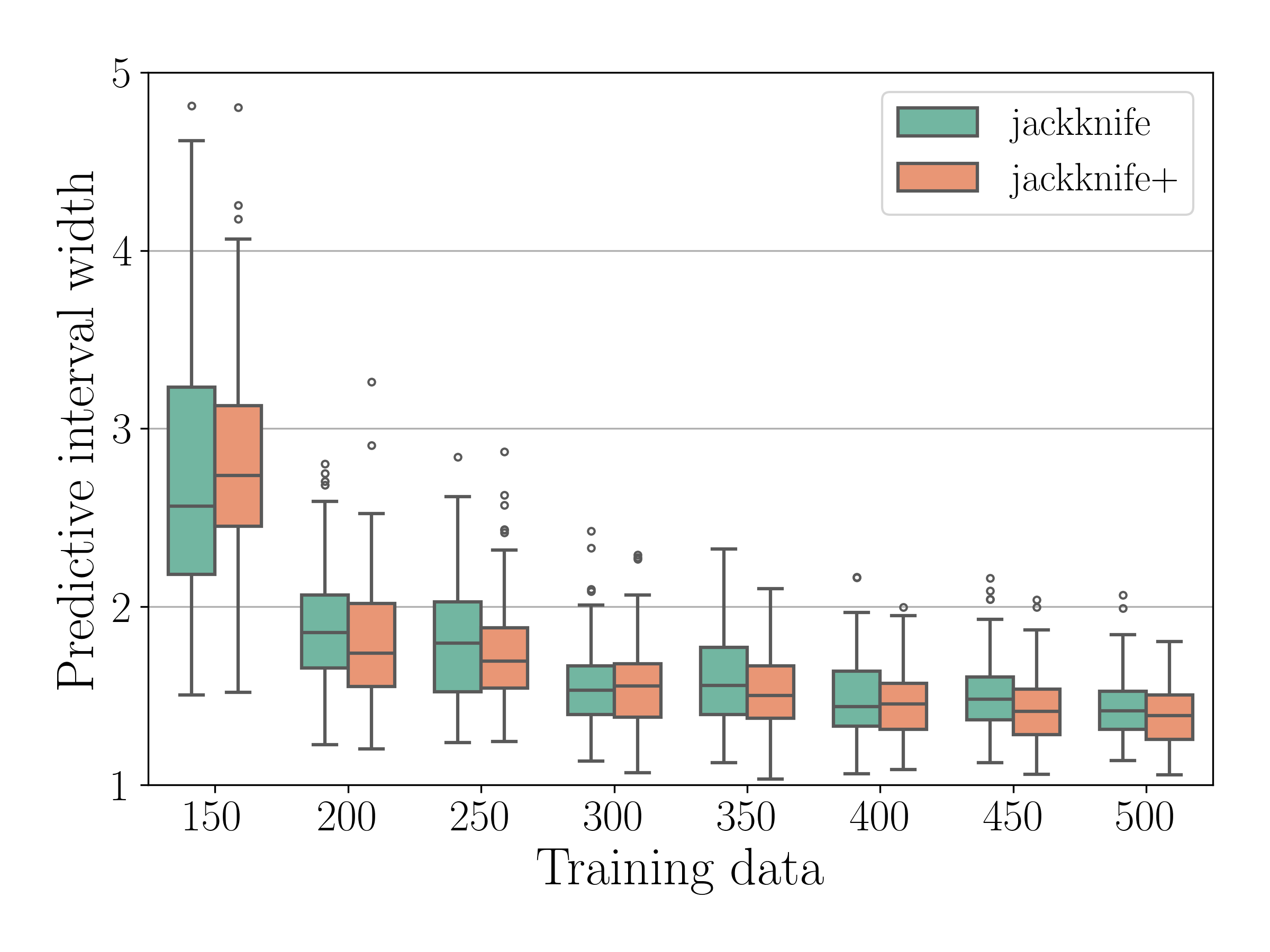}
    \caption{$P=2$, $s=0.01$.}
\end{subfigure}
\caption{Box plots of the prediction interval widths provided by conformalized \gls{pce} surrogates of the Stern-Gerlach magnet model, for different combinations of polynomial degree $P$, training dataset size $M$, and significance level $s$.}
\label{fig:stern-gerlach-interval-boxplots}
\end{figure}

Figures~\ref{fig:stern-gerlach-parity-plots-005} and \ref{fig:stern-gerlach-parity-plots-001} compare ground truth $\overline{\tau}_\text{b}$ values provided by the \gls{iga} model against conformalized \gls{pce} predictions for the original dataset without reshuffling, for the significance levels $s=0.05$ and $s=0.01$, respectively.
Training dataset sizes $M \in \left\{ 150, 300, 450 \right\}$ are considered in these figures.
Figures~\ref{fig:stern-gerlach-coverage-boxplots} and \ref{fig:stern-gerlach-interval-boxplots} show boxplots that summarize the statistics concerning empirical coverage and prediction interval width over the $100$ dataset reshuffles and for all considered training dataset sizes.
The conformalized \gls{pce} provides the required coverage, even for the inaccurate $P=1$ models and the more demanding significance level $s=0.01$.
As would be expected, the prediction intervals become narrower for $P=2$, while still preserving the required coverage.
Similar to the previous test-cases, the empirical coverages and the prediction interval widths present high variations for smaller experimental designs that progressively stabilize as more training data become available. 
The jackknife method is once again the more conservative option.

\section{Summary, conclusions, and outlook}
\label{sec:conclusion}
%
This work presented a new method that combines regression-based \glspl{pce} with jackknife-based \gls{cp} for constructing uncertainty-aware, data-driven surrogate models.
Using this method, data-driven \glspl{pce} are complemented with prediction intervals that quantify the uncertainty in their predictions.
The prediction intervals are estimated with very low computational overhead, by leveraging the linearity of the \gls{pce} regression model. 
In particular, analytical, closed-form expressions can be used for computing the \gls{loo} residuals and \gls{loo} predictions employed by the jackknife and jackknife+ \gls{cp} methods for prediction interval estimation. 

The conformalized \gls{pce} method was demonstrated on benchmark models and one electrical engineering application concerning the design of a high-gradient-field electromagnet. 
Different conformalized \gls{pce} configurations were examined, based on maximum polynomial degree, experimental design size, and \gls{cp} method (jackknife/jackknife+). 
Based on the numerical results, the conformalized \gls{pce} is in general able to provide empirical coverages close to the required target level. 
Training data availability, equivalently, experimental design size is the most important factor for providing the required coverage with the minimum possible interval width. 
Moreover, considering \gls{pce} models trained with different datasets, larger training datasets result in reduced variations in coverage and prediction interval width. 
The jackknife method typically results in more conservative predictive uncertainty estimates, with wider intervals and empirical coverages above the target level. 
The jackknife+ method yields coverages closer to the target level. 

The conformalized \gls{pce} method developed in this work shows definite promise as uncertainty-aware, data-driven, surrogate modeling method. 
Nonetheless, it can be improved in several aspects.
\textcolor{black}{
First, this work has considered total-degree \glspl{pce} computed by means of \gls{ols}. 
Therefore, the selected benchmark models and engineering use-cases were limited to low-to-moderate dimensions and polynomial degrees.
The integration of \gls{cp} within sparse \gls{pce} algorithms \cite{luethen2021sparse, luethen2022automatic} is a natural extension that would allow addressing more challenging problem settings. 
In fact, first attempts towards this goal have already emerged \cite{hatstatt2026conformal}.
}
Second, in its current form, the conformalized \gls{pce} method is applicable to scalar outputs only. 
In principle, it can be applied elementwise to address multidimensional outputs. 
However, this would be a suboptimal approach, as it does not account for correlations among the output's elements and would incur high computational costs with increasing output dimension.
The development of a conformalized \gls{pce} method suitable for multivariate outputs seems to be a natural follow-up, for example based on the combination of multi-output \glspl{pce} \cite{hawchar2017principal, bhattacharyya2020uncertainty, loukrezis2025multivariate} and multi-output \gls{cp} \cite{zhang2023improved, xu2023conformal}.
Third, the current approach provides symmetric prediction intervals that do not adapt to heteroskedastic data. 
Combinations with locally weighted non-conformity measures \cite{lei2018distribution} or with conformalized quantile regression \cite{romano2019conformalized} could offer possible remedies.
We aim to address these limitations in dedicated follow-up works.

\textcolor{black}{Finally, a rigorous comparative study assessing and benchmarking the currently available uncertainty-aware, \gls{pce}-based surrogate modeling approaches, e.g., conformalized \glspl{pce}, Bayesian \glspl{pce} \cite{shao2017bayesian, li2025new, rumsey2026bayesian}, and combined \gls{pce}-\gls{gp} models \cite{schobi2015polynomial, kersaudy2015new, manfredi2025hybrid, manfredi2026polynomial}, could provide valuable insights.}

\subsubsection*{Funding statement}
Dimitris G. Giovanis was supported by the Department of Energy (DOE) under Grant No. DE-SC0024162.

\subsubsection*{Conflict of interest statement}
The authors have nothing to disclose.

\subsubsection*{Author contributions}
\textbf{Dimitrios Loukrezis}: Conceptualization, Methodology, Software, Validation, Data curation, Writing - original draft, Writing - review \& editing. \\ \textbf{Dimitris G. Giovanis}: Validation, Data curation, Writing - original draft, Writing - review \& editing.

\printbibliography

@article{barber2021predictive,
  title={Predictive inference with the jackknife+},
  author={Barber, Rina Foygel and Candes, Emmanuel J and Ramdas, Aaditya and Tibshirani, Ryan J},
  journal={The Annals of Statistics},
  volume={49},
  number={1},
  pages={486--507},
  year={2021},
  publisher={JSTOR}
}

@article{bhattacharyya2020uncertainty,
  title={Uncertainty quantification of stochastic impact dynamic oscillator using a proper orthogonal decomposition-polynomial chaos expansion technique},
  author={Bhattacharyya, Biswarup and Jacquelin, Eric and Brizard, Denis},
  journal={Journal of Vibration and Acoustics},
  volume={142},
  number={6},
  pages={061013},
  year={2020},
  publisher={American Society of Mechanical Engineers}
}

@article{blatman2011adaptive,
  title={Adaptive sparse polynomial chaos expansion based on least angle regression},
  author={Blatman, G{\'e}raud and Sudret, Bruno},
  journal={Journal of Computational Physics},
  volume={230},
  number={6},
  pages={2345--2367},
  year={2011},
  publisher={Elsevier}
}

@phdthesis{blatman2009phd,
  title={Adaptive sparse polynomial chaos expansions for uncertainty propagation and sensitivity analysis},
  author={Blatman, G{\'e}raud},
  year={2009},
  school={Universite Blaise Pascal-Clermont-Ferrand II}
}

@inproceedings{burnaev2016conformalized,
  title={Conformalized kernel ridge regression},
  author={Burnaev, Evgeny and Nazarov, Ivan},
  booktitle={2016 15th IEEE International Conference on Machine Learning and Applications (ICMLA)},
  pages={45--52},
  year={2016},
  organization={IEEE}
}

@article{buzzard2012global,
  title={Global sensitivity analysis using sparse grid interpolation and polynomial chaos},
  author={Buzzard, Gregery T},
  journal={Reliability Engineering \& System Safety},
  volume={107},
  pages={82--89},
  year={2012},
  publisher={Elsevier}
}

@article{campos2023polynomial,
  title={Polynomial chaos expansion surrogate modeling of passive cardiac mechanics using the Holzapfel--Ogden constitutive model},
  author={Campos, Joventino Oliveira and Guedes, Rafael Moreira and Werneck, Yan Barbosa and Barra, Luis Paulo S and dos Santos, Rodrigo Weber and Rocha, Bernardo M},
  journal={Journal of Computational Science},
  volume={71},
  pages={102039},
  year={2023},
  publisher={Elsevier}
}

@article{chernick2012jackknife,
  title={The jackknife: a resampling method with connections to the bootstrap},
  author={Chernick, Michael R},
  journal={Wiley Interdisciplinary Reviews: Computational Statistics},
  volume={4},
  number={2},
  pages={224--226},
  year={2012},
  publisher={Wiley Online Library}
}

@article{conrad2013adaptive,
  title={Adaptive {Smolyak} pseudospectral approximations},
  author={Conrad, Patrick R and Marzouk, Youssef M},
  journal={SIAM Journal on Scientific Computing},
  volume={35},
  number={6},
  pages={A2643--A2670},
  year={2013},
  publisher={SIAM}
}

@article{constantine2012sparse,
  title={Sparse pseudospectral approximation method},
  author={Constantine, Paul G and Eldred, Michael S and Phipps, Eric T},
  journal={Computer Methods in Applied Mechanics and Engineering},
  volume={229},
  pages={1--12},
  year={2012},
  publisher={Elsevier}
}

@article{dai2025cost,
  title={A cost surrogate model for TSO-DSO coordination based on polynomial chaos expansion},
  author={Dai, Wei and Li, Dewen and Liu, Hui and Liu, Yuelin},
  journal={IEEE Transactions on Power Systems},
  year={2025},
  publisher={IEEE}
}

@article{deutschmann2024adaptive,
  title={Adaptive conformal regression with split-jackknife+ scores},
  author={Deutschmann, Nicolas and Rigotti, Mattia and Martinez, Maria Rodriguez},
  journal={Transactions on Machine Learning Research},
  year={2024}
}

@article{fontana2023conformal,
  title={Conformal prediction: a unified review of theory and new challenges},
  author={Fontana, Matteo and Zeni, Gianluca and Vantini, Simone},
  journal={Bernoulli},
  volume={29},
  number={1},
  pages={1--23},
  year={2023},
  publisher={Bernoulli Society for Mathematical Statistics and Probability}
}

@article{galetzka2023hp,
  title={An hp-adaptive multi-element stochastic collocation method for surrogate modeling with information re-use},
  author={Galetzka, Armin and Loukrezis, Dimitrios and Georg, Niklas and De Gersem, Herbert and R{\"o}mer, Ulrich},
  journal={International Journal for Numerical Methods in Engineering},
  volume={124},
  number={12},
  pages={2902--2930},
  year={2023},
  publisher={Wiley Online Library}
}

@book{ghanem2003stochastic,
  title={Stochastic finite elements: a spectral approach},
  author={Ghanem, Roger G and Spanos, Pol D},
  year={2003},
  publisher={Courier Corporation}
}

@article{giovanis2024polynomial,
  title={Polynomial chaos expansions on principal geodesic Grassmannian submanifolds for surrogate modeling and uncertainty quantification},
  author={Giovanis, Dimitris G and Loukrezis, Dimitrios and Kevrekidis, Ioannis G and Shields, Michael D},
  journal={Journal of Computational Physics},
  volume={519},
  pages={113443},
  year={2024},
  publisher={Elsevier}
}

@article{gopakumar2024uncertainty,
  title={Uncertainty quantification of surrogate models using conformal prediction},
  author={Gopakumar, Vignesh and Gray, Ander and Oskarsson, Joel and Zanisi, Lorenzo and Pamela, Stanislas and Giles, Daniel and Kusner, Matt and Deisenroth, Marc Peter},
  journal={arXiv preprint arXiv:2408.09881},
  year={2024}
}

@article{hadigol2018least,
  title={Least squares polynomial chaos expansion: A review of sampling strategies},
  author={Hadigol, Mohammad and Doostan, Alireza},
  journal={Computer Methods in Applied Mechanics and Engineering},
  volume={332},
  pages={382--407},
  year={2018},
  publisher={Elsevier}
}

@article{haghi2022surrogate,
  title={Surrogate models for the blade element momentum aerodynamic model using non-intrusive polynomial chaos expansions},
  author={Haghi, Rad and Crawford, Curran},
  journal={Wind Energy Science},
  volume={7},
  number={3},
  pages={1289--1304},
  year={2022},
  publisher={Copernicus Publications G{\"o}ttingen, Germany}
}

@article{hatstatt2026conformal,
  title={Conformal prediction for full and sparse polynomial chaos expansions},
  author={Hatstatt, Arthur and Zhu, Xujia and Sudret, Bruno},
  journal={arXiv preprint arXiv:2601.16636},
  year={2026}
}

@article{hawchar2017principal,
  title={Principal component analysis and polynomial chaos expansion for time-variant reliability problems},
  author={Hawchar, Lara and El Soueidy, Charbel-Pierre and Schoefs, Franck},
  journal={Reliability Engineering \& System Safety},
  volume={167},
  pages={406--416},
  year={2017},
  publisher={Elsevier}
}

@article{huang2023uncertainty,
  title={Uncertainty quantification over graph with conformalized graph neural networks},
  author={Huang, Kexin and Jin, Ying and Candes, Emmanuel and Leskovec, Jure},
  journal={Advances in Neural Information Processing Systems},
  volume={36},
  pages={26699--26721},
  year={2023}
}

@article{jaber2025conformal,
  title={Conformal approach to Gaussian process surrogate evaluation with marginal coverage guarantees},
  author={Jaber, Edgar and Blot, Vincent and Brunel, Nicolas JB and Chabridon, Vincent and Remy, Emmanuel and Iooss, Bertrand and Lucor, Didier and Mougeot, Mathilde and Leite, Alessandro},
  journal={Journal of Machine Learning for Modeling and Computing},
  volume={6},
  number={3},
  year={2025},
  publisher={Begel House Inc.}
}

@article{jospin2022hands,
  title={Hands-on Bayesian neural networks—A tutorial for deep learning users},
  author={Jospin, Laurent Valentin and Laga, Hamid and Boussaid, Farid and Buntine, Wray and Bennamoun, Mohammed},
  journal={IEEE Computational Intelligence Magazine},
  volume={17},
  number={2},
  pages={29--48},
  year={2022},
  publisher={IEEE}
}

@article{kersaudy2015new,
  title={A new surrogate modeling technique combining Kriging and polynomial chaos expansions--Application to uncertainty analysis in computational dosimetry},
  author={Kersaudy, Pierric and Sudret, Bruno and Varsier, Nad{\`e}ge and Picon, Odile and Wiart, Joe},
  journal={Journal of Computational Physics},
  volume={286},
  pages={103--117},
  year={2015},
  publisher={Elsevier}
}

@article{kontolati2022manifold,
  title={Manifold learning-based polynomial chaos expansions for high-dimensional surrogate models},
  author={Kontolati, Katiana and Loukrezis, Dimitrios and Dos Santos, Ketson RM and Giovanis, Dimitrios G and Shields, Michael D},
  journal={International Journal for Uncertainty Quantification},
  volume={12},
  number={4},
  year={2022},
  publisher={Begel House Inc.}
}

@article{kontolati2022survey,
  title={A survey of unsupervised learning methods for high-dimensional uncertainty quantification in black-box-type problems},
  author={Kontolati, Katiana and Loukrezis, Dimitrios and Giovanis, Dimitrios G and Vandanapu, Lohit and Shields, Michael D},
  journal={Journal of Computational Physics},
  volume={464},
  pages={111313},
  year={2022},
  publisher={Elsevier}
}

@article{lee2025surrogate,
  title={Surrogate models for multiregime flow problems},
  author={Lee, Jiyoung and Chan, Leon and Zahtila, Tony and Lu, Wilson and Iaccarino, Gianluca and Ooi, Andrew},
  journal={Physical Review Fluids},
  volume={10},
  number={2},
  pages={024703},
  year={2025},
  publisher={APS}
}

@article{lei2018distribution,
  title={Distribution-free predictive inference for regression},
  author={Lei, Jing and G’Sell, Max and Rinaldo, Alessandro and Tibshirani, Ryan J and Wasserman, Larry},
  journal={Journal of the American Statistical Association},
  volume={113},
  number={523},
  pages={1094--1111},
  year={2018},
  publisher={Taylor \& Francis}
}

@article{li2025new,
  title={A new Bayesian sparse polynomial chaos expansion based on Gaussian scale mixture prior model},
  author={Li, Yangtian and Luo, Yangjun and Zhong, Zheng and Yu, Jing and Wang, Ruiting},
  journal={Mechanical Systems and Signal Processing},
  volume={229},
  pages={112511},
  year={2025},
  publisher={Elsevier}
}

@article{loukrezis2025multivariate,
  title={Multivariate sensitivity-adaptive polynomial chaos expansion for high-dimensional surrogate modeling and uncertainty quantification},
  author={Loukrezis, Dimitrios and Diehl, Eric and De Gersem, Herbert},
  journal={Applied Mathematical Modelling},
  volume={137},
  pages={115746},
  year={2025},
  publisher={Elsevier}
}

@article{loukrezis2022power,
  title={Power module heat sink design optimization with ensembles of data-driven polynomial chaos surrogate models},
  author={Loukrezis, Dimitrios and De Gersem, Herbert},
  journal={e-Prime-Advances in Electrical Engineering, Electronics and Energy},
  volume={2},
  pages={100059},
  year={2022},
  publisher={Elsevier}
}

@article{loukrezis2019assessing,
  title={Assessing the performance of Leja and Clenshaw-Curtis collocation for computational electromagnetics with random input data},
  author={Loukrezis, Dimitrios and R{\"o}mer, Ulrich and De Gersem, Herbert},
  journal={International Journal for Uncertainty Quantification},
  volume={9},
  number={1},
  year={2019},
  publisher={Begel House Inc.}
}

@article{luethen2021sparse,
  title={Sparse polynomial chaos expansions: Literature survey and benchmark},
  author={L{\"u}then, Nora and Marelli, Stefano and Sudret, Bruno},
  journal={SIAM/ASA Journal on Uncertainty Quantification},
  volume={9},
  number={2},
  pages={593--649},
  year={2021},
  publisher={SIAM}
}

@article{luethen2022automatic,
  title={Automatic selection of basis-adaptive sparse polynomial chaos expansions for engineering applications},
  author={L{\"u}then, Nora and Marelli, Stefano and Sudret, Bruno},
  journal={International Journal for Uncertainty Quantification},
  volume={12},
  number={3},
  year={2022},
  publisher={Begel House Inc.}
}

@article{manfredi2024nonparametric,
  title={Nonparametric formulation of polynomial chaos expansion based on least-square support-vector machines},
  author={Manfredi, Paolo and Trinchero, Riccardo},
  journal={Engineering Applications of Artificial Intelligence},
  volume={133},
  pages={108182},
  year={2024},
  publisher={Elsevier}
}

@article{manfredi2025hybrid,
  title={A hybrid polynomial chaos expansion--Gaussian process regression method for Bayesian uncertainty quantification and sensitivity analysis},
  author={Manfredi, Paolo},
  journal={Computer Methods in Applied Mechanics and Engineering},
  volume={436},
  pages={117693},
  year={2025},
  publisher={Elsevier}
}

@article{manfredi2026polynomial,
  title={Polynomial chaos vs kernel machine learning methods for uncertainty quantification: A comparative study and benchmarking with the hybrid PCE-GPR approach},
  author={Manfredi, Paolo},
  journal={Computer Methods in Applied Mechanics and Engineering},
  volume={449},
  pages={118523},
  year={2026},
  publisher={Elsevier}
}

@article{marelli2018active,
  title={An active-learning algorithm that combines sparse polynomial chaos expansions and bootstrap for structural reliability analysis},
  author={Marelli, Stefano and Sudret, Bruno},
  journal={Structural Safety},
  volume={75},
  pages={67--74},
  year={2018},
  publisher={Elsevier}
}

@article{masschaele2011design,
  title={Design of a strong gradient magnet for the deflection of nanoclusters},
  author={Masschaele, Bert and Roggen, Toon and De Gersem, Herbert and Jannsens, Ewald and Nguyen, Tung Thang},
  journal={IEEE Transactions on Applied Superconductivity},
  volume={22},
  number={3},
  pages={3700604--3700604},
  year={2011},
  publisher={IEEE}
}

@article{migliorati2013approximation,
  title={Approximation of quantities of interest in stochastic PDEs by the random discrete $L^2$ projection on polynomial spaces},
  author={Migliorati, Giovanni and Nobile, Fabio and von Schwerin, Erik and Tempone, Ra{\'u}l},
  journal={SIAM Journal on Scientific Computing},
  volume={35},
  number={3},
  pages={A1440--A1460},
  year={2013},
  publisher={SIAM}
}

@article{migliorati2014analysis,
  title={Analysis of discrete $L^2$ projection on polynomial spaces with random evaluations},
  author={Migliorati, Giovanni and Nobile, Fabio and Von Schwerin, Erik and Tempone, Ra{\'u}l},
  journal={Foundations of Computational Mathematics},
  volume={14},
  number={3},
  pages={419--456},
  year={2014},
  publisher={Springer}
}

@article{novak2024physics,
  title={Physics-informed polynomial chaos expansions},
  author={Nov{\'a}k, Luk{\'a}{\v{s}} and Sharma, Himanshu and Shields, Michael D},
  journal={Journal of Computational Physics},
  volume={506},
  pages={112926},
  year={2024},
  publisher={Elsevier}
}

@inproceedings{papadopoulos2002inductive,
  title={Inductive confidence machines for regression},
  author={Papadopoulos, Harris and Proedrou, Kostas and Vovk, Volodya and Gammerman, Alex},
  booktitle={European Conference on Machine Learning},
  pages={345--356},
  year={2002},
  organization={Springer}
}

@article{pels2015optimization,
  title={Optimization of a Stern--Gerlach magnet by magnetic field--circuit coupling and isogeometric analysis},
  author={Pels, Andreas and Bontinck, Zeger and Corno, Jacopo and De Gersem, Herbert and Sch{\"o}ps, Sebastian},
  journal={IEEE Transactions on Magnetics},
  volume={51},
  number={12},
  pages={1--7},
  year={2015},
  publisher={IEEE}
}

@article{podina2024conformalized,
  title={Conformalized physics-informed neural networks},
  author={Podina, Lena and Rad, Mahdi Torabi and Kohandel, Mohammad},
  journal={arXiv preprint arXiv:2405.08111},
  year={2024}
}

@article{porta2014inverse,
  title={Inverse modeling of geochemical and mechanical compaction in sedimentary basins through Polynomial Chaos Expansion},
  author={Porta, Giovanni and Tamellini, Lorenzo and Lever, Valentina and Riva, Monica},
  journal={Water Resources Research},
  volume={50},
  number={12},
  pages={9414--9431},
  year={2014},
  publisher={Wiley Online Library}
}

@inproceedings{ranftl2021bayesian,
  title={Bayesian surrogate analysis and uncertainty propagation},
  author={Ranftl, Sascha and von der Linden, Wolfgang and MaxEnt 2021 Scientific Committee},
  booktitle={Physical Sciences Forum},
  volume={3},
  number={1},
  pages={6},
  year={2021},
  organization={MDPI}
}

@book{rao2006probability,
  title={Probability Theory with Applications},
  author={Rao, Malempati Madhusudana and Swift, Randall J},
  year={2006},
  publisher={Springer}
}

@article{romano2019conformalized,
  title={Conformalized quantile regression},
  author={Romano, Yaniv and Patterson, Evan and Candes, Emmanuel},
  journal={Advances in Neural Information Processing Systems},
  volume={32},
  year={2019}
}

@article{rumsey2026bayesian,
  title={Bayesian adaptive polynomial chaos expansions},
  author={Rumsey, Kellin N and Francom, Devin and Gibson, Graham and Derek Tucker, J and Huerta, Gabriel},
  journal={Stat},
  volume={15},
  number={1},
  pages={e70151},
  year={2026},
  publisher={Wiley Online Library}
}

@article{schobi2015polynomial,
  title={Polynomial-chaos-based Kriging},
  author={Schobi, Roland and Sudret, Bruno and Wiart, Joe},
  journal={International Journal for Uncertainty Quantification},
  volume={5},
  number={2},
  year={2015},
  publisher={Begel House Inc.}
}

@book{seber2003linear,
  title={Linear regression analysis},
  author={Seber, George A F and Lee, Alan J},
  year={2003},
  publisher={John Wiley \& Sons}
}

@article{shang2024active,
  title={Active learning of ensemble polynomial chaos expansion method for global sensitivity analysis},
  author={Shang, Xiaobing and Wang, Lipeng and Fang, Hai and Lu, Lingyun and Zhang, Zhi},
  journal={Reliability Engineering \& System Safety},
  volume={249},
  pages={110226},
  year={2024},
  publisher={Elsevier}
}

@article{shao2017bayesian,
  title={Bayesian sparse polynomial chaos expansion for global sensitivity analysis},
  author={Shao, Qian and Younes, Anis and Fahs, Marwan and Mara, Thierry A},
  journal={Computer Methods in Applied Mechanics and Engineering},
  volume={318},
  pages={474--496},
  year={2017},
  publisher={Elsevier}
}

@article{sharma2025polynomial,
  title={Polynomial chaos expansion for operator learning},
  author={Sharma, Himanshu and Nov{\'a}k, Luk{\'a}{\v{s}} and Shields, Michael D},
  journal={arXiv preprint arXiv:2508.20886},
  year={2025}
}

@article{sharma2024physics,
  title={Physics-constrained polynomial chaos expansion for scientific machine learning and uncertainty quantification},
  author={Sharma, Himanshu and Nov{\'a}k, Luk{\'a}{\v{s}} and Shields, Michael},
  journal={Computer Methods in Applied Mechanics and Engineering},
  volume={431},
  pages={117314},
  year={2024},
  publisher={Elsevier}
}

@article{soize2017polynomial,
  title={Polynomial chaos representation of databases on manifolds},
  author={Soize, Christian and Ghanem, Roger G},
  journal={Journal of Computational Physics},
  volume={335},
  pages={201--221},
  year={2017},
  publisher={Elsevier}
}

@article{spanos1989stochastic,
  title={Stochastic finite element expansion for random media},
  author={Spanos, Pol D and Ghanem, Roger G},
  journal={Journal of engineering mechanics},
  volume={115},
  number={5},
  pages={1035--1053},
  year={1989},
  publisher={American Society of Civil Engineers}
}

@misc{surjanovic2013virtual,
  title={Virtual library of simulation experiments: test functions and datasets},
  author={Surjanovic, Sonja and Bingham, Derek},
  year={2013},
  note={Retrieved October 22, 2025, from http://www.sfu.ca/~ssurjano.}
}

@article{torre2019data,
  title={Data-driven polynomial chaos expansion for machine learning regression},
  author={Torre, Emiliano and Marelli, Stefano and Embrechts, Paul and Sudret, Bruno},
  journal={Journal of Computational Physics},
  volume={388},
  pages={601--623},
  year={2019},
  publisher={Elsevier}
}

@book{vovk2005algorithmic,
  title={Algorithmic learning in a random world},
  author={Vovk, Vladimir and Gammerman, Alexander and Shafer, Glenn},
  year={2005},
  publisher={Springer}
}

@article{vovk2015cross,
  title={Cross-conformal predictors},
  author={Vovk, Vladimir},
  journal={Annals of Mathematics and Artificial Intelligence},
  volume={74},
  number={1},
  pages={9--28},
  year={2015},
  publisher={Springer}
}

@inproceedings{vovk2018cross,
  title={Cross-conformal predictive distributions},
  author={Vovk, Vladimir and Nouretdinov, Ilia and Manokhin, Valery and Gammerman, Alexander},
  booktitle={Conformal and Probabilistic Prediction and Applications},
  pages={37--51},
  year={2018},
  organization={PMLR}
}

@book{williams2006gaussian,
  title     = {Gaussian Processes for Machine Learning},
  author    = {Williams, Christopher K. I. and Rasmussen, Carl Edward},
  year      = {2006},
  publisher = {The MIT Press},
  address   = {Cambridge, MA, USA},
  isbn      = {026218253X}
}

@article{xiu2002wiener,
  title={The Wiener--Askey polynomial chaos for stochastic differential equations},
  author={Xiu, Dongbin and Karniadakis, George Em},
  journal={SIAM Journal on Scientific Computing},
  volume={24},
  number={2},
  pages={619--644},
  year={2002},
  publisher={SIAM}
}

@article{xu2023conformal,
  title={Conformal prediction for time series},
  author={Xu, Chen and Xie, Yao},
  journal={IEEE Transactions on Pattern Analysis and Machine Intelligence},
  volume={45},
  number={10},
  pages={11575--11587},
  year={2023},
  publisher={IEEE}
}

@inproceedings{zaffran2022adaptive,
  title={Adaptive conformal predictions for time series},
  author={Zaffran, Margaux and F{\'e}ron, Olivier and Goude, Yannig and Josse, Julie and Dieuleveut, Aymeric},
  booktitle={International Conference on Machine Learning},
  pages={25834--25866},
  year={2022},
  organization={PMLR}
}

@article{zhang2023improved,
  title={Improved Copula-based conformal prediction for uncertainty quantification of multi-output regression},
  author={Zhang, Ruiyao and Zhou, Ping and Chai, Tianyou},
  journal={Journal of Process Control},
  volume={129},
  pages={103036},
  year={2023},
  publisher={Elsevier}
}

@article{zhu2023stochastic,
  title={Stochastic polynomial chaos expansions to emulate stochastic simulators},
  author={Zhu, Xujia and Sudret, Bruno},
  journal={International Journal for Uncertainty Quantification},
  volume={13},
  number={2},
  year={2023},
  publisher={Begel House Inc.}
}

\appendix 
\appendixpage
\section{Leave-one-out residuals in closed form}
\label{sec:appendixA}
Let $\widehat{\mu}\left(\mathbf{x}\right)$ denote a \gls{pce} regression model as in section \ref{sec:pce}. 
Removing the $m$-th sample from the training data, we get the \gls{loo} design matrix $\mathbf{D}_{\sim m} \in \mathbb{R}^{(M-1) \times K}$ and the \gls{loo} output vector $\mathbf{y}_{\sim m} \in \mathbb{R}^{M-1}$.
The omitted $m$-th row of the original design matrix $\mathbf{D}$ is denoted with  $\mathbf{d}_m^\top \in \mathbb{R}^{1 \times K}$, where $\mathbf{d}_m^\top = \left(\Psi_1\left(\mathbf{x}^{(m)}\right), \dots, \Psi_K\left(\mathbf{x}^{(m)}\right)\right)$. 
Then, the \gls{loo} equivalent to the least-squares solution \eqref{eq:regression-solution} is 
\begin{equation}
\label{eq:regression-solution-loo}
\mathbf{c}_{\sim m} = \left( \mathbf{D}_{\sim m}^\top \mathbf{D}_{\sim m}\right)^{-1} \mathbf{D}_{\sim m}^\top \mathbf{y}_{\sim m},
\end{equation}
where it also holds that 
\begin{align}
\mathbf{D}_{\sim m}^\top \mathbf{D}_{\sim m} &= \mathbf{D}^\top \mathbf{D} - \mathbf{d}_m \mathbf{d}_m^\top, \label{eq:holds1}\\
\mathbf{D}_{\sim m}^\top \mathbf{y}_{\sim m} &= \mathbf{D}^\top \mathbf{y} - \mathbf{d}_m y^{(m)}. \label{eq:holds2}
\end{align}
Let $\mathbf{A} = \left( \mathbf{D}^\top \mathbf{D}\right)^{-1} \in \mathbb{R}^{K \times K}$. 
Using the Sherman-Morrison formula we get 
\begin{equation}
\label{eq:sherman-morrison}
\left( \mathbf{D}^\top \mathbf{D} - \mathbf{d}_m \mathbf{d}_m^\top \right)^{-1} = \mathbf{A} + \frac{\mathbf{A} \mathbf{d}_m \mathbf{d}_m^\top \mathbf{A}}{1 - \mathbf{d}_m^\top \mathbf{A} \mathbf{d}_m},
\end{equation}
assuming that $1 - \mathbf{d}_m^\top \mathbf{A} \mathbf{d}_m \neq 0$.
Note that $\mathbf{d}_m^\top \mathbf{A} \mathbf{d}_m = h_{mm}$, where $h_{mm}$ are the diagonal elements of the hat matrix $\mathbf{H} \in \mathbb{R}^{M \times M}$ defined in \eqref{eq:hat-matrix}. 
Using expressions \eqref{eq:holds1}, \eqref{eq:holds2}, and \eqref{eq:sherman-morrison}, expression \eqref{eq:regression-solution-loo} can now be written as
\begin{align}
\mathbf{c}_{\sim m} &= \left( \mathbf{A} + \frac{\mathbf{A} \mathbf{d}_m \mathbf{d}_m^\top \mathbf{A}}{1 - h_{mm}} \right) \left( \mathbf{D}^\top \mathbf{y} - \mathbf{d}_m y^{(m)} \right) \nonumber \\
&= \mathbf{A} \mathbf{D}^\top \mathbf{y} - \mathbf{A} \mathbf{d}_m y^{(m)} + \frac{\mathbf{A} \mathbf{d}_m \mathbf{d}_m^\top \mathbf{A} \mathbf{D}^\top \mathbf{y}}{1 - h_{mm}} - \frac{\mathbf{A} \mathbf{d}_m \mathbf{d}_m^\top \mathbf{A} \mathbf{d}_m y^{(m)}}{1 - h_{mm}} \nonumber \\
&= \mathbf{c} - \mathbf{A} \mathbf{d}_m y^{(m)} + \frac{\mathbf{A} \mathbf{d}_m \mathbf{d}_m^\top \mathbf{c}}{1 - h_{mm}} - \frac{\mathbf{A} \mathbf{d}_m h_{mm} y^{(m)}}{1 - h_{mm}} \nonumber \\
&= \mathbf{c} + \mathbf{A} \mathbf{d}_m \left( \frac{\mathbf{d}_m^\top \mathbf{c} - h_{mm} y^{(m)}}{1 - h_{mm}} - y^{(m)} \right) \nonumber \\
&= \mathbf{c} + \mathbf{A} \mathbf{d}_m  \frac{\mathbf{d}_m^\top \mathbf{c} - y^{(m)}}{1-h_{mm}} \nonumber \\
&= \mathbf{c} + \mathbf{A} \mathbf{d}_m \frac{\widehat{y}^{(m)} - y^{(m)}}{1 - h_{mm}} \nonumber \\
&= \mathbf{c} - \mathbf{A} \mathbf{d}_m \frac{y^{(m)} - \widehat{y}^{(m)}}{1 - h_{mm}},
\end{align}
where we have used that $\mathbf{c} = \mathbf{A} \mathbf{D}^\top \mathbf{y}$ and $\widehat{y}^{(m)} = \widehat{\mu}\left(\mathbf{x}^{(m)}\right) = \mathbf{d}_m^\top \mathbf{c}$.
The \gls{loo} model evaluation at $\mathbf{x}^{(m)}$ is then given by
\begin{align} 
\widehat{\mu}_{\sim m}\left(\mathbf{x}^{(m)}\right) = \mathbf{d}_m^\top \mathbf{c}_{\sim m} = \mathbf{d}_m^\top \mathbf{c} - \mathbf{d}_m^\top \mathbf{A} \mathbf{d}_m \frac{y^{(m)} - \widehat{y}^{(m)}}{1 - h_{mm}} = \widehat{y}^{(m)} - h_{mm} \frac{y^{(m)} - \widehat{y}^{(m)}}{1 - h_{mm}}.
\end{align}
Therefore, the \gls{loo} residual defined in \eqref{eq:loo-residual} is given by 
\begin{align}
\label{eq:loo-residual-analytical-appendix}
r_m^{\text{LOO}} &= y^{(m)} - \widehat{\mu}_{\sim m}\left( \mathbf{x}^{(m)} \right) \nonumber \\
&= y^{(m)} - \widehat{y}^{(m)} + h_{mm} \frac{\widehat{y}^{(m)} - y^{(m)}}{1 - h_{mm}} \nonumber \\
&= \frac{y^{(m)} - \widehat{y}^{(m)}}{1 - h_{mm}} \nonumber \\
&= \frac{y^{(m)} - \widehat{\mu}\left(\mathbf{x}^{(m)}\right)}{1 - h_{mm}}.
\end{align}

\section{Leave-one-out predictions in closed form}
\label{sec:appendixB}
Using similar notation as in appendix \ref{sec:appendixA}, 
the \gls{loo} prediction at a test input $\mathbf{x}^*$ is given by 
\begin{equation}
\label{eq:loo-prediction-appendixB} 
\widehat{\mu}_{\sim m}\left(\mathbf{x}^*\right) = \mathbf{d}_*^\top \mathbf{c}_{\sim m},
\end{equation} 
where $\mathbf{d}_*^\top = \left(\Psi_1\left(\mathbf{x}^*\right), \dots, \Psi_K\left(\mathbf{x}^*\right)\right)$. 
Substituting the \gls{loo} \gls{pce} coefficients from \eqref{eq:regression-solution-loo} into \eqref{eq:loo-prediction-appendixB}, we get
\begin{align} 
\label{eq:loo-prediction-appendixB-cont} 
\widehat{\mu}_{\sim m}\left(\mathbf{x}^*\right) &= \mathbf{d}_*^\top \left( \mathbf{c} - \mathbf{A} \mathbf{d}_m \frac{y^{(m)} - \widehat{y}^{(m)}}{1 - h_{mm}} \right) \nonumber \\
&= \mathbf{d}_*^\top \mathbf{c} - \mathbf{d}_*^\top \mathbf{A} \mathbf{d}_m \frac{y^{(m)} - \widehat{y}^{(m)}}{1 - h_{mm}} \nonumber \\
&= \widehat{\mu}\left(\mathbf{x}^*\right) - \mathbf{d}_*^\top \mathbf{A} \mathbf{d}_m \frac{y^{(m)} - \widehat{y}^{(m)}}{1 - h_{mm}} \nonumber \\
&= \widehat{\mu}\left(\mathbf{x}^*\right) - \mathbf{d}_*^\top \mathbf{A} \mathbf{d}_m \frac{y^{(m)} - \widehat{\mu}\left(\mathbf{x}^{(m)}\right)}{1 - h_{mm}} \nonumber \\
&= \widehat{\mu}\left(\mathbf{x}^*\right) - \mathbf{d}_*^\top \left(\mathbf{D}^\top \mathbf{D}\right)^{-1} \mathbf{d}_m \frac{y^{(m)} - \widehat{\mu}\left(\mathbf{x}^{(m)}\right)}{1 - h_{mm}} \nonumber \\
&= \widehat{\mu}\left(\mathbf{x}^*\right) - \mathbf{d}_*^\top \left(\mathbf{D}^\top \mathbf{D}\right)^{-1} \mathbf{d}_m r_m^{\text{LOO}},
\end{align} 
where we used that $\mathbf{A} = \left(\mathbf{D}^\top \mathbf{D}\right)^{-1}$, $\widehat{\mu}\left(\mathbf{x}^*\right) = \mathbf{d}_*^\top \mathbf{c}$, and the analytical expression \eqref{eq:loo-residual-analytical-appendix} for the \gls{loo} residual.

\section{\textcolor{black}{Validation on additional benchmark models}}
\label{sec:appendixC}

\subsection{OTL circuit function}
\label{sec:otl-circuit}
We consider the response of an output transformerless (OTL) push-pull circuit, given as 
\begin{equation}
\mu(\mathbf{x}) = 
\frac{\left(\tfrac{12\,R_{b2}}{R_{b1}+R_{b2}} + 0.74\right)\,\beta\,\left(R_{c2}+9\right)}
{\beta\,(R_{c2}+9) + R_f}
\;+\;
\frac{11.35\,R_f}{\beta\,\left(R_{c2}+9\right) + R_f}
\;+\;
\frac{0.74\,R_f\,\beta\,(R_{c2}+9)}
{\left(\beta\,\left(R_{c2}+9\right) + R_f\right)\,R_{c1}},
\end{equation}
where the input vector $\mathbf{x} \in \mathbb{R}^6$ consists of the resistances $R_{b1}$, $R_{b2}$, $R_{f}$, $R_{c1}$, $R_{c2}$, and the current gain $\beta$.
The response corresponds to the midpoint voltage. 
The input parameters are assumed to be uniformly distributed within the value ranges given in Table~\ref{tab:otl-circuit-parameters}.
\begin{table}[h!]
\centering
\caption{Input parameters of the OTL circuit function.}
\label{tab:otl-circuit-parameters}
\begin{threeparttable}
\begin{tabular}{c c l}
\toprule
Parameter & Units & Range \\
\midrule 
$R_{b1}$ & k$\Omega$ & $\left[50, 150\right]$ \\  
$R_{b2}$ & k$\Omega$ & $\left[25, 70\right]$ \\
$R_{f}$  & k$\Omega$ & $\left[0.5, 30\right]$ \\
$R_{c1}$ & k$\Omega$ & $\left[1.2, 2.50\right]$ \\
$R_{c2}$ & k$\Omega$ & $\left[0.25, 1.2\right]$ \\
$\beta$  & $A$      & $\left[50, 300\right]$ \\
\bottomrule 
\end{tabular}
\end{threeparttable}
\end{table}

\begin{figure}[t!]
\centering
\begin{subfigure}[b]{0.32\textwidth}
    \centering
    \includegraphics[width=\textwidth]{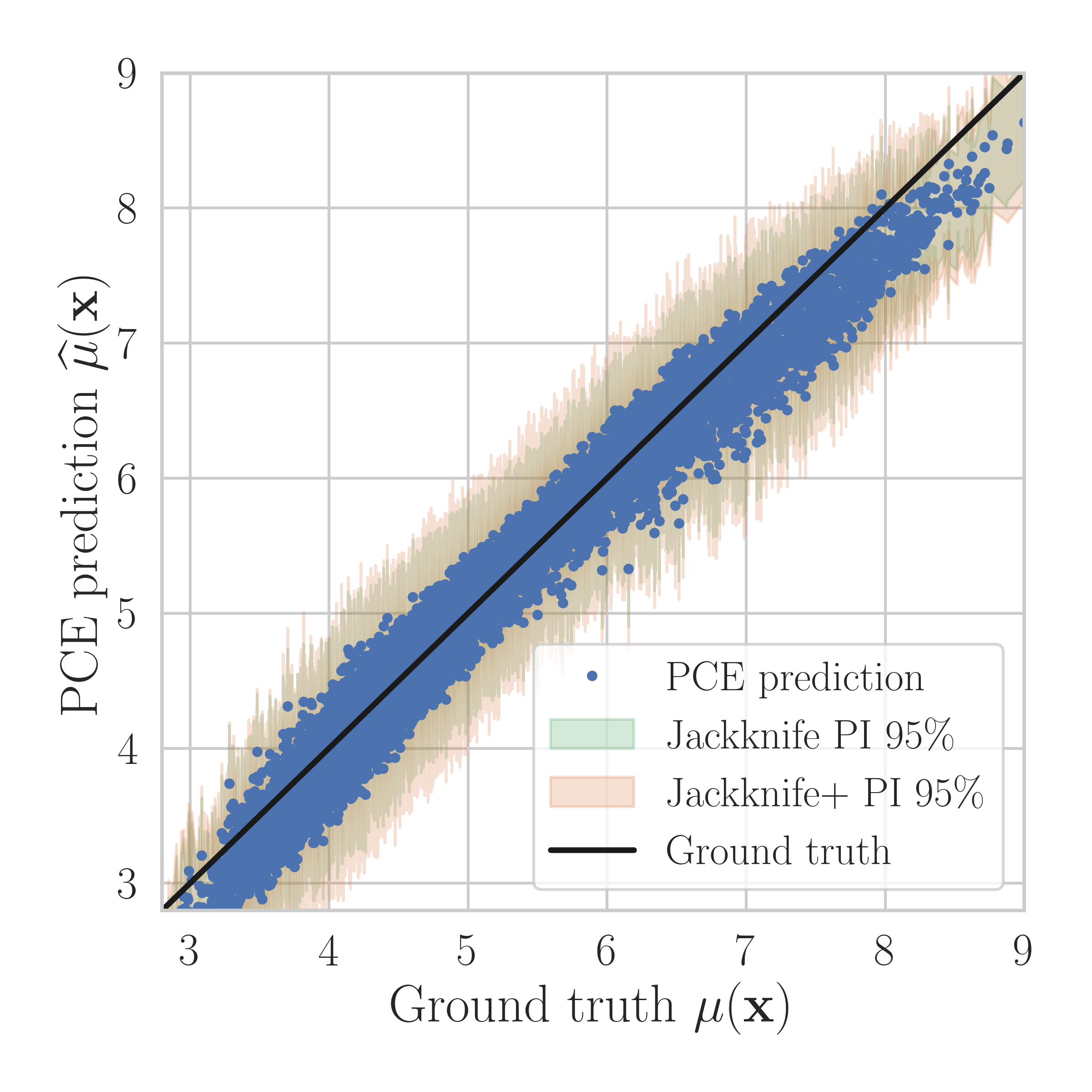}
    \caption{ $P=1$, $C=2$.}
\end{subfigure}
\hfill
\begin{subfigure}[b]{0.32\textwidth}
    \centering
    \includegraphics[width=\textwidth]{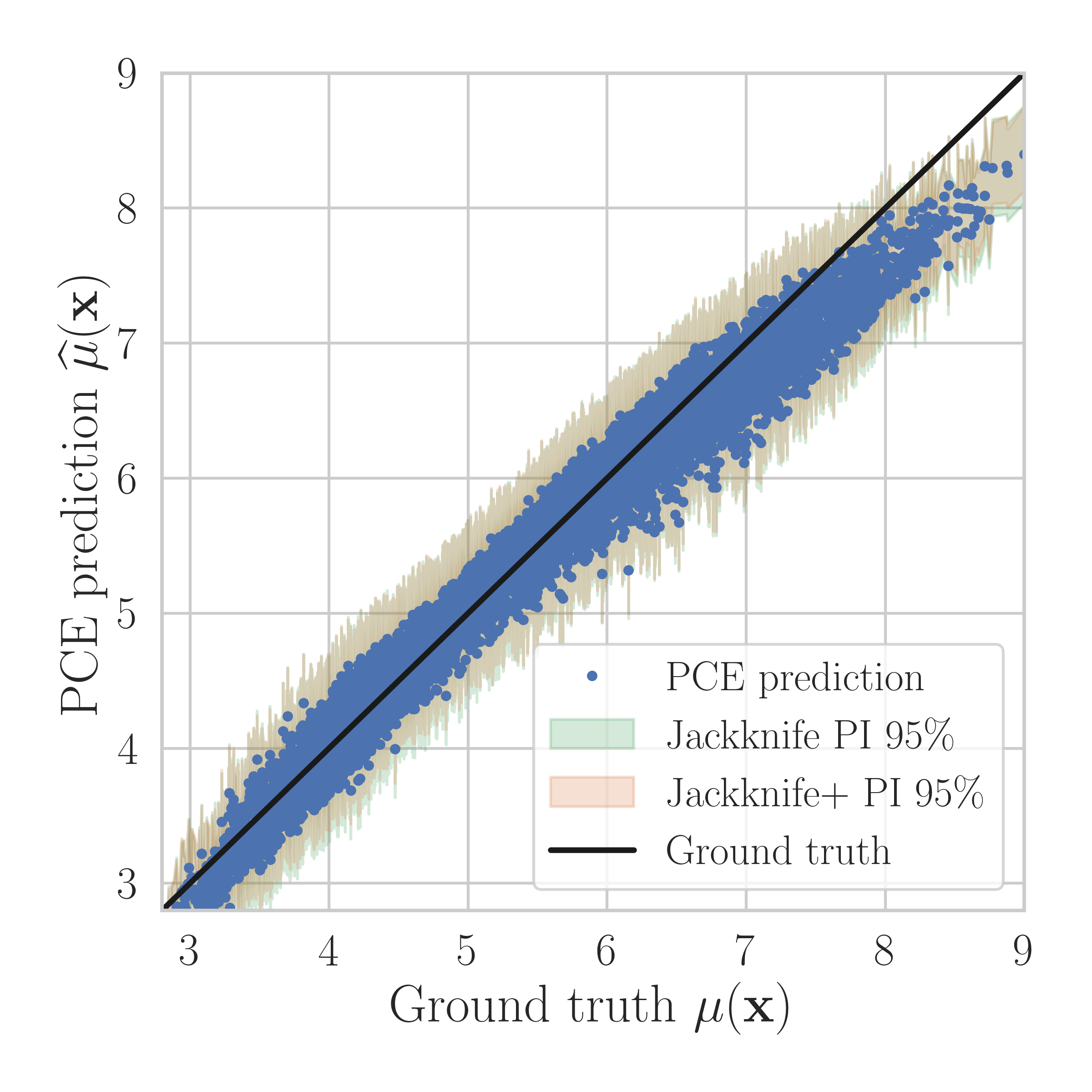}
    \caption{ $P=1$, $C=5$.}
\end{subfigure}
\hfill
\begin{subfigure}[b]{0.32\textwidth}
    \centering
    \includegraphics[width=\textwidth]{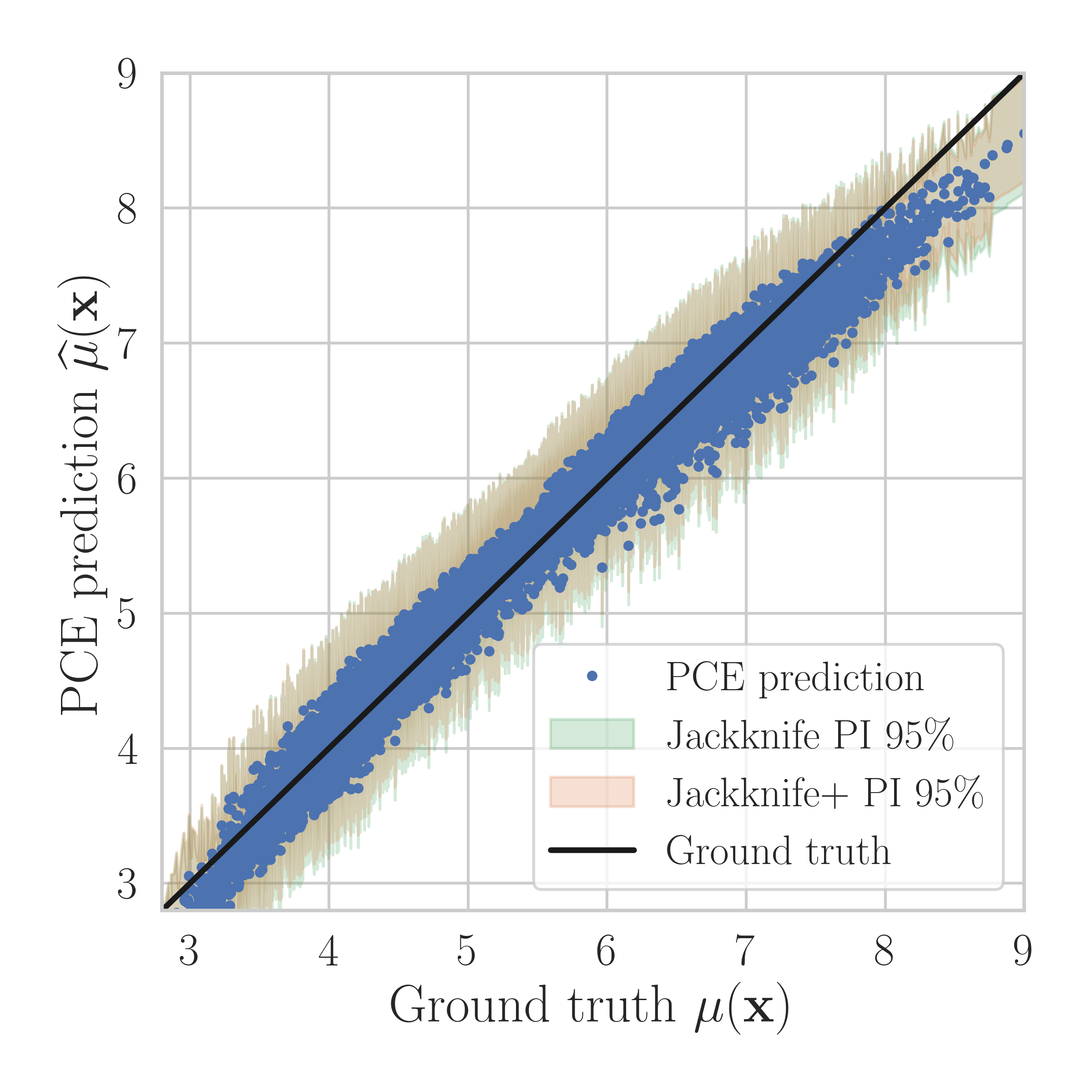}
    \caption{ $P=1$, $C=10$.}
\end{subfigure}
\\
\begin{subfigure}[b]{0.32\textwidth}
    \centering
    \includegraphics[width=\textwidth]{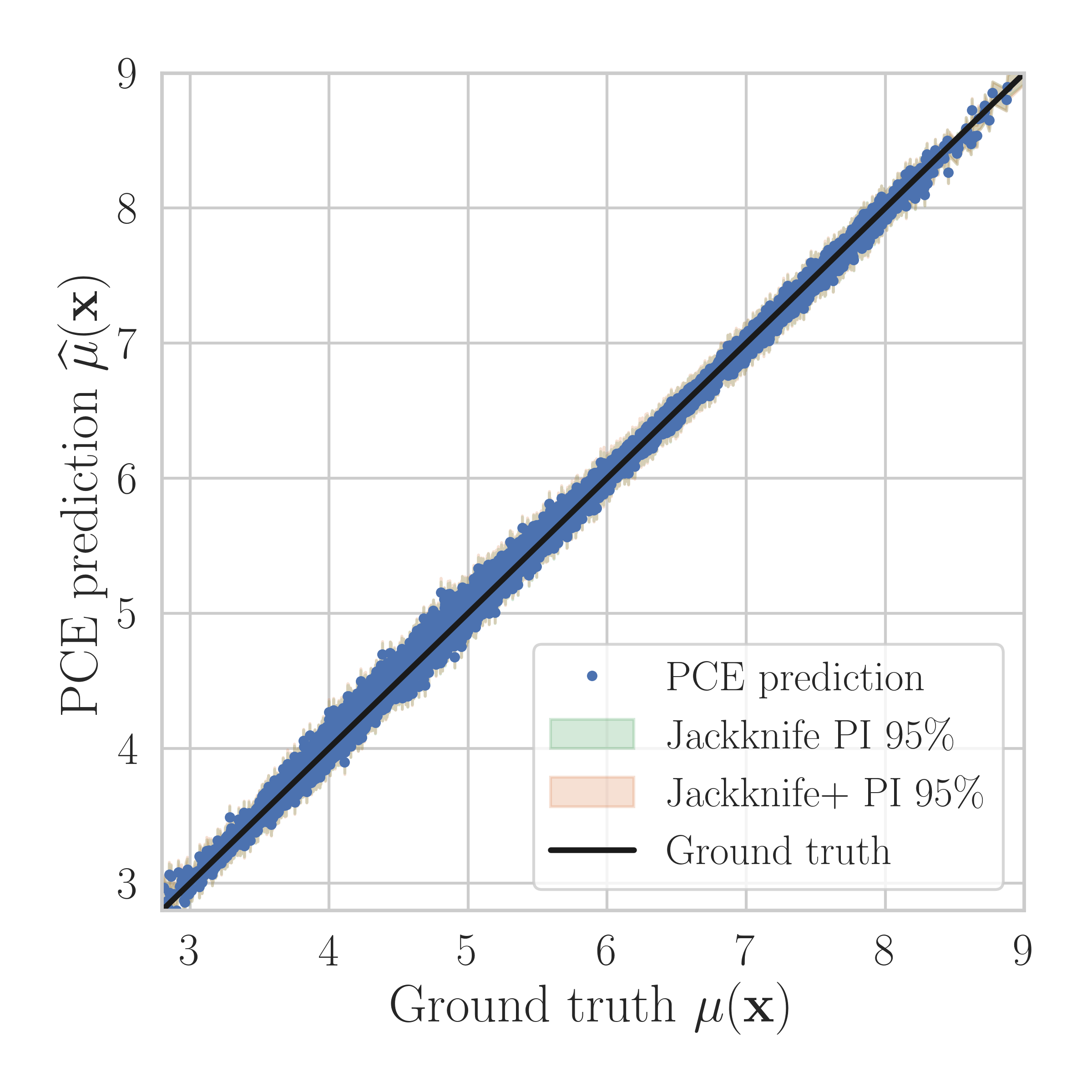}
    \caption{ $P=2$, $C=2$.}
\end{subfigure}
\hfill
\begin{subfigure}[b]{0.32\textwidth}
    \centering
    \includegraphics[width=\textwidth]{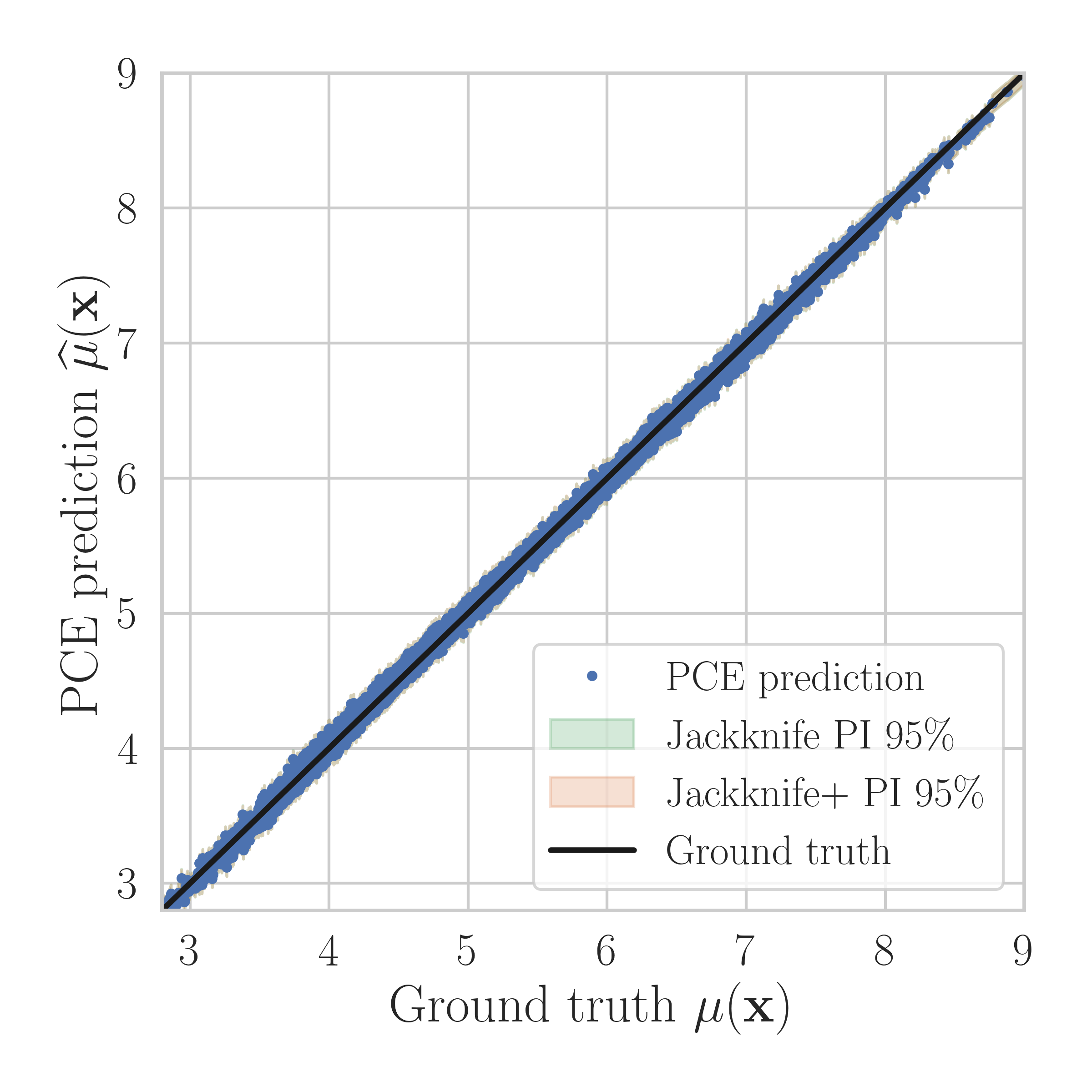}
    \caption{ $P=2$, $C=5$.}
\end{subfigure}
\hfill
\begin{subfigure}[b]{0.32\textwidth}
    \centering
    \includegraphics[width=\textwidth]{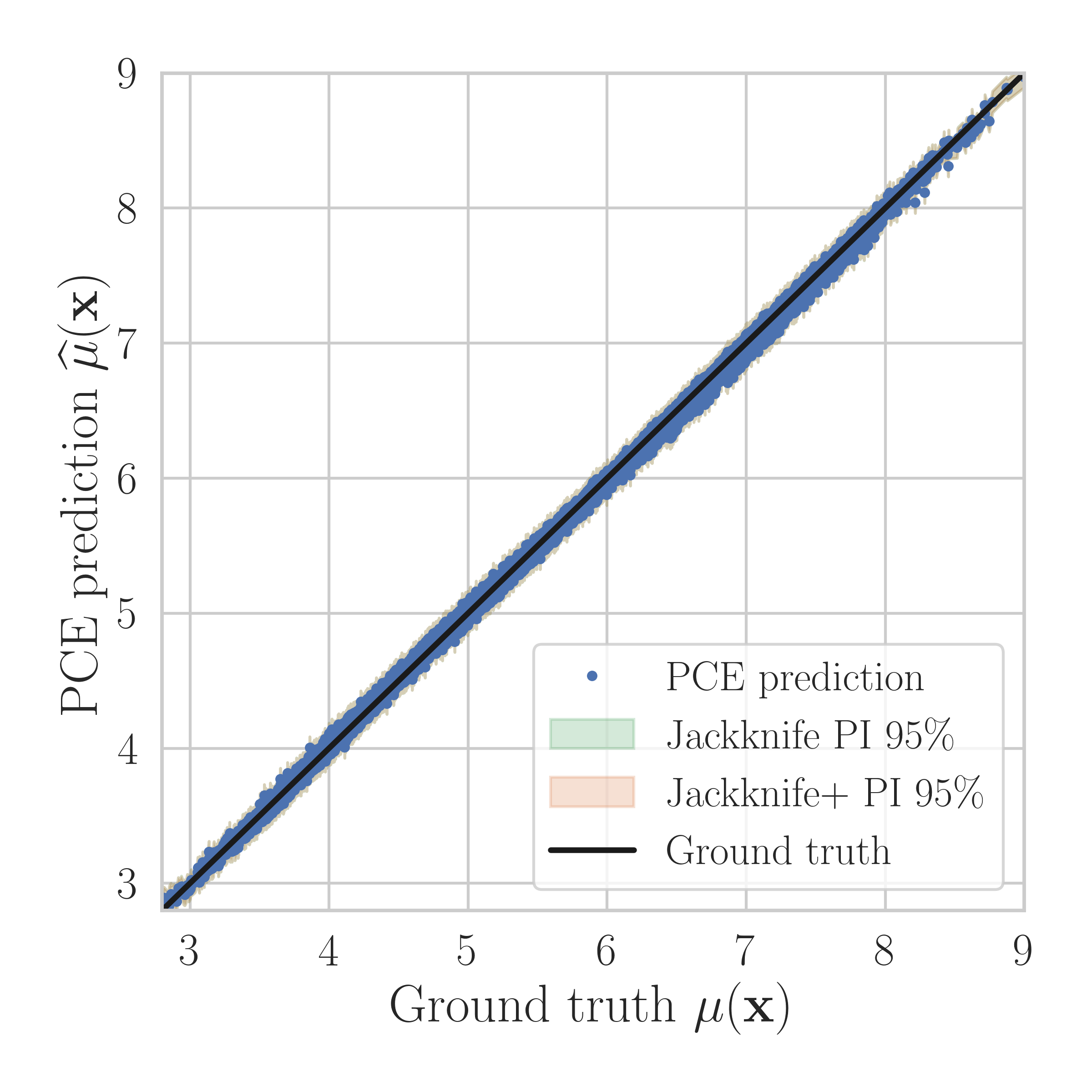}
    \caption{ $P=2$, $C=10$.}
\end{subfigure}
\\
\begin{subfigure}[b]{0.32\textwidth}
    \centering
    \includegraphics[width=\textwidth]{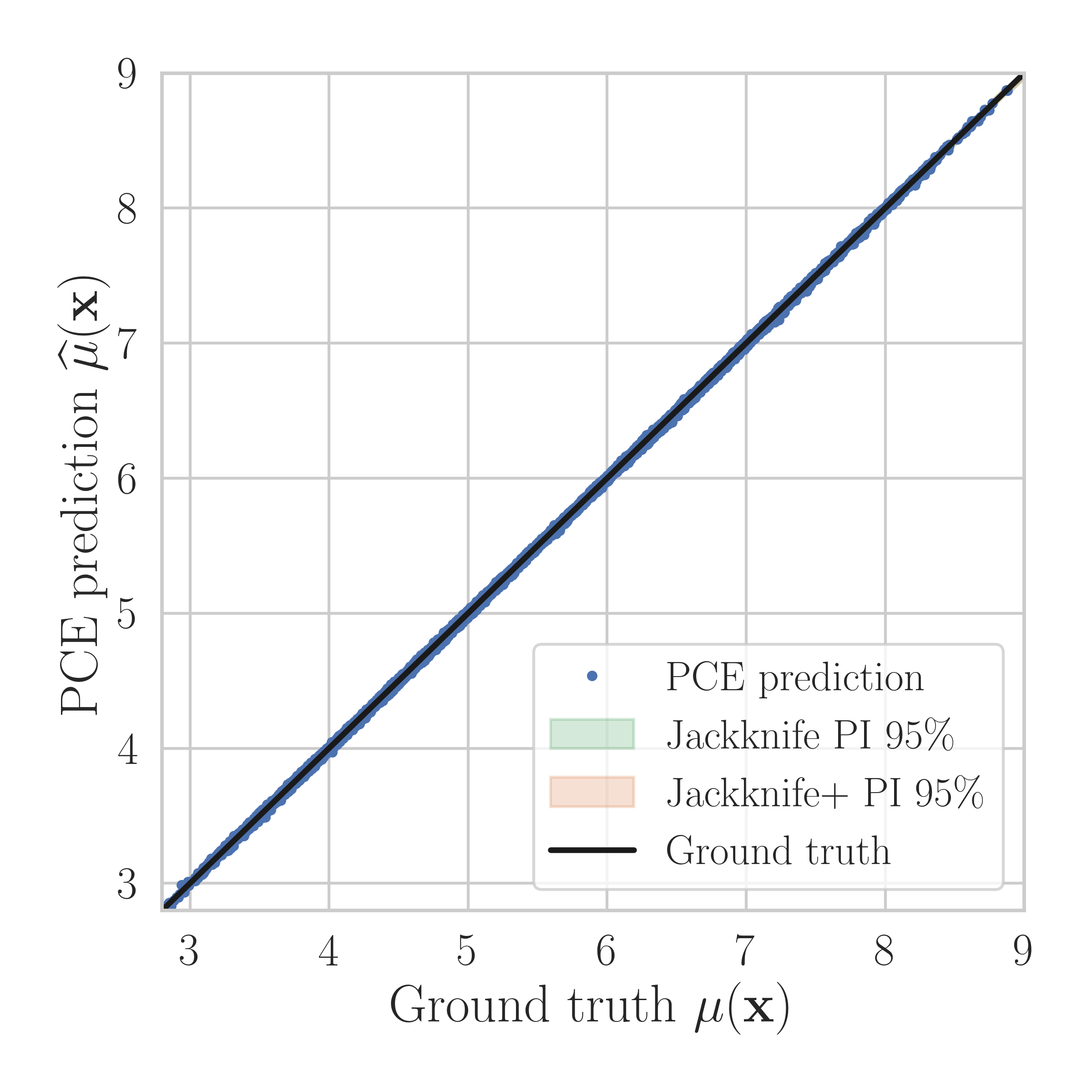}
    \caption{ $P=3$, $C=2$.}
\end{subfigure}
\hfill
\begin{subfigure}[b]{0.32\textwidth}
    \centering
    \includegraphics[width=\textwidth]{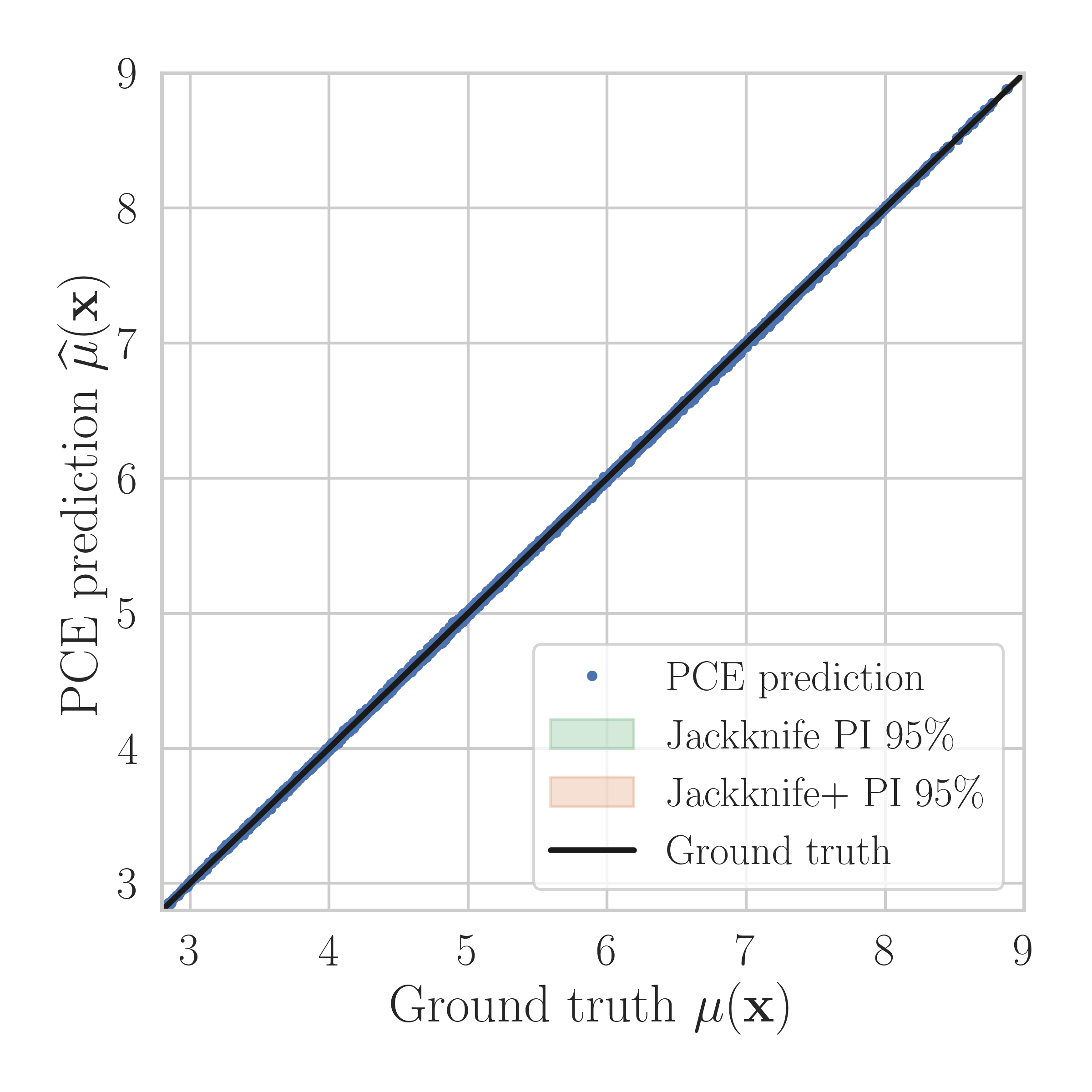}
    \caption{ $P=3$, $C=5$.}
\end{subfigure}
\hfill
\begin{subfigure}[b]{0.32\textwidth}
    \centering
    \includegraphics[width=\textwidth]{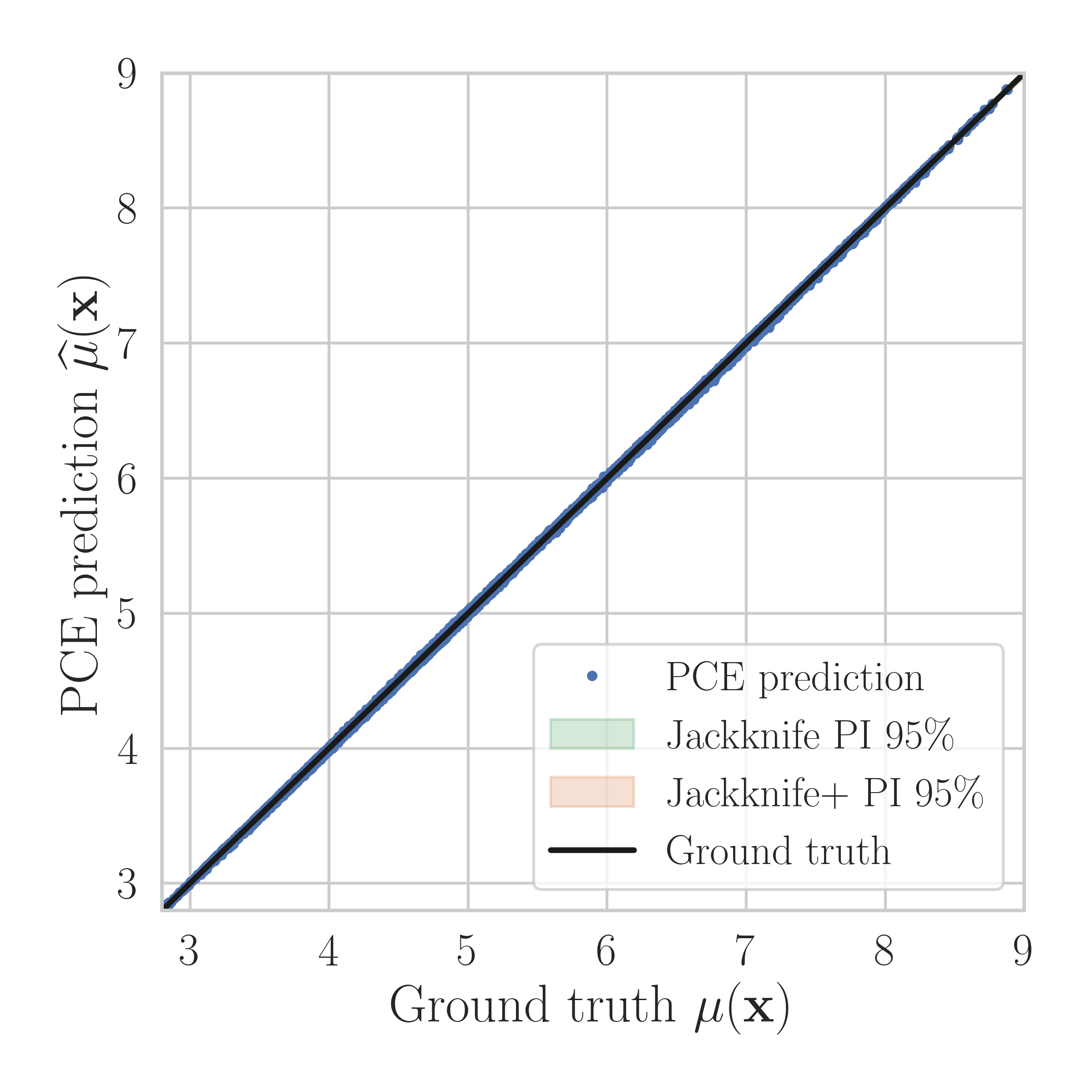}
    \caption{ $P=3$, $C=10$.}
\end{subfigure}
\caption{Parity plots comparing ground truth values of the OTL circuit function against conformalized \gls{pce} predictions for different combinations of polynomial degree $P$ and oversampling coefficient $C$. The results correspond to a single random seed. \textcolor{black}{The target coverage is fixed at $95\%$, corresponding to a significance level of $s=0.05$.}}
\label{fig:otl-circuit-parity-plots}
\end{figure}

\begin{figure}[t!]
\centering
\begin{subfigure}[b]{0.32\textwidth}
    \centering
    \includegraphics[width=\textwidth]{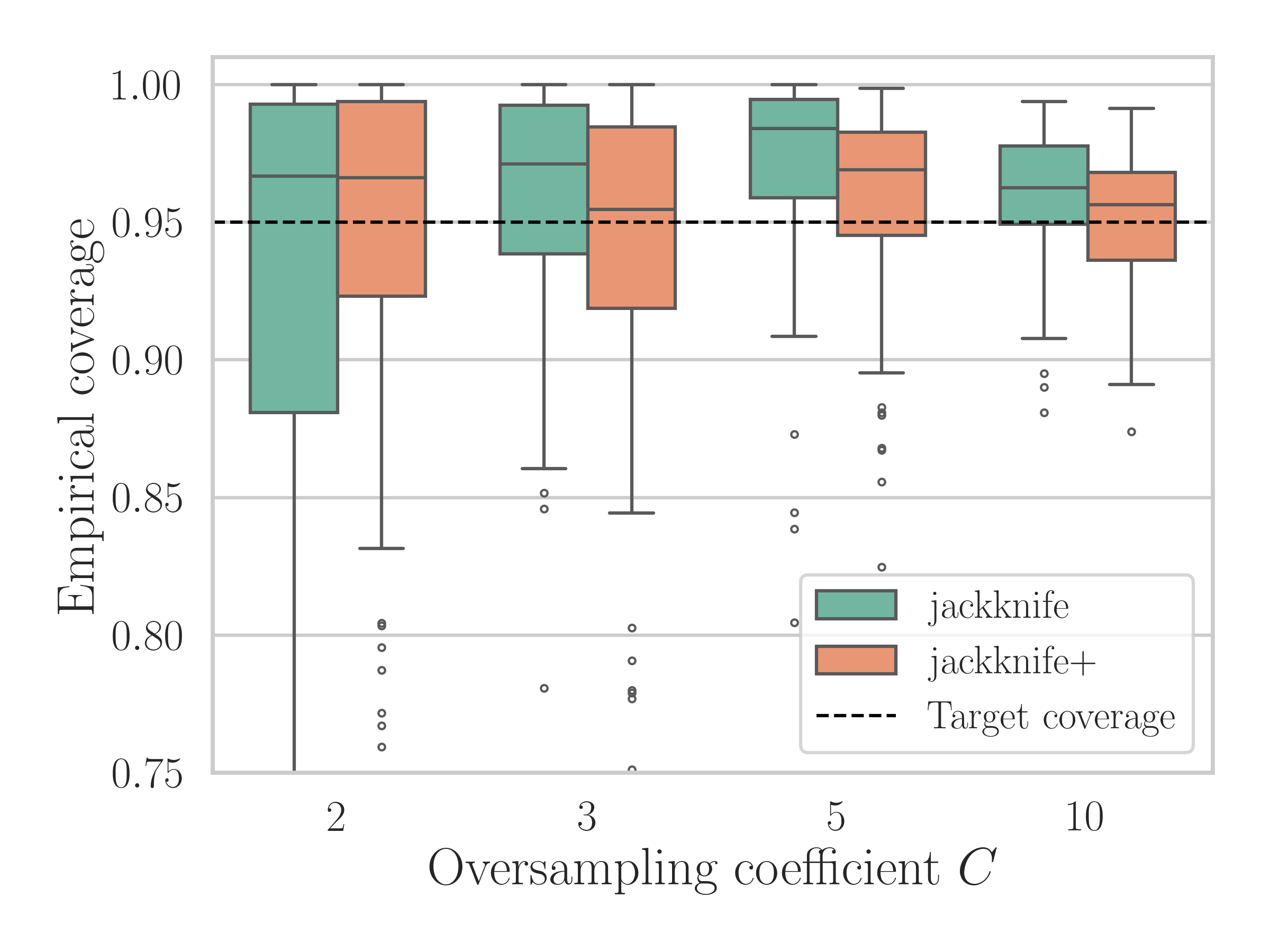}
    \caption{$P=1$.}
\end{subfigure}
\hfill 
\begin{subfigure}[b]{0.32\textwidth}
    \centering
    \includegraphics[width=\textwidth]{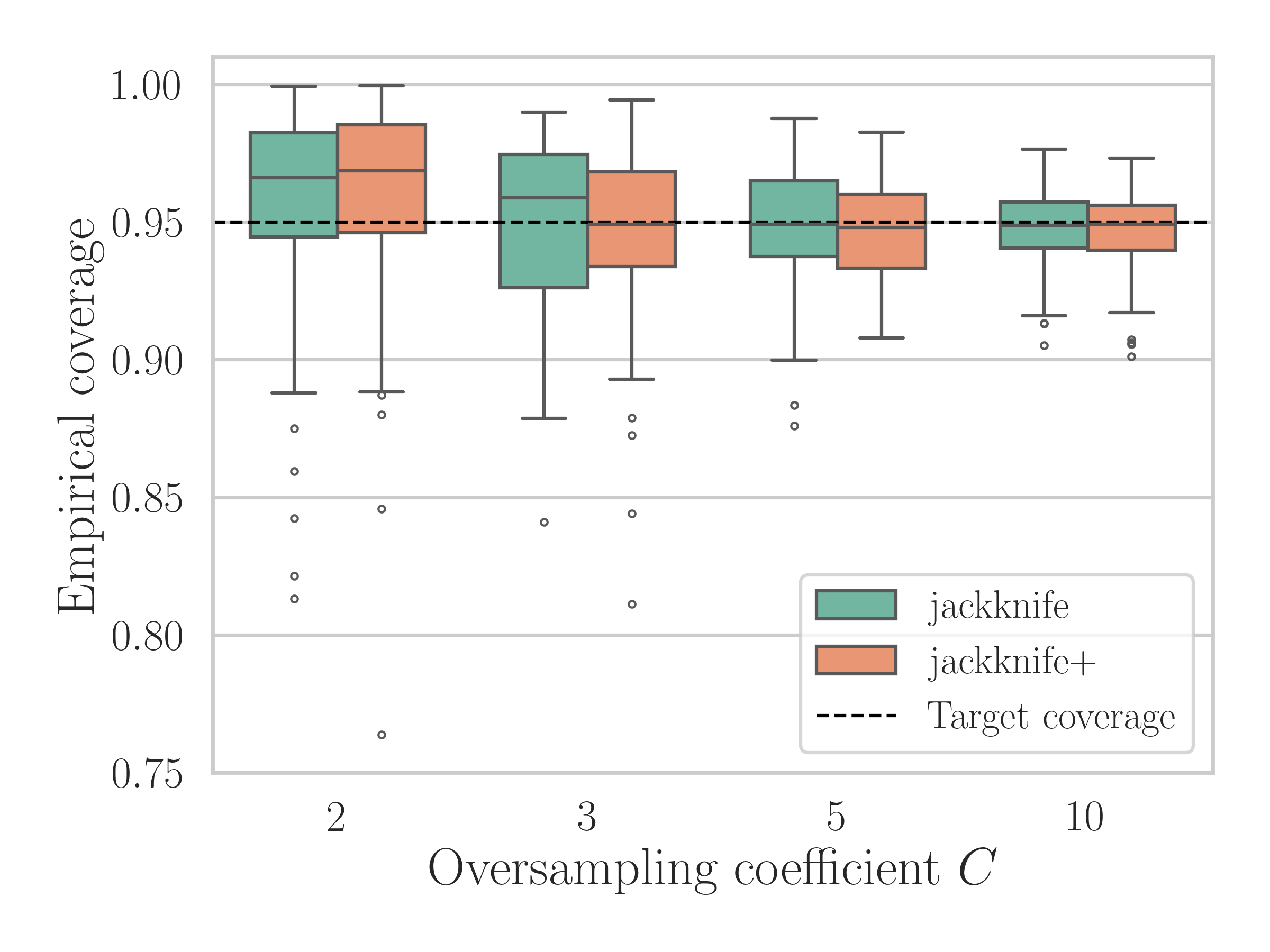}
    \caption{$P=2$.}
\end{subfigure}
\hfill
\begin{subfigure}[b]{0.32\textwidth}
    \centering
    \includegraphics[width=\textwidth]{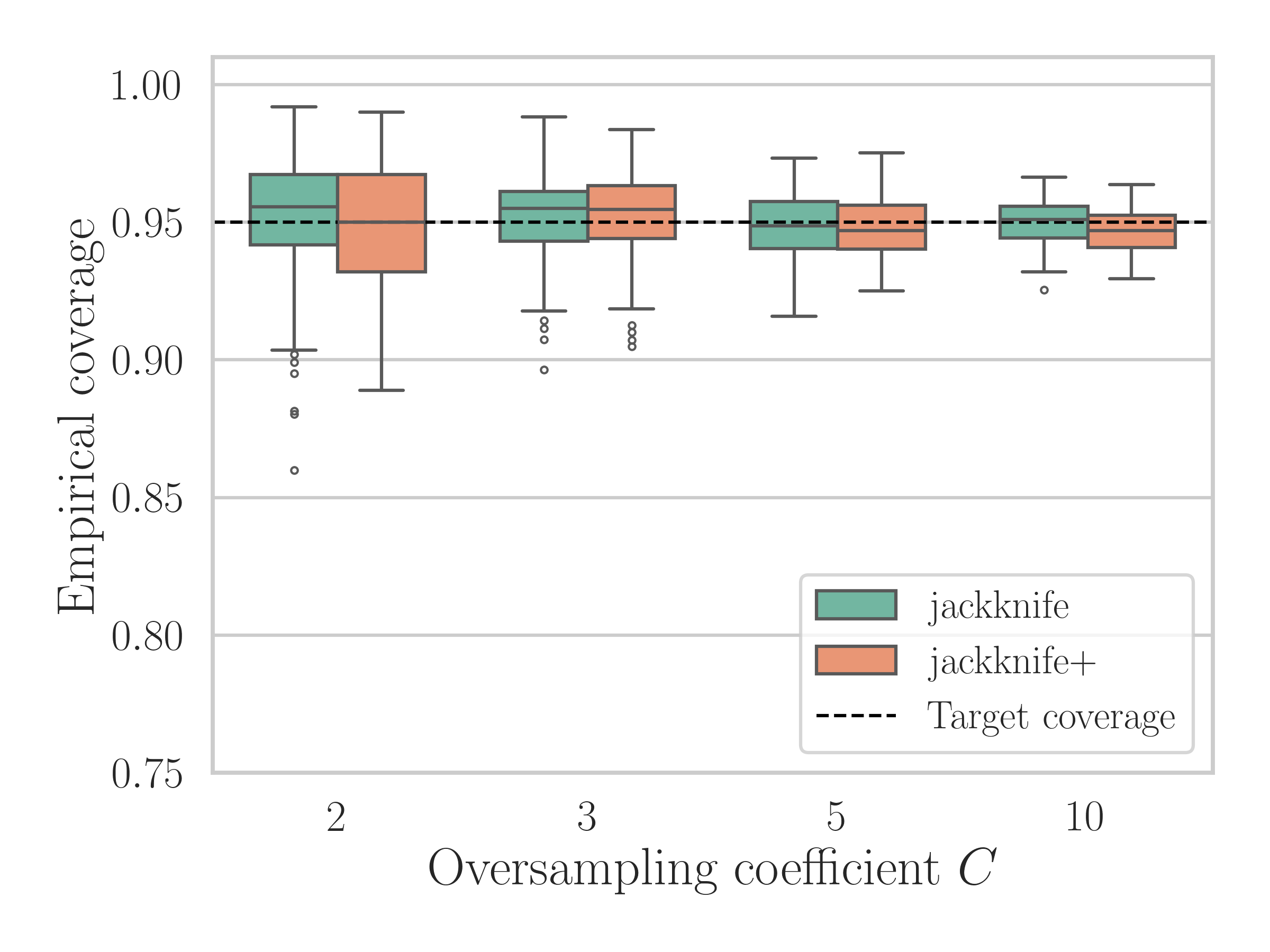}
    \caption{$P=3$.}
\end{subfigure}
\caption{Box plots of the empirical coverage provided by conformalized \gls{pce} surrogates of the OTL circuit function, for different combinations of polynomial degree $P$, and oversampling coefficient $C$. \textcolor{black}{The target coverage is fixed at $95\%$, corresponding to a significance level of $s=0.05$.}}
\label{fig:otl-circuit-coverage-boxplots}
\end{figure}

\begin{figure}[t!]
\centering
\begin{subfigure}[b]{0.32\textwidth}
    \centering
    \includegraphics[width=\textwidth]{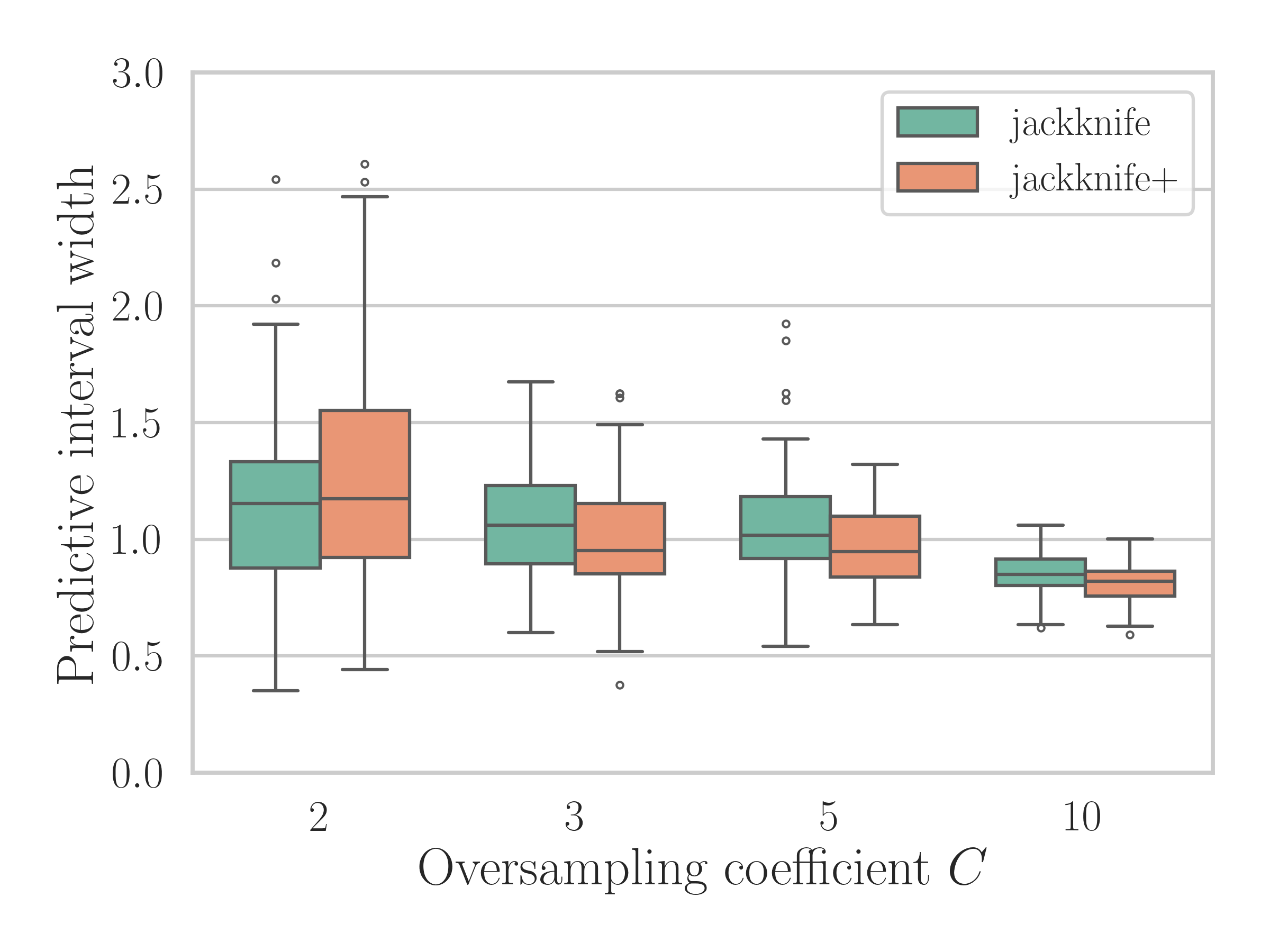}
    \caption{$P=1$.}
\end{subfigure}
\hfill
\begin{subfigure}[b]{0.32\textwidth}
    \centering
    \includegraphics[width=\textwidth]{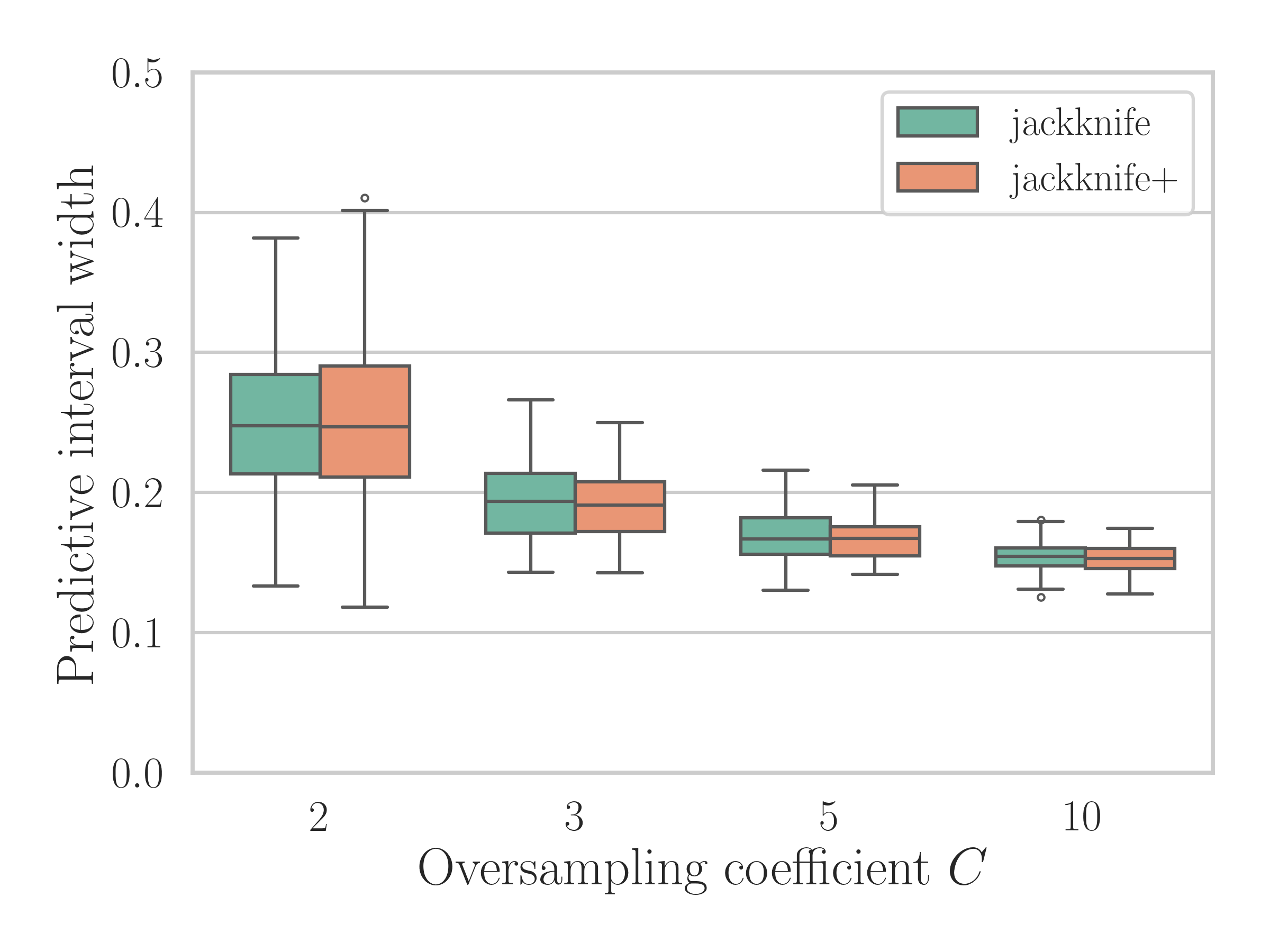}
    \caption{$P=2$.}
\end{subfigure}
\hfill
\begin{subfigure}[b]{0.32\textwidth}
    \centering
    \includegraphics[width=\textwidth]{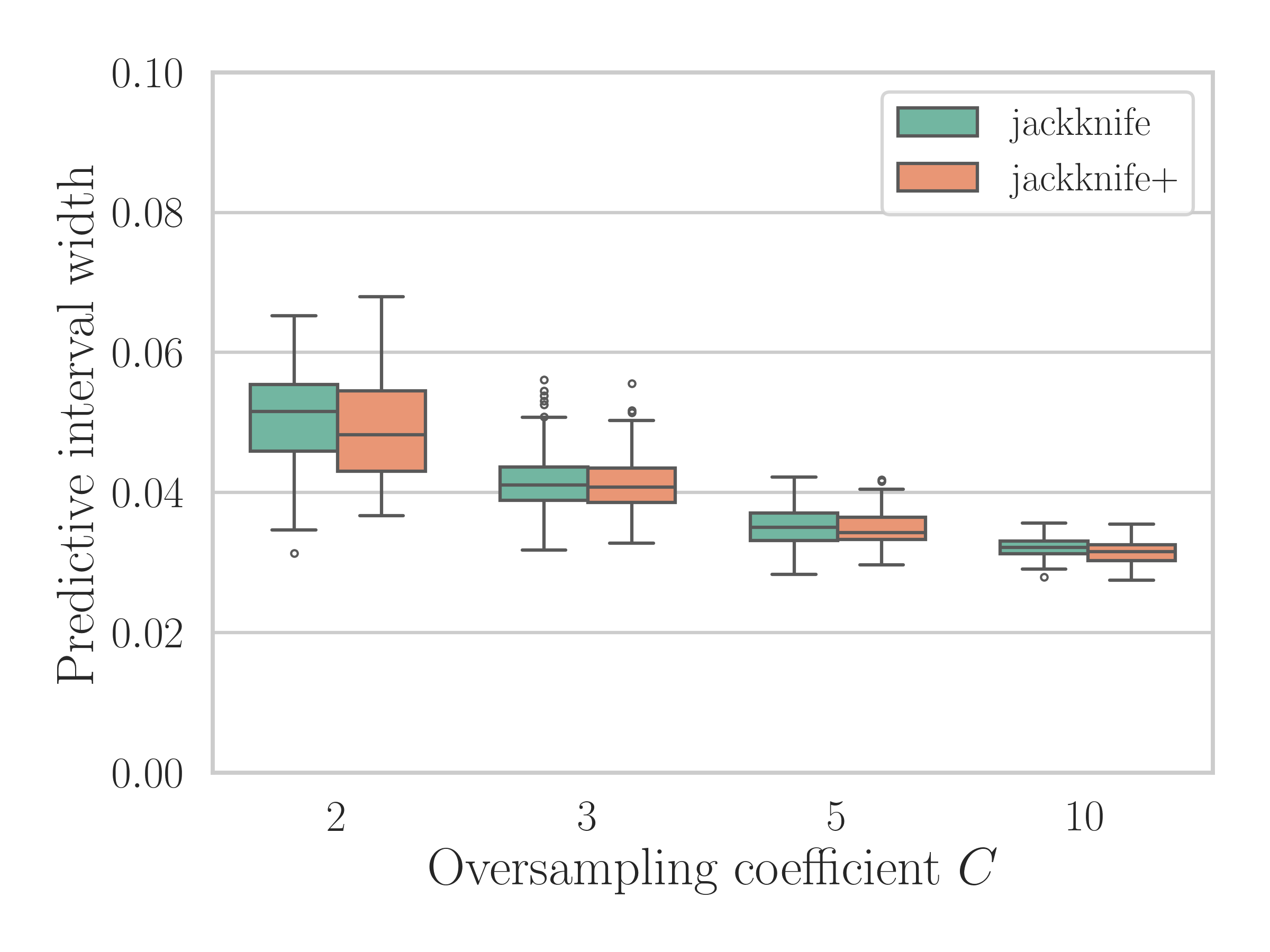}
    \caption{$P=3$.}
\end{subfigure}
\caption{Box plots of the prediction interval widths provided by conformalized \gls{pce} surrogates of the OTL circuit function, for different combinations of polynomial degree $P$, and oversampling coefficient $C$. \textcolor{black}{The target coverage is fixed at $95\%$, corresponding to a significance level of $s=0.05$.}}
\label{fig:otl-circuit-interval-boxplots}
\end{figure}

Total-degree conformalized \glspl{pce} with maximum polynomial degrees $P \in \left\{1,2,3\right\}$ and oversampling coefficients $C \in \left\{2,3,5,10\right\}$ are employed. 
Following Migliorati et al. \cite{migliorati2014analysis}, the experimental design now scales linearly with the size of the \gls{pce} basis, such that $M = C K = C \#\Lambda$. 
The different conformalized \gls{pce} configurations are compared against the true OTL circuit function for a single random seed in Figure~\ref{fig:otl-circuit-parity-plots}, where the results for $C=3$ have been omitted for better readability. 
Figures~\ref{fig:otl-circuit-coverage-boxplots} and \ref{fig:otl-circuit-interval-boxplots} show box plots regarding the distributions of the conformalized \glspl{pce}' empirical coverages and prediction intervals over 100 random seeds.

\subsection{Piston simulation function}
\label{sec:piston}
We consider the circular motion of a piston within a cylinder, the cycle-time of which is given by the function
\begin{align} 
&\mu(\mathbf{x}) = 2\pi \sqrt{ \frac{M}{k + S^2 \frac{P_0 V_0}{T_0} \frac{T_0}{V^2}} }, \\
&\text{where } V = \frac{S}{2k} \left( \sqrt{A^2 + 4k \frac{P_0 V_0}{T_0} T_a} - A \right) \text{ and }
A = P_0\,S + 19.62\,M - \frac{k\,V_0}{S}. \nonumber
\end{align}
The input vector $\mathbf{x} \in \mathbb{R}^7$ consists of the parameters listed in Table~\ref{tab:piston-parameters} along with their value ranges.
All parameters are assumed to be uniformly distributed within their ranges.
\begin{table}[h!]
\centering
\caption{Input parameters of the piston simulation function.}
\label{tab:piston-parameters}
\begin{threeparttable}
\begin{tabular}{c c c l}
\toprule
Parameter & Description & Units & Range \\
\midrule 
$M$     & piston weight & kg  & $\left[30, 60\right]$ \\  
$S$     & piston surface area & m$^2$ & $\left[0.005, 0.02\right]$ \\
$V_{0}$ & initial gas volume & m$^3$ & $\left[0.002, 0.01\right]$ \\
$k$     & spring coefficient & N/m & $\left[1000, 5000\right]$ \\
$P_{0}$ & atmospheric pressure & N/m$^2$ & $\left[90000, 110000\right]$ \\
$T_a$   & ambient temperature & K & $\left[290, 296\right]$ \\
$T_0$   & filling gas temperature & K & $\left[340, 360\right]$\\ 
\bottomrule 
\end{tabular}
\end{threeparttable}
\end{table} 

\begin{figure}[t!]
\centering
\begin{subfigure}[b]{0.32\textwidth}
    \centering
    \includegraphics[width=\textwidth]{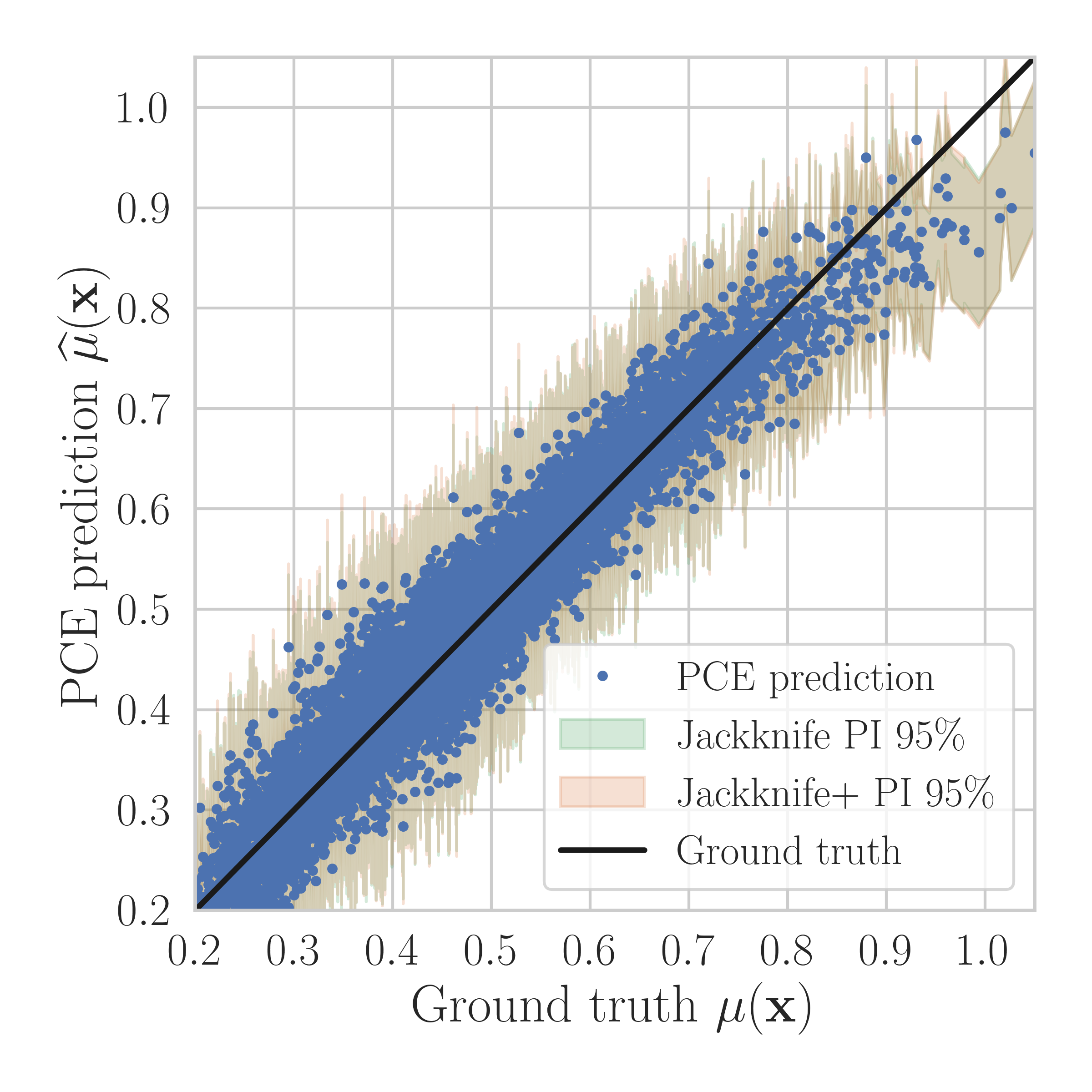}
    \caption{ $P=2$, $C=2$.}
\end{subfigure}
\hfill
\begin{subfigure}[b]{0.32\textwidth}
    \centering
    \includegraphics[width=\textwidth]{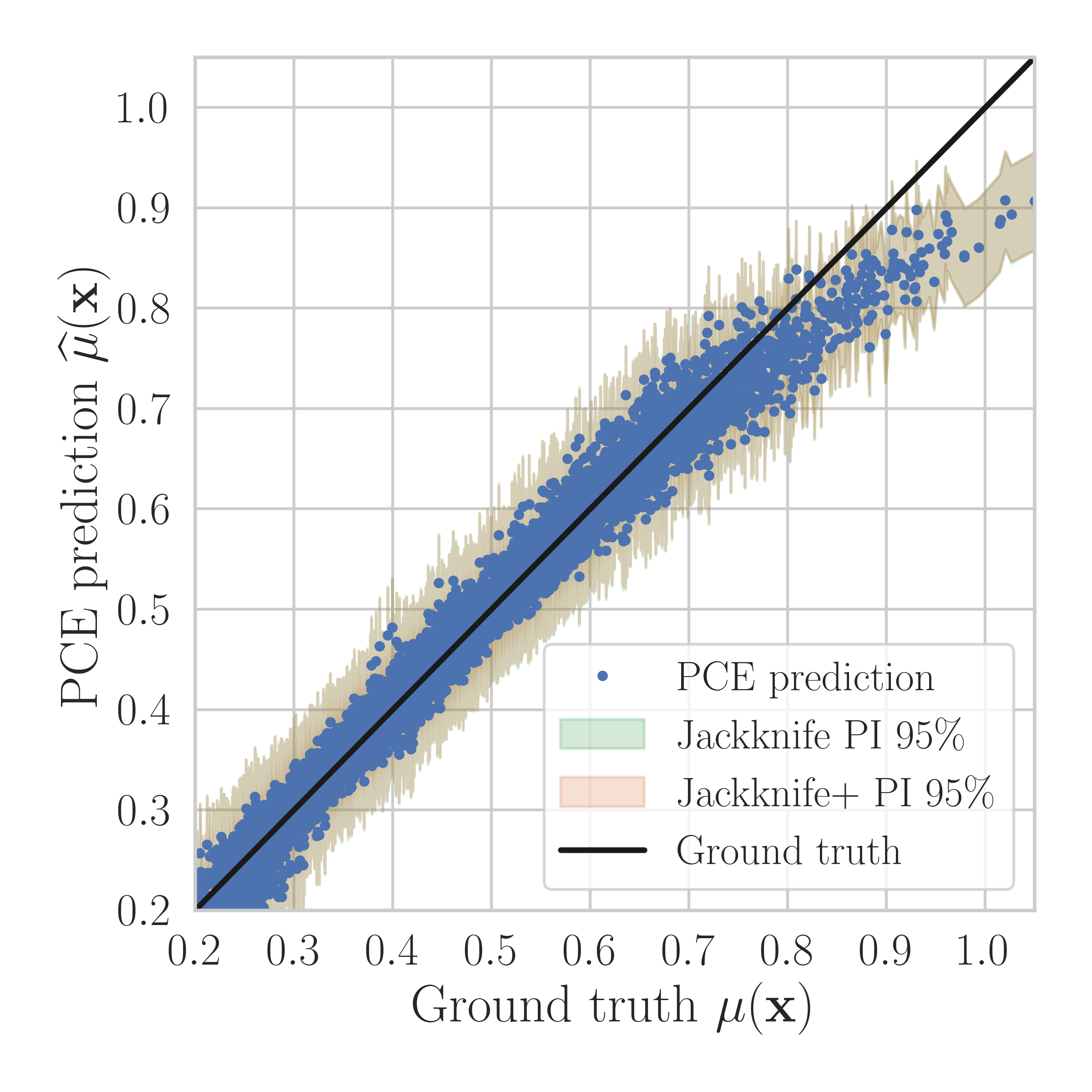}
    \caption{ $P=2$, $C=5$.}
\end{subfigure}
\hfill
\begin{subfigure}[b]{0.32\textwidth}
    \centering
    \includegraphics[width=\textwidth]{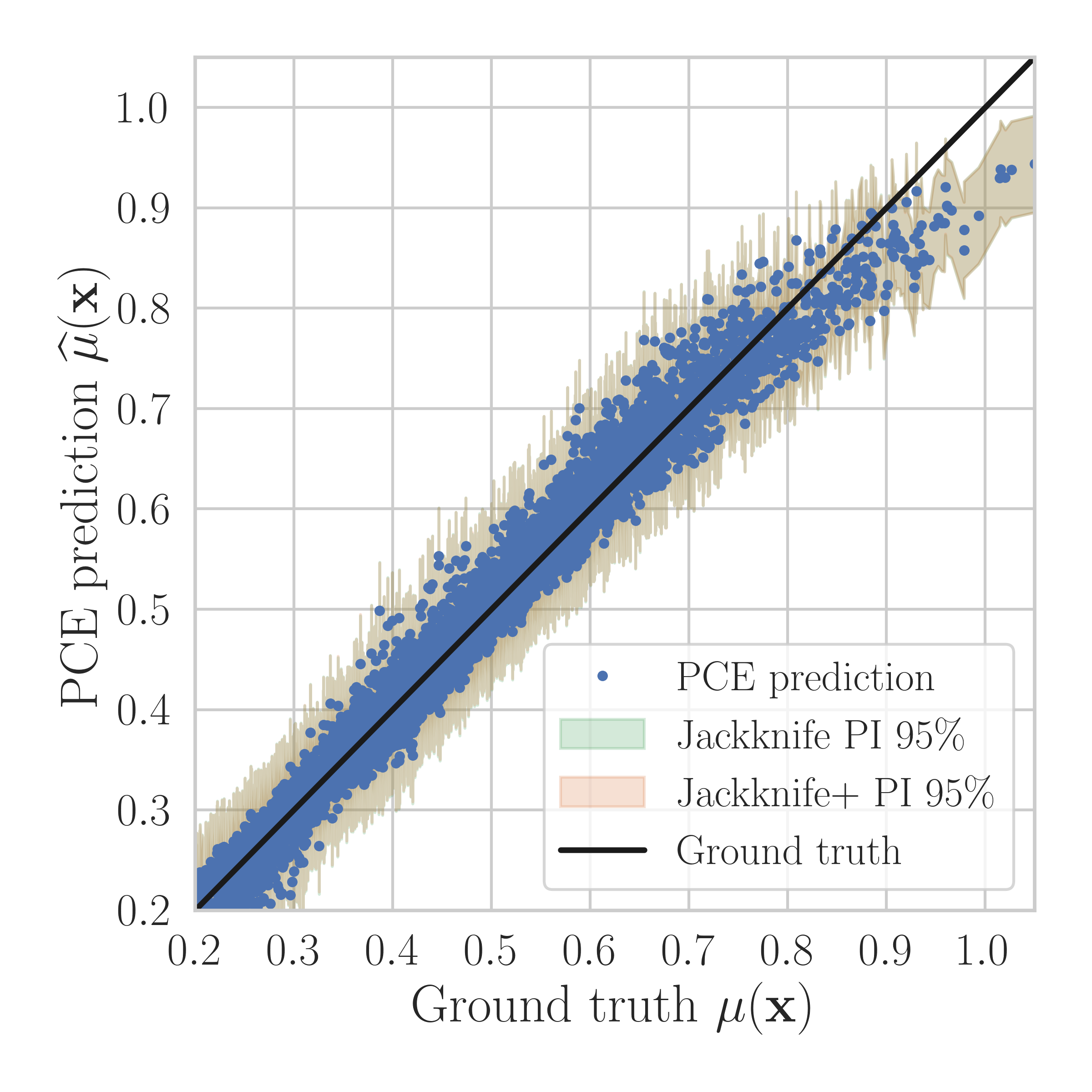}
    \caption{ $P=2$, $C=10$.}
\end{subfigure}
\\
\begin{subfigure}[b]{0.32\textwidth}
    \centering
    \includegraphics[width=\textwidth]{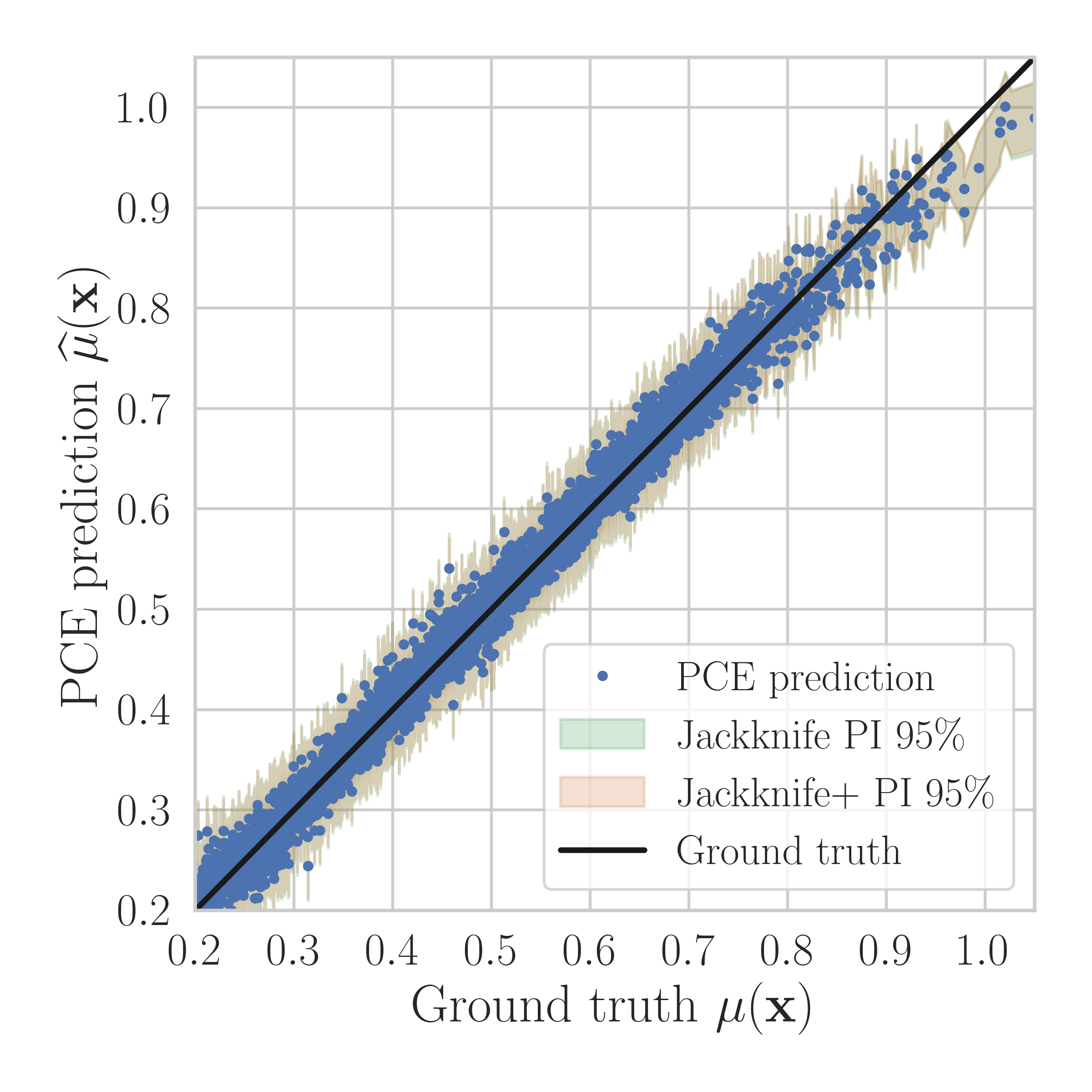}
    \caption{ $P=3$, $C=2$.}
\end{subfigure}
\hfill
\begin{subfigure}[b]{0.32\textwidth}
    \centering
    \includegraphics[width=\textwidth]{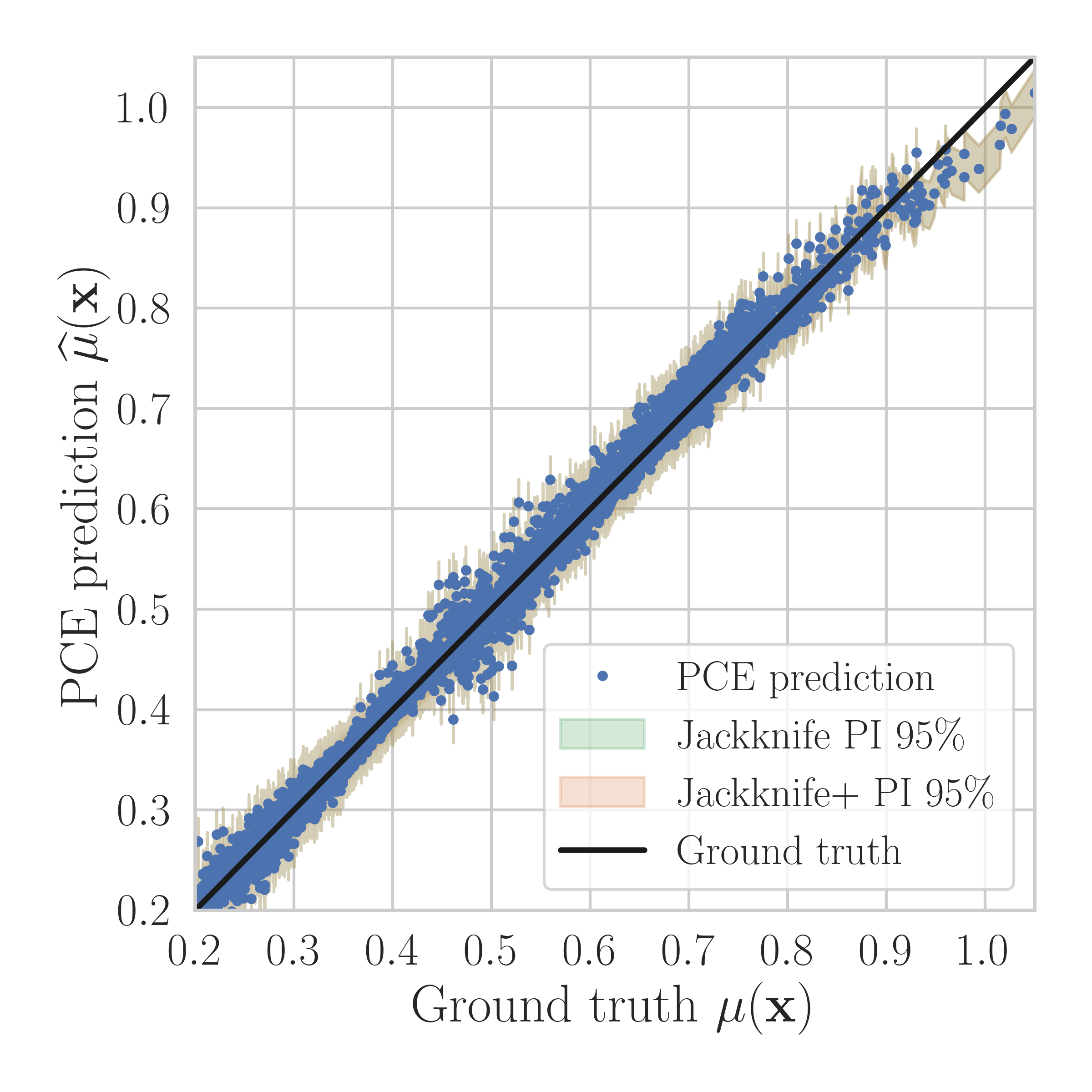}
    \caption{ $P=3$, $C=5$.}
\end{subfigure}
\hfill
\begin{subfigure}[b]{0.32\textwidth}
    \centering
    \includegraphics[width=\textwidth]{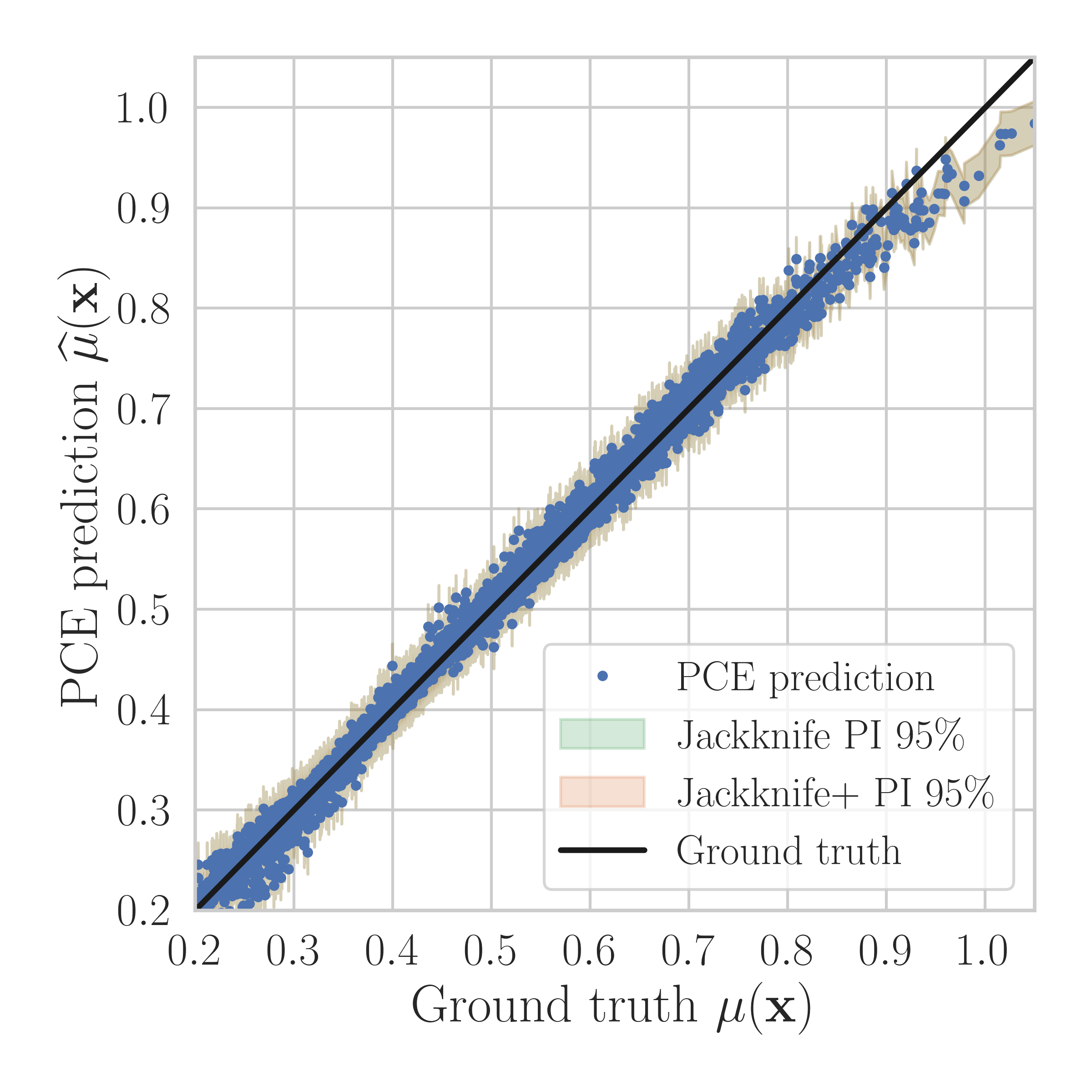}
    \caption{ $P=3$, $C=10$.}
\end{subfigure}
\\
\begin{subfigure}[b]{0.32\textwidth}
    \centering
    \includegraphics[width=\textwidth]{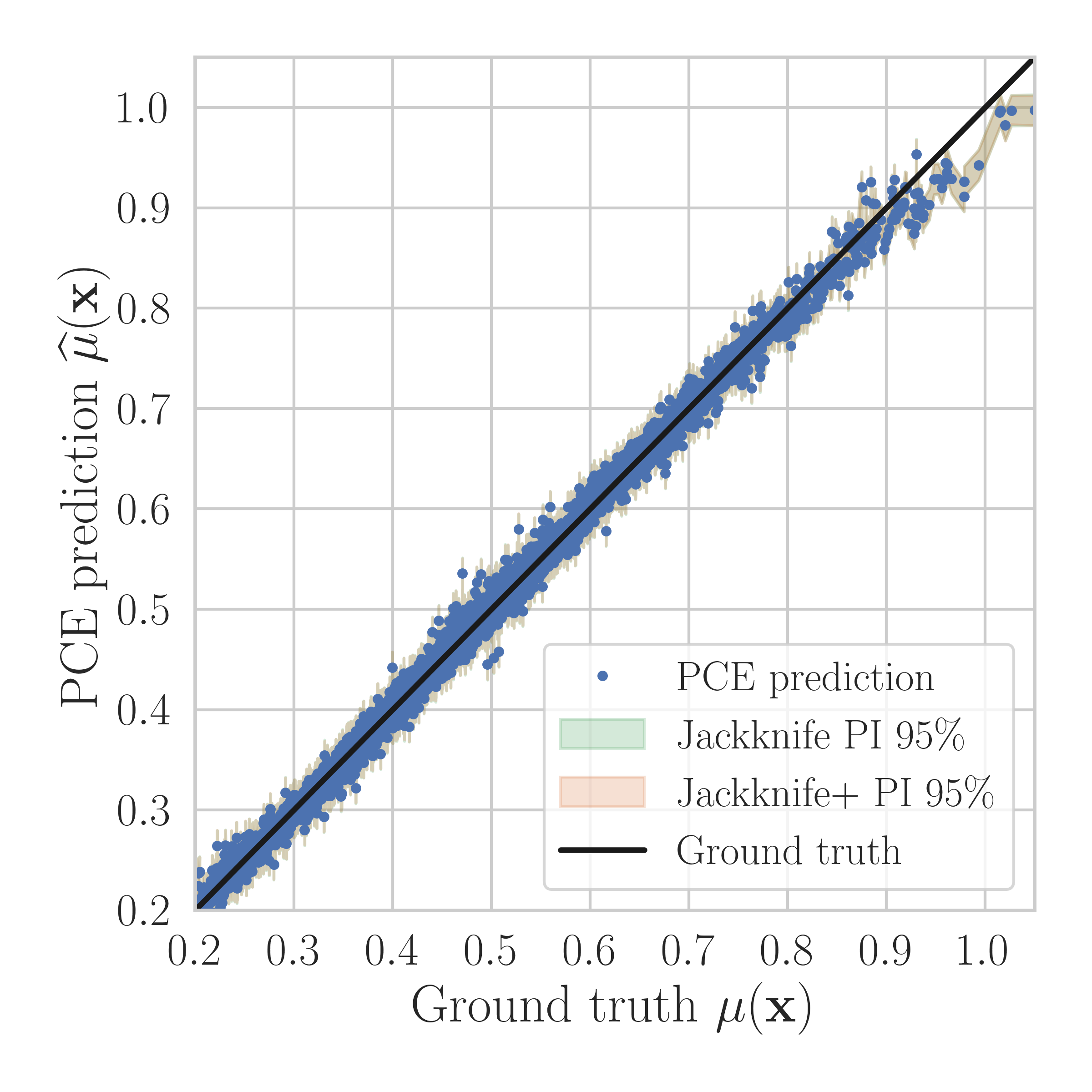}
    \caption{ $P=4$, $C=2$.}
\end{subfigure}
\hfill
\begin{subfigure}[b]{0.32\textwidth}
    \centering
    \includegraphics[width=\textwidth]{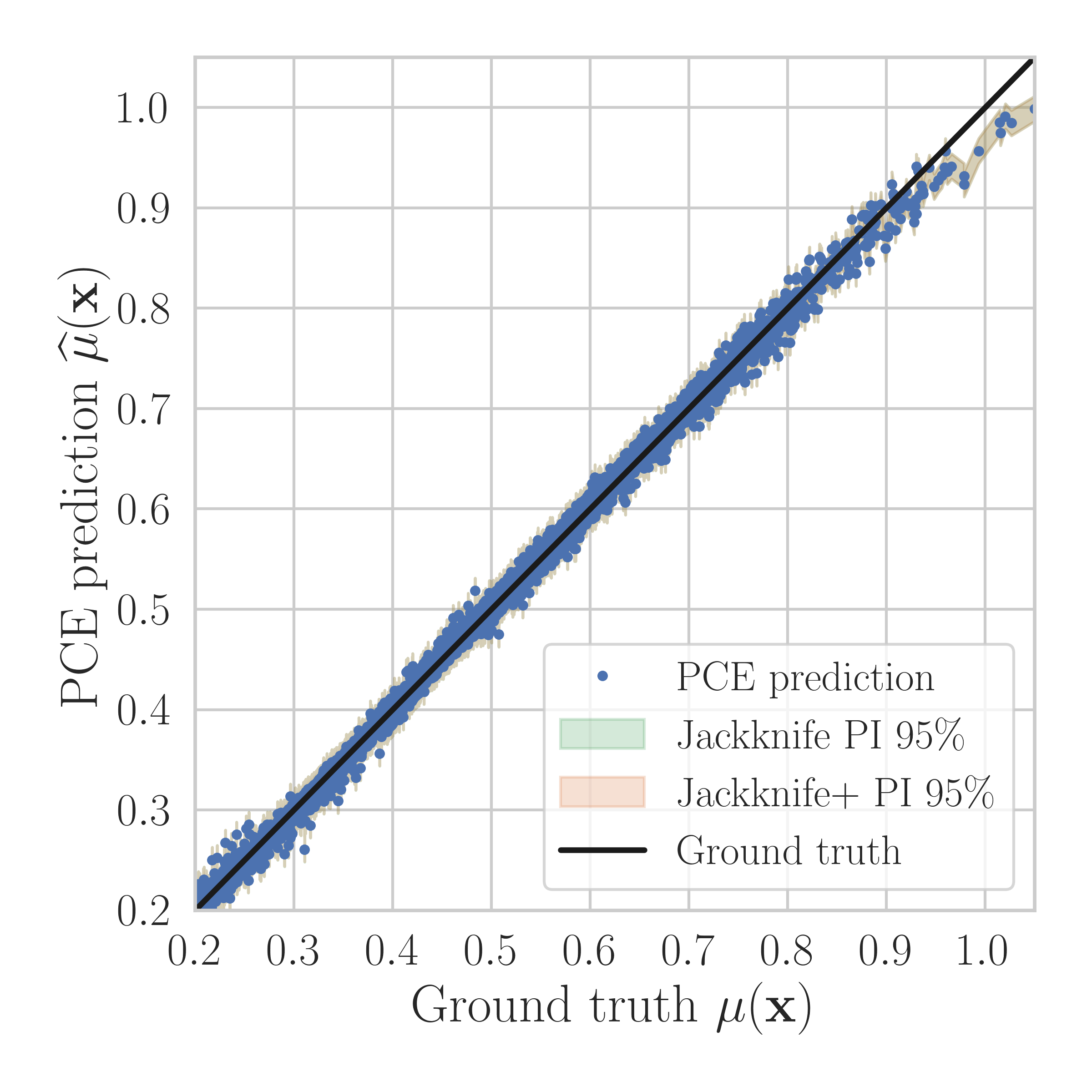}
    \caption{ $P=4$, $C=5$.}
\end{subfigure}
\hfill
\begin{subfigure}[b]{0.32\textwidth}
    \centering
    \includegraphics[width=\textwidth]{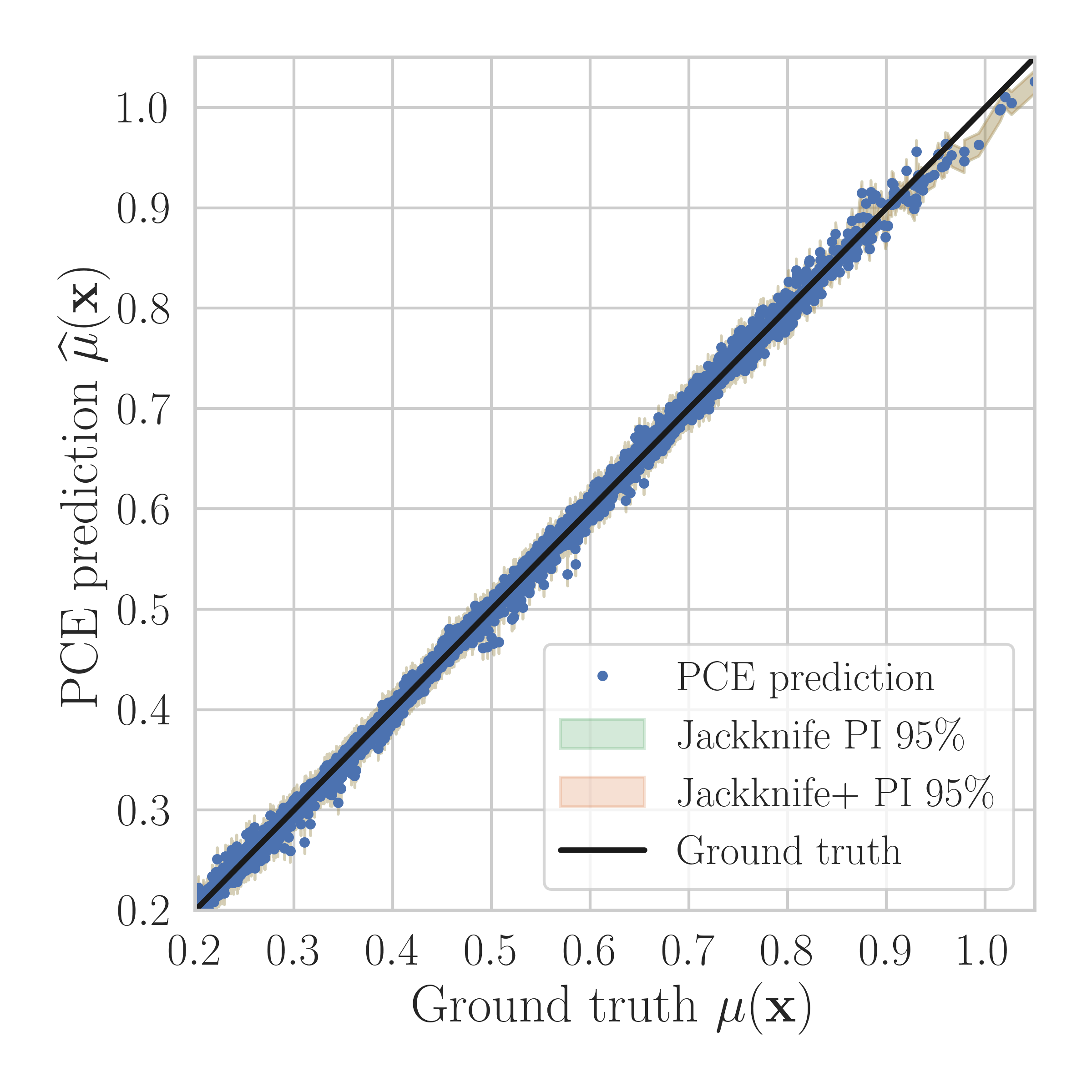}
    \caption{ $P=4$, $C=10$.}
\end{subfigure}
\caption{Parity plots comparing ground truth values of the piston simulation function against conformalized \gls{pce} predictions for different combinations of polynomial degree $P$ and oversampling coefficient $C$. The results correspond to a single random seed. \textcolor{black}{The target coverage is fixed at $95\%$, corresponding to a significance level of $s=0.05$.}}
\label{fig:piston-parity-plots}
\end{figure}

\begin{figure}[t!]
\centering
\begin{subfigure}[b]{0.32\textwidth}
    \centering
    \includegraphics[width=\textwidth]{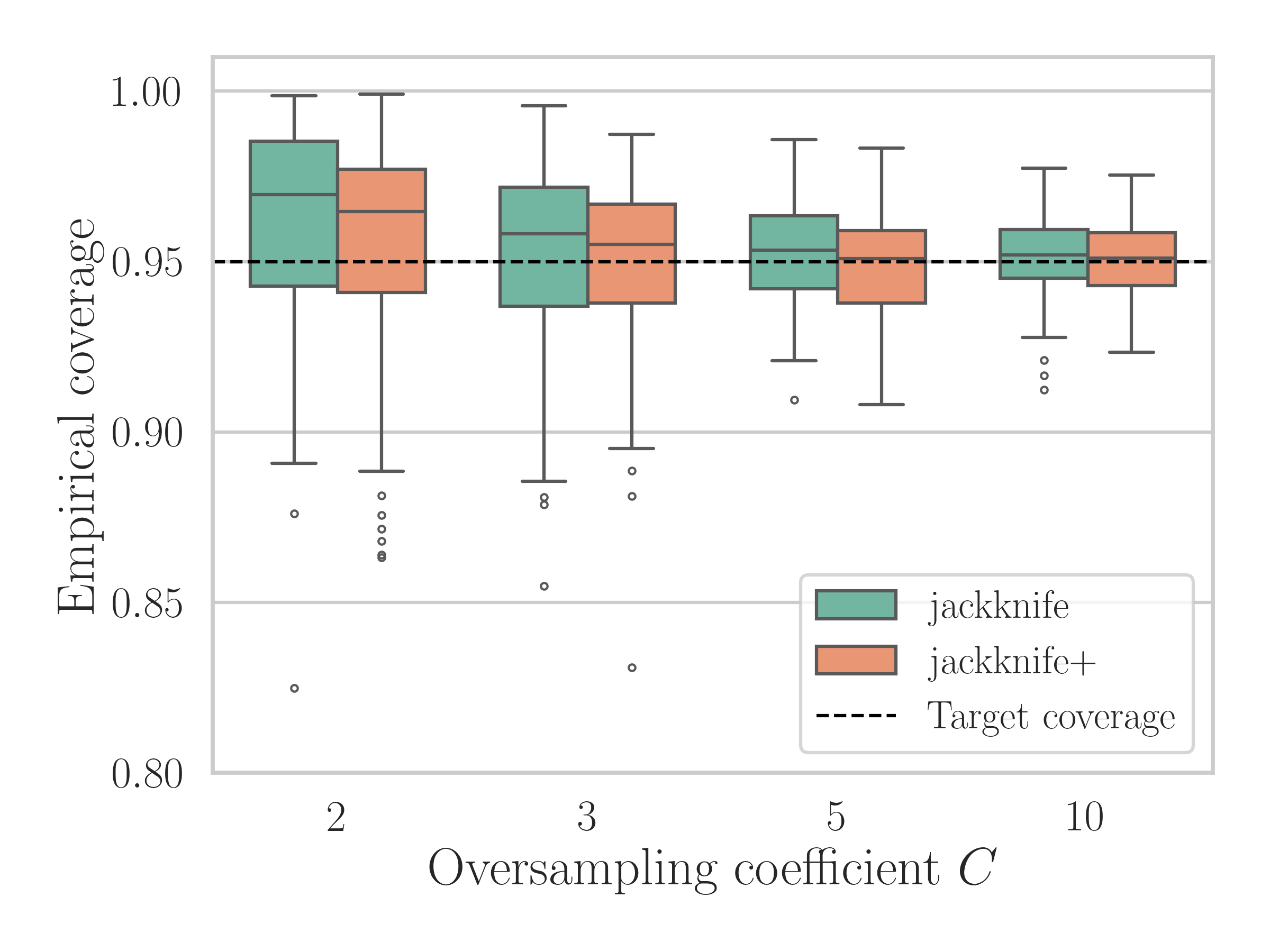}
    \caption{$P=2$.}
\end{subfigure}
\hfill 
\begin{subfigure}[b]{0.32\textwidth}
    \centering
    \includegraphics[width=\textwidth]{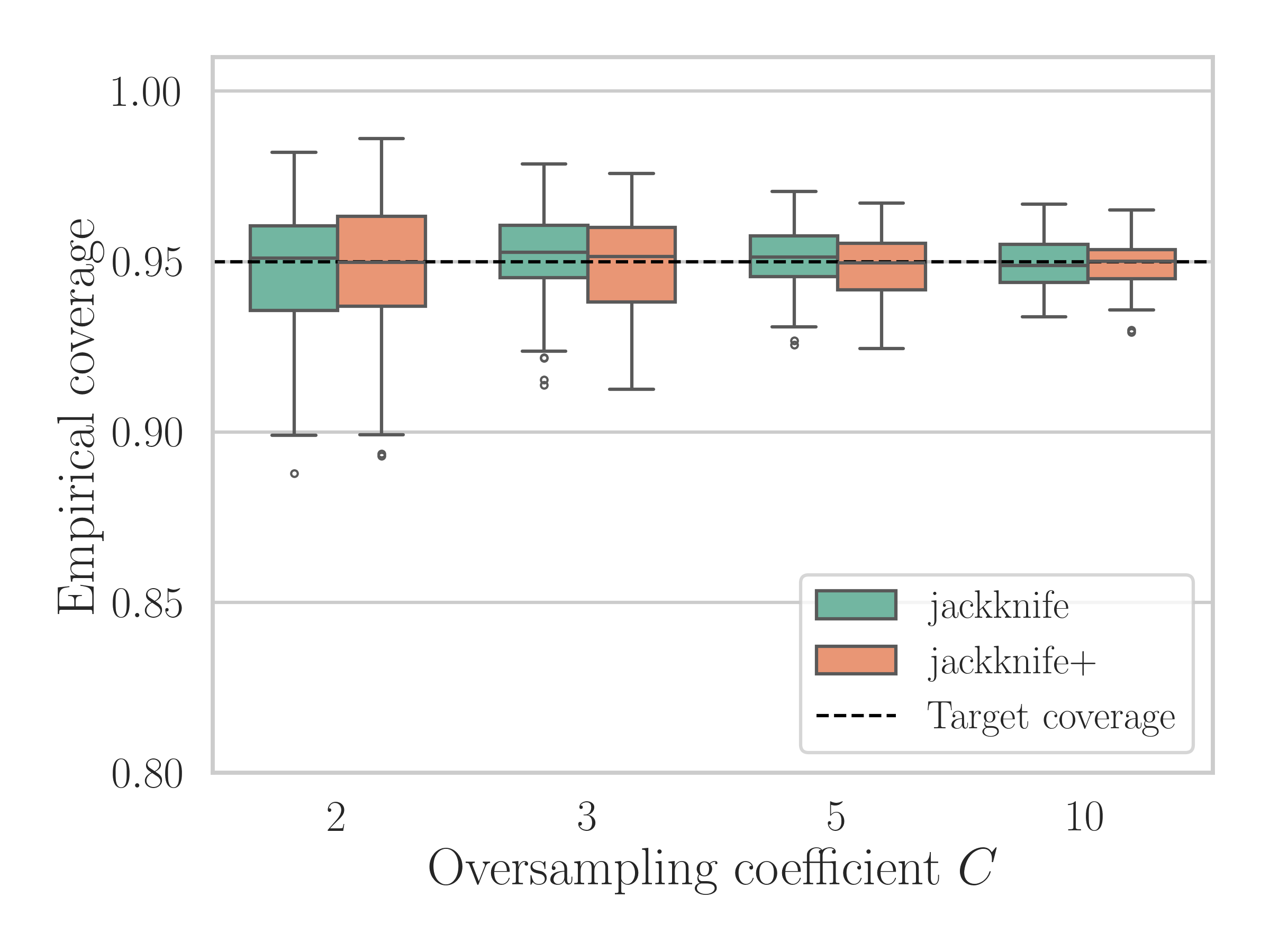}
    \caption{$P=3$.}
\end{subfigure}
\hfill
\begin{subfigure}[b]{0.32\textwidth}
    \centering
    \includegraphics[width=\textwidth]{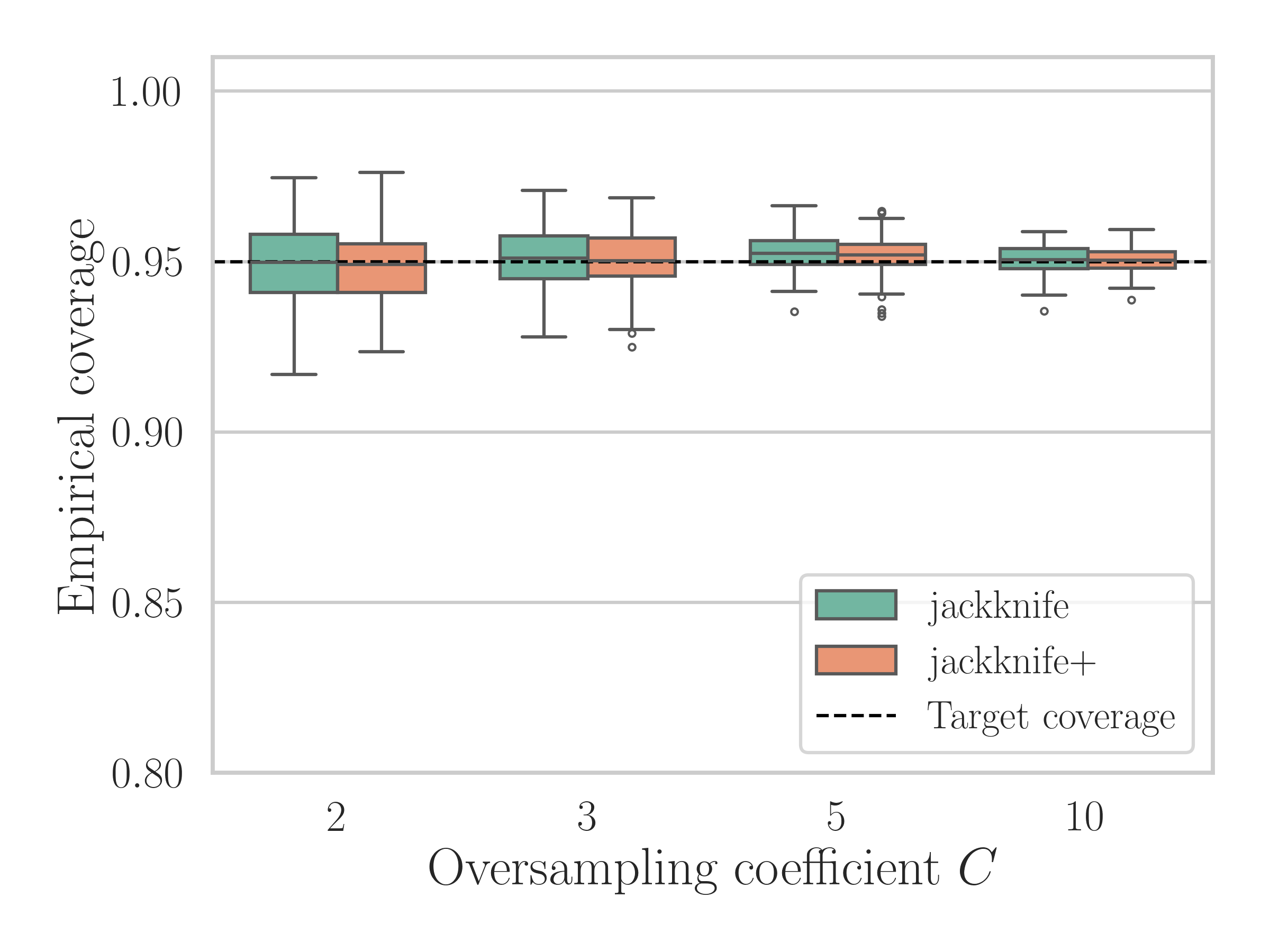}
    \caption{$P=4$.}
\end{subfigure}
\caption{Box plots of the empirical coverage provided by conformalized \gls{pce} surrogates of the piston simulation function, for different combinations of polynomial degree $P$, and oversampling coefficient $C$. \textcolor{black}{The target coverage is fixed at $95\%$, corresponding to a significance level of $s=0.05$.}}
\label{fig:piston-coverage-boxplots}
\end{figure}

\begin{figure}[t!]
\centering
\begin{subfigure}[b]{0.32\textwidth}
    \centering
    \includegraphics[width=\textwidth]{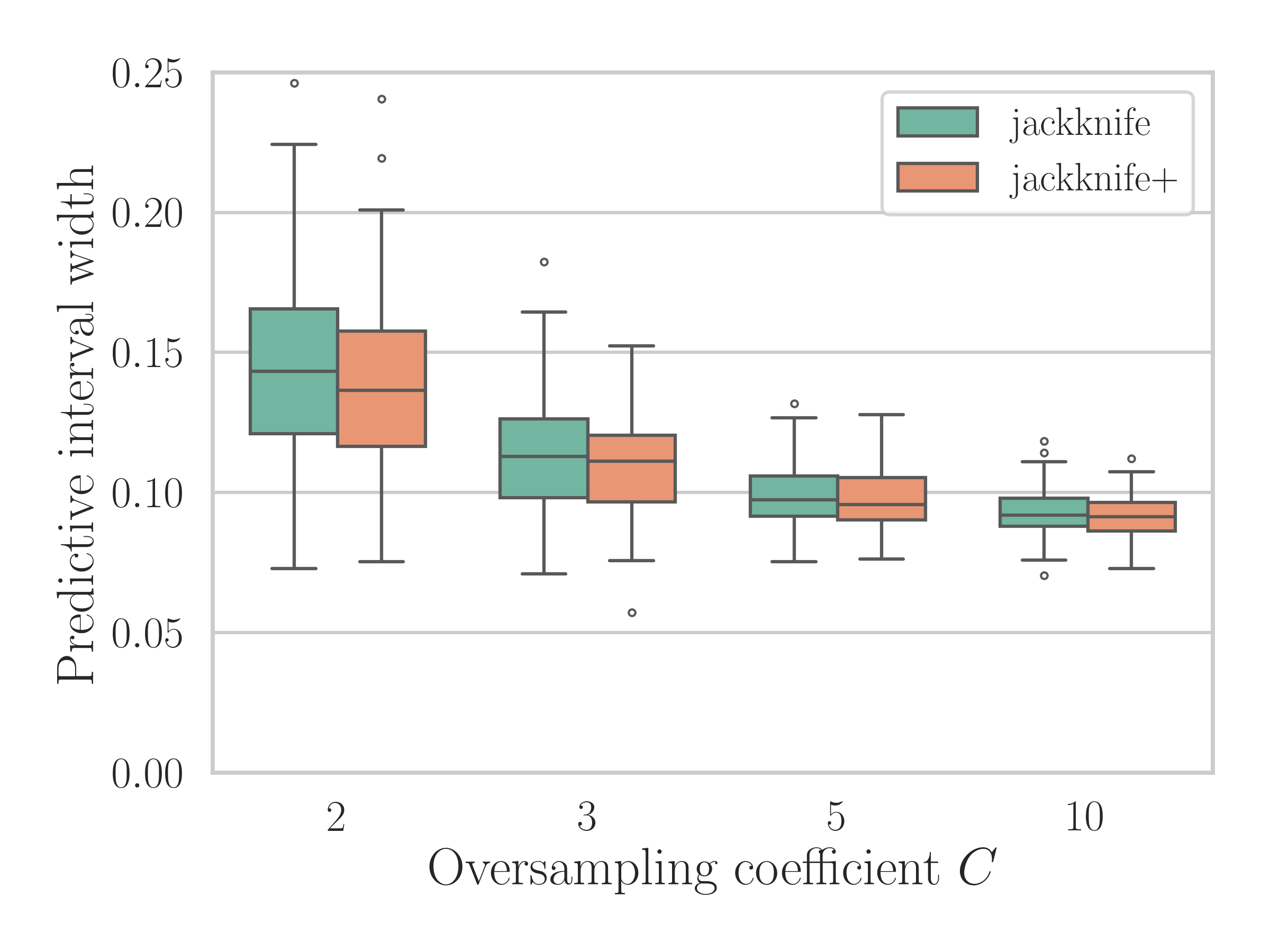}
    \caption{$P=2$.}
\end{subfigure}
\hfill
\begin{subfigure}[b]{0.32\textwidth}
    \centering
    \includegraphics[width=\textwidth]{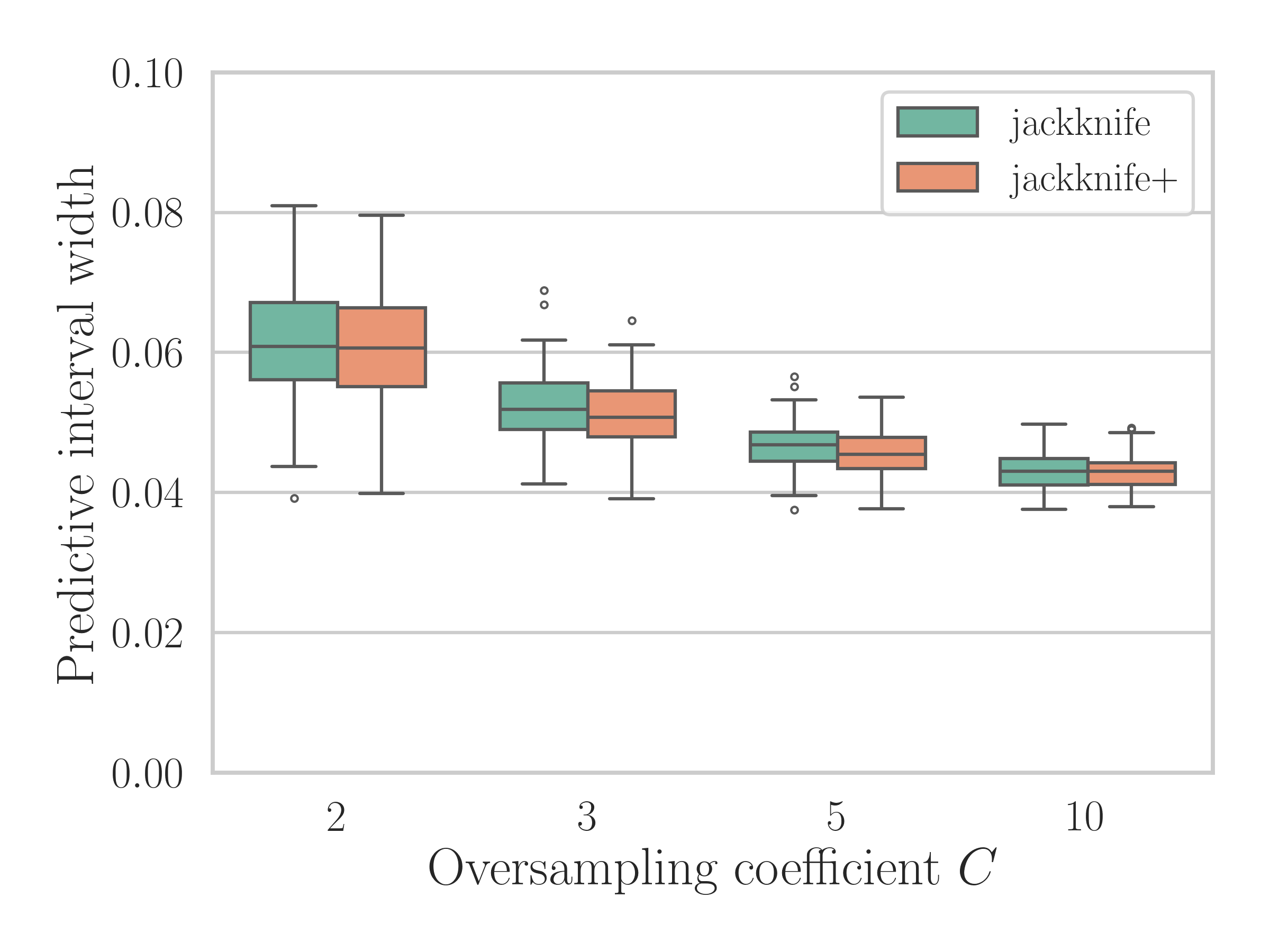}
    \caption{$P=3$.}
\end{subfigure}
\hfill
\begin{subfigure}[b]{0.32\textwidth}
    \centering
    \includegraphics[width=\textwidth]{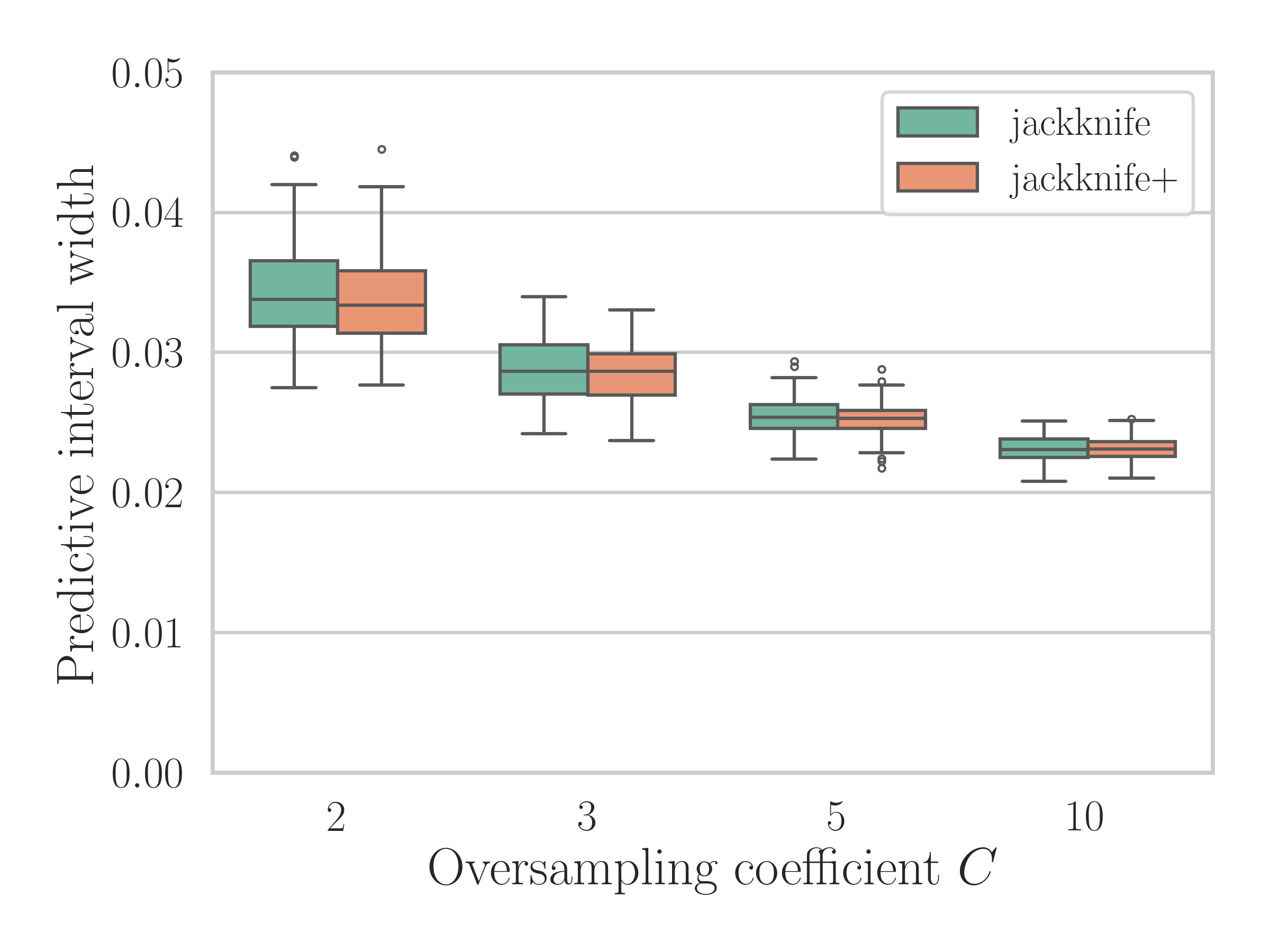}
    \caption{$P=4$.}
\end{subfigure}
\caption{Box plots of the prediction interval widths provided by conformalized \gls{pce} surrogates of the piston simulation function, for different combinations of polynomial degree $P$, and oversampling coefficient $C$. \textcolor{black}{The target coverage is fixed at $95\%$, corresponding to a significance level of $s=0.05$.}}
\label{fig:piston-interval-boxplots}
\end{figure}

Total-degree conformalized \glspl{pce} with maximum polynomial degrees $P \in \left\{2,3,4\right\}$ and oversampling coefficients $C \in \left\{2,3,5,10\right\}$ are employed. 
Similar to section~\ref{sec:otl-circuit}, the experimental design scales linearly with the size of the \gls{pce} basis, i.e., $M = C K = C \#\Lambda$.
Figure~\ref{fig:piston-parity-plots} compares the various configurations of the conformalized \gls{pce} against the true piston simulation function for a single random seed, where the results for $C=3$ have been omitted for better readability. 
Figures~\ref{fig:piston-coverage-boxplots} and \ref{fig:piston-interval-boxplots} show coverage and prediction interval statistics over 100 random seeds.

\end{document}